\RequirePackage[OT1]{fontenc}

\documentclass[journal]{IEEEtran}

\usepackage{siunitx}
\usepackage{hyperref}

\usepackage{cite}

\usepackage{graphicx}
\usepackage{framed}
\usepackage{tcolorbox}
\usepackage{float}

\usepackage{amsmath}
\usepackage{amssymb}
\usepackage{amsthm}
\usepackage{bm}
\usepackage{enumitem}

\usepackage{algorithmic}

\usepackage{stfloats}

\usepackage{url}

\usepackage{tikz}
\usepackage{textcomp}

\DeclareMathOperator*{\argmin}{arg\,min}
\DeclareMathOperator*{\argmax}{arg\,max}
\DeclareMathOperator{\mse}{mse}

\newtheorem{criterion}{Criterion}
\newtheorem{theorem}{Theorem}
\newtheorem{definition}{Definition}

\newcommand\copyrighttext{%
  \footnotesize \textcopyright 2021 IEEE. Personal use of this material is permitted.
  Permission from IEEE must be obtained for all other uses, in any current or future
  media, including reprinting/republishing this material for advertising or promotional
  purposes, creating new collective works, for resale or redistribution to servers or
  lists, or reuse of any copyrighted component of this work in other works.
  DOI: \href{https://doi.org/10.1109/TSP.2021.3061298}{10.1109/TSP.2021.3061298}
}

\newcommand\copyrightnotice{%
	\begin{tikzpicture}[remember picture,overlay]
		\node[anchor=south, yshift=3pt] at (current page.south) {\fbox{\parbox{\dimexpr\textwidth-\fboxsep-\fboxrule\relax}{\copyrighttext}}};
	\end{tikzpicture}%
}

\begin{document}
%%%%%%%%%%%%%%%%%%%%%%%%%%%%%%%%%%%%%%%%%%%%%%%%%%%%%%%%%%%%%%%%%%%%%%%%%%%%%%%%

\title{Minimax Robust Detection: \\ Classic Results and Recent Advances}

\author{Michael~Fau\ss{},~\IEEEmembership{Member,~IEEE,}
        Abdelhak~M.~Zoubir,~\IEEEmembership{Fellow,~IEEE,}
        and~H.~Vincent~Poor,~\IEEEmembership{Fellow,~IEEE}%
\thanks{M.~Fau\ss{} and H.~V.~Poor are with the Department of Electrical and Computer Engineering, Princeton University, Princeton, NJ, USA. E-mail: \{mfauss, poor\}@princeton.edu}% 
\thanks{Abdelhak~M.~Zoubir is with the Signal Processing Group, Technische Universit\"at Darmstadt, Darmstadt, Germany. E-mail: zoubir@spg.tu-darmstadt.de}%
\thanks{The work of M.~Fau{\ss} was supported by the German Research Foundation (DFG) under Grant 424522268.}%
\thanks{H.~V.~Poor gratefully acknowledges financial support from the Schmidt Data X Fund at Princeton University made possible through a major gift from the Schmidt Futures Foundation.}
}

\markboth{}{}

\maketitle

\copyrightnotice

%%%%%%%%%%%%%%%%%%%%%%%%%%%%%%%%%%%%%%%%%%%%%%%%%%%%%%%%%%%%%%%%%%%%%%%%%%%%%%%%
\begin{abstract}
  This paper provides an overview of results and concepts in minimax robust hypothesis testing for two and multiple hypotheses. It starts with an introduction to the subject, highlighting its connection to other areas of robust statistics and giving a brief recount of the most prominent developments. Subsequently, the minimax principle is introduced and its strengths and limitations are discussed. The first part of the paper focuses on the two-hypothesis case. After briefly reviewing the basics of statistical hypothesis testing, uncertainty sets are introduced as a generic way of modeling distributional uncertainty. The design of minimax detectors is then shown to reduce to the problem of determining a pair of least favorable distributions, and different criteria for their characterization are discussed. Explicit expressions are given for least favorable distributions under three types of uncertainty: $\varepsilon$-contamination, probability density bands, and $f$-divergence balls. Using examples, it is shown how the properties of these least favorable distributions translate to properties of the corresponding minimax detectors. The second part of the paper deals with the problem of robustly testing multiple hypotheses, starting with a discussion of why this is fundamentally different from the binary problem. Sequential detection is then introduced as a technique that enables the design of strictly minimax optimal tests in the multi-hypothesis case. Finally, the usefulness of robust detectors in practice is showcased using the example of ground penetrating radar. The paper concludes with an outlook on robust detection beyond the minimax principle and a brief summary of the presented material.
\end{abstract}
%%%%%%%%%%%%%%%%%%%%%%%%%%%%%%%%%%%%%%%%%%%%%%%%%%%%%%%%%%%%%%%%%%%%%%%%%%%%%%%%


%%%%%%%%%%%%%%%%%%%%%%%%%%%%%%%%%%%%%%%%%%%%%%%%%%%%%%%%%%%%%%%%%%%%%%%%%%%%%%%%
\section{Introduction}
\label{sec:introduction}
%%%%%%%%%%%%%%%%%%%%%%%%%%%%%%%%%%%%%%%%%%%%%%%%%%%%%%%%%%%%%%%%%%%%%%%%%%%%%%%%

\IEEEPARstart{A}{fundamental} task in signal processing, data science, and machine learning is to extract useful information from noisy data. In more and more applications, signal processing algorithms are being employed that have not been designed by experts, but whose behavior was learned exclusively from large data sets \cite{Vapnik1999, Bengio2013, Mohri2018}. In practice, learning from data often means choosing a (probabilistic) model such that the behavior of a system following this model is close to that of its real-world counterpart. Here ``choosing a model'' is meant in a wide sense and can range from simply fitting a distribution to training a deep neural network. In any case, the information contained in the original data set is condensed into a preferably small number of model parameters.

This approach is common practice and can provide excellent results if the training data adequately capture the dynamics of the underlying system. However, it has been shown to be susceptible to errors caused by unrepresentative samples, be it because of noisy measurements, corrupted or mislabeled entries in data bases, or simply an insufficient number of data points \cite{Zoubir2012, RussellDeweyTegmark2015, Fawzi2018, Zhang2019}. In general, every data set is subject to uncertainty about which aspects of it represent useful, generalizable information and which are spurious, random artifacts. As a consequence, there typically is a \emph{mismatch} between reality and the model that is fitted to the training data. 

Robust statistical signal processing provides the tools to deal with many commonly encountered types of model mismatch and distributional uncertainty in a systematic and rigorous manner. It offers methods and algorithms that do not rely on strict assumptions, but allow for deviations within well-defined boundaries. In this way, robust signal processing makes it possible to combine the efficiency of parametric, model-based methods with the reliability and flexibility of non-parametric, model-free methods. Therefore, robust statistics is often argued to provide a middle ground between both approaches \cite{Zoubir2012}. 

This paper deals with a particular area of robust statistics, namely robust detection and hypothesis testing.%
\footnote{
  Some authors use ``hypothesis testing'' to refer to the general problem of deciding from which population a sample was drawn and reserve ``detection'' for the problem of establishing the presence or absence of a signal in (additive) noise. Since this distinction is not always clear and often merely a matter of interpretation, both terms are used interchangeably in this paper. 
}% 
This topic tends to exist in the shadow of its bigger, more prominent sibling, robust estimation. One of the goals of this paper is to convey that this does not do robust detection justice, but that it is an interesting and useful area in its own right. Moreover, it has seen notable progress within recent years, including more flexible uncertainty models and novel results in robust multiple and sequential hypothesis testing. Taking into account that many readers might not be closely familiar with the subject, a thorough review of classic results is also provided.

%------------------------------------------------------------------------------%
\subsection{Robust Detection in the Context of Robust Statistics}
\label{ssec:context}
%------------------------------------------------------------------------------%
While robust estimation has long been an active area of research and comprehensive treatments of the subject have been released regularly since its inception in the early 1960s \cite{Andrews1972, Rey1978, Huber1981, Hampel1986, Staudte1990, Maronna2006, Zoubir2018}, robust detection has not received comparable attention. But what sets robust detection apart from robust estimation? Why does it require a separate treatment in the first place?

Conceptually speaking, robust estimation deals with the problem of inferring parameters or statistics of a distribution, such as its location or scale. The parameters are typically real-valued vectors so that there exist natural measures for ``how far off'' an estimate is. Commonly used examples are the square error, absolute deviation and other, often geometrically motivated, distances and norms. The robustness of an estimator can then be quantified by studying how sensitive to a model mismatch these accuracy measures are, that is, how large the (expected) distance between the estimated and true parameter value can grow under deviations from the nominal model. In principle, robust detection can be embedded in this framework by assigning each hypothesis a numeric value, which in turn can be \emph{estimated} from the data. However, in many detection problems, there is neither a natural choice for the numeric values associated with the hypotheses, nor a meaningful measure for the distance between them, in particular in the multi-hypothesis case. This makes it difficult, although not impossible \cite{Sadowsky1986, Barton1990, Toma2011}, to apply common techniques from robust estimation directly to robust detection.

Nevertheless, there do exist useful connections between the two areas. In particular, two concepts that emerge naturally in both robust detection and robust estimation are maximum likelihood (ML) or maximum \emph{a posteriori} probability (MAP) estimation and divergence measures between distributions. ML or MAP estimation offers an approach that does not require the notion of a distance between estimates, while divergence measures allow for replacing \emph{geometric} distances between parameters with \emph{statistical} distances between distributions. This connection will be made clearer and more explicit in the course of the paper.

In addition to robust estimation, there are many other areas of robust statistics that intersect in one form or another with robust detection. These include robust optimization and chance constrained programming \cite{CharnesCooper1959, BenTal2009, Gabrel2014, Rahimian2019}, robust decision making \cite{Marchau2019, Chamberlain2020}, robust control  \cite{Speyer1974, Whittle1990, ZhouDoyle1997, HansenSargent2001, HansenSargent2005, HansenSargent2007, SpeyerChung2008}, constrained Bayesian optimization \cite{Gelbart2015, HernandezLobato2016, Letham2019}, robust dynamic programming \cite{Iyengar2005, NilimGhaoui2005}, and imprecise probability theory \cite{Walley1991, Weichselberger2000, GuoTanaka2010, Bradley2019}, to name just a few. Doing all these topics justice is well beyond the scope of any single article, or even a book. However, in order to position robust detection in this vast landscape, it is often sufficient to keep just one fundamental definition in mind: robust detectors are insensitive to \emph{changes in the similarity of probability distributions with respect to each other}. This statement may sound trivial, but it summarizes some important characteristics of robust detection. First, it implies that \emph{multiple} distributions are subject to uncertainty. This is in contrast to many problems in robust statistics in which a \emph{single} distribution is subject to uncertainty, as is typically the case in robust estimation, robust control, or robust signaling. Second, robust detection is about the similarity of \emph{distributions}, not individual realizations or functions of realizations. This also implies that there is no intrinsic notion of ``good'' and ``bad'' observations or even outliers. The usefulness of an observation is entirely determined by the certainty with which one can state that it was drawn from one and not from another distribution. Finally, changing the emphasis, robust detection is about the \emph{similarity} of distributions, which is reflected in the fact that statistical distance or divergence measures arise naturally in the vast majority of problem formulations.

The notions of similarity, cost, distance, usefulness etc.\ will shortly be made explicit. Before going into the technicalities, however, it is worthwhile to have a look at the origins of minimax robust detection and its history. This will help to paint a broader picture of the subject, although in admittedly rough strokes, and to put more recent results into perspective

%------------------------------------------------------------------------------%
\subsection{A Brief Historical Account}
\label{ssec:history}
%------------------------------------------------------------------------------%
The birth of robust detection as a self-contained branch of robust statistics can be dated to a seminal paper by Huber \cite{Huber1965}, published in 1965. In \cite{Huber1965}, Huber showed that a statistical test based on a clipped version of the likelihood ratio is minimax optimal when the observations are contaminated by a fraction of arbitrary outliers. To the present day, this result is one of the most fundamental and influential findings in robust detection. Interestingly, Huber's paper would remain an ``outlier'' in the robust statistics literature until several years later.

In the early 1970s, the idea of robust detection was picked up by researchers and practitioners in various areas of applied statistics, with the information theory, communications, and signal processing communities arguably leading the way \cite{MartinSchwartz1971, MartinMcGath1974, KassamThomas1976, MillerThomas1976, El-SawyVandeLinde1977, MillerThomas1977, Kassam1982}. In particular, robust detectors were identified as a practical yet rigorous way of dealing with non-Gaussian and impulsive noise environments, a problem that remains relevant to the present day \cite{Zoubir2012, Shongwe2015}. While the earliest works in applied robust detection were variations of Huber's original problem and used the same outlier model, it soon became clear that it should be possible to design minimax detectors for a much larger class of uncertainty sets.

This conjecture was proved true in 1973, when Huber and Strassen showed that a sufficient and, in a certain sense, also necessary condition for the existence of minimax robust detectors is the existence of least favorable distributions that attain a 2-alternating Choquet capacity \cite{Choquet1954} over the uncertainty sets of feasible distributions \cite{HuberStrassen1973}. This result, which will be discussed in more detail in Section~\ref{ssec:characterizing_lfds}, was significant for several reasons. First, it showed that the problem of designing a minimax robust detector can, under mild assumptions, be reduced to finding a pair of least favorable distributions. Second, it provided a characterization of least favorable distributions that is independent of a particular cost function, but only depends on the uncertainty sets. Finally, the characterization via Choquet capacities established a strong connection between robust detection and convex divergence measures. In fact, recent results on this connection have partially motivated this overview paper.

In the years after the publication of \cite{HuberStrassen1973}, a substantial number of notable papers had been written on the form and characteristics of least favorable distributions under various types of uncertainty. Besides the classic outlier models, these included Prokhorov neighborhoods \cite{Hafner1982}, density band models \cite{Kassam1981, Hafner1993}, p-point classes \cite{El-SawyVandeLinde1977, VastolaPoor1984}, and many more \cite{Rieder1977, Oesterreicher1978, Bednarski1981}. In fact, the period between the late 1970s and the early 1990s can be considered the most prolific in the history of robust detection. In addition to deepening the understanding of the classic minimax test, the scope of robust detection also widened. Topics that came into focus include, for example, robust detection for dependent observations \cite{Poor1982, Sadowsky1986, Moustakides1987}, robust sequential detection \cite{El-SawyVandeLinde1979, Dwyer1980}, robust distributed detection \cite{GeraniotisChau1990}, robust filtering \cite{Kassam1977, Poor1980_Wiener, Whittle1981, Poor1983, VerduPoor1983, VerduPoor1983_com, Boel2002}, robust quantization \cite{Poor1978}, and alternatives to the minimax approach, such as locally robust detection \cite{Kassam1978, Poor1980, MoustakidesThomas1984, GerlachSangston1993, Brown2000} and robust detection based on robust estimators \cite{El-SawyVandeLinde1977}, distance criteria \cite{Poor1980_Distance}, or extreme-value theory \cite{Milstein1969}. It was also the time when the first systematic reviews of the growing body of literature on the subject were published, most notably the surveys by Poor and Kassam \cite{Kassam1983, KassamPoor1985, Poor1986, Poor1987}. However, to the best of our knowledge, there has been no survey article covering robust detection in a signal processing context since then.

In recent years, the trend of diversification in robust detection has continued, with new topics emerging, such as robust change detection \cite{Unnikrishnan2011, Dasu2011, Molloy2017, MolloyKennedy2017}, robust detection of adversarial attacks \cite{Zheng2018RobustDO, Grosse2017OnT}, and robust Bayesian filtering \cite{LevyNikoukhah2004, LevyNikoukhah2013, Zorzi2017, Zorzi2017b}. Also, new combinations of existing problems have been investigated, such as robust distributed sequential detection \cite{HouLeonard2017, Leonard2018} and robust joint detection and estimation \cite{Reinhard2016}. Some of these advanced topics will be picked up in later sections, after the necessary foundations have been introduced. However, giving a comprehensive overview of all flavors and areas of application of robust detection is not the main goal of this paper.

%%%%%%%%%%%%%%%%%%%%%%%%%%%%%%%%%%%%%%%%%%%%%%%%%%%%%%%%%%%%%%%%%%%%%%%%%%%%%%%%
\section{Scope and Outline of the Paper}
\label{sec:scope}
%%%%%%%%%%%%%%%%%%%%%%%%%%%%%%%%%%%%%%%%%%%%%%%%%%%%%%%%%%%%%%%%%%%%%%%%%%%%%%%%

As the title suggests, the objective of this paper is two-fold. On the one hand, it is supposed to give a self-contained, in-depth overview of the fundamental concepts underlying minimax robust detection. Most of these can of course be found in the literature, however, they are scattered over different books and articles, with arguably none of them proving a complete and transparent picture. With this paper, our aim is to aggregate these sources into a coherent, tutorial style treatment that can serve as a unified reference and as a starting point for future researchers interested in the topic. 

On the other hand, the second goal of this paper is to present a recent line of research that builds on and generalizes classic results. It is based on the aforementioned fundamental connection between minimax robust detectors and convex similarity measures. This connection turned out to be fruitful in terms of both a more comprehensive theory of robust detectors as well as more flexible algorithms for their design. In particular, recent results include

\begin{itemize}
	\item an alternative criterion for the characterization of least favorable distributions, which is typically easier to evaluate than the existing criteria by Huber and Strassen;
	\item a unified approach to the construction of least favorable distributions for several uncertainty sets that have so far been considered separately in the literature;
	\item an extension of minimax detectors to uncertainty sets for which no capacity achieving distributions in the sense of Huber and Strassen exist;
	\item useful insights into what characteristics robust detectors for \emph{multiple} hypotheses need to admit and why their design is significantly harder or even impossible;
	\item a proof of existence and insights into the working-principles of strictly minimax \emph{sequential} detectors, whose existence had not been established rigorously before;
	\item efficient algorithms for the design of the latter;
	\item a promising route towards a more unified theory of robust and sequential detection for two and multiple hypotheses;
	\item in extension, a possible route towards a more unified theory of robust detection and estimation in general.
\end{itemize}

Most of these findings have been published as standalone technical papers \cite{Fauss2015, Fauss2016_thesis, Fauss2016_old_bands, Gul2015_composite_distances, Gul2016_alpha_divergence, Gul2017_minimax_robust, Gul2017_decentralized, Fauss2017_ISI, Fauss2018_tsp, Fauss2019_aos, Fauss2018_icassp} and the reader will be referred to these for details. The additional value provided by this overview paper is a coherent presentation, with a strong emphasize on conceptual insights, as well as a more comprehensive discussion. Ultimately, we are convinced that a solid understanding of the concepts presented here equips the reader with powerful tools that will remain useful and relevant in robust signal processing, communications, and data science for the foreseeable future.

The remainder of the paper is structured as follows: 

\begin{description}
  \item[Section~\ref{sec:minimax_principle}] introduces the minimax principle as a design approach to robust detectors and discusses its implications, limitations, and areas of applicability.
  \item[Section~\ref{sec:two_hypotheses_optimal}] revisits optimal detectors for two hypotheses, introduces common jargon, and fixes some notations.
  \item[Section~\ref{sec:two_hypotheses}] is the main section of the paper. It covers how distributional uncertainty is modeled via uncertainty sets of feasible distributions, how the least favorable among these distributions can be identified, and introduces three types of uncertainty sets for which the latter are guaranteed to exist and can be calculated efficiently. In addition, it presents tangible examples and discusses some intricacies of robust detectors, such as the need for randomized decision rules.
  \item[Section~\ref{sec:multiple_hypotheses}] enters the more advanced topic of robustly testing \emph{multiple} hypotheses and shows why this problem cannot be solved in analogy to the two-hypothesis case.
  \item[Section~\ref{sec:sequential_detection}] identifies minimax robust \emph{sequential} detectors as a useful generalization of regular detectors, offering an increased efficiency on the one hand, and enabling the design of minimax optimal tests for multiple hypotheses on the other hand. However, this comes at the expense of a more complex design and, in a sense, a violation of the ideas underlying the minimax principle.
  \item[Section~\ref{sec:application}] uses the example of ground penetrating radar (GPR) to illustrate how robust detectors can improve the performance of real-world systems while introducing little to no extra costs and complexity.
  \item[Section~\ref{sec:beyond_minimax}] gives an outlook on the future of robust detection. In particular, it argues that for future applications a more unified framework of robust detection and estimation will be required that goes beyond the traditional minimax approach, yet is informed by the underlying concepts and insights. 
  \item[Section~\ref{sec:conclusion}] summarizes and concludes the paper.
\end{description}

%%%%%%%%%%%%%%%%%%%%%%%%%%%%%%%%%%%%%%%%%%%%%%%%%%%%%%%%%%%%%%%%%%%%%%%%%%%%%%%%
\section{The Minimax Principle}
\label{sec:minimax_principle}
%%%%%%%%%%%%%%%%%%%%%%%%%%%%%%%%%%%%%%%%%%%%%%%%%%%%%%%%%%%%%%%%%%%%%%%%%%%%%%%%

The discussion in this paper is limited to the  \emph{minimax} approach \cite{VerduPoor1984, Minimax1995}. That is, the worst-case performance over a given set of feasible scenarios is used as an objective function. This section provides a conceptual introduction to the minimax principle, highlights its strengths and limitations, and compares it to alternative ways of handling model uncertainty. 

\begin{figure}
	\centering
	\includegraphics{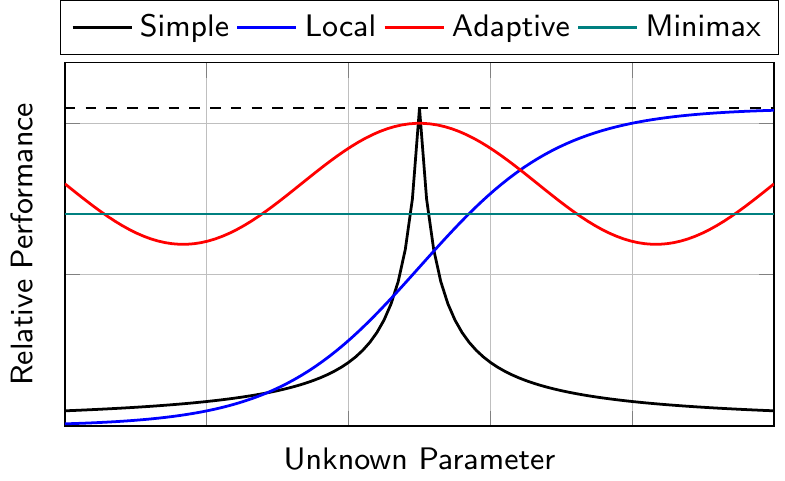}
	\caption{Typical performance profiles of common approaches to dealing with distributional uncertainty, here in form of an unknown parameter. The dashed line indicates the optimal performance when the true parameter value is known, that is, when there is no uncertainty.}
	\label{fig:minimax_illustration}
\end{figure}

As an introductory example, consider the problem of detecting a deterministic signal in additive noise with an unknown shape parameter. A qualitative illustration of different ways of dealing with this uncertainty is given in Fig.~\ref{fig:minimax_illustration}. The simplest approach is to \emph{ignore} the uncertainty and to assume that the noise distribution is known and fixed. A common example is a detector that is designed under the assumption of normally distributed noise. This results in a procedure that is optimal at the assumed parameter value, but whose performance can deteriorate rapidly when the true noise distribution starts to differ from the assumed one. A more practical approach is to design the detector such that it works well over a certain range of parameter values. A typical example is a locally optimal detector \cite{Poor1994, Song2002}, which achieves close to optimal performance at the assumed parameter value, while also performing well in a neighborhood around it. Other common examples are methods based on low or high signal-to-noise ratio (SNR) assumptions \cite{TandraSahai2005, Matthaiou2013}, which perform close to optimal only in the respective SNR regime. 

Approaches that aim to provide good performance over the \emph{entire} uncertainty set typically achieve this by adapting to the true scenario, that is, by estimating the unknown parameters and thereby \emph{reducing} the uncertainty \cite{Kelly1986, Poor1994}. In the context of detection, commonly used examples are the generalized likelihood ratio test \cite{VanTrees1968, Zeitouni1992} or Bayesian detectors \cite{Poor1994, Candy2016}. A downside of adaptive techniques is that they are often difficult to analyze so that performance guarantees are only available in the asymptotic regime \cite{LeangJohnson1997, Shapiro2009}. Moreover, their implementation, in particular that of Bayesian detectors, can be prohibitively complex. The minimax approach also aims to provide good performance over the entire uncertainty set, however, it is stricter in the sense that a minimax procedure \emph{guarantees} a certain performance, \emph{independent} of the true value of the unknown parameter. In order to achieve this independence, a minimax detector does not \emph{reduce} the uncertainty, but \emph{tolerates} it. That is, it is designed \emph{a priori} such that it works sufficiently well under all feasible scenarios. As a consequence, minimax procedures often turn out to be \emph{equalizers} over the uncertainty set, meaning that the performance is (almost) identical under all feasible distributions. This is indicated by the flat performance profile in Fig.~\ref{fig:minimax_illustration}.

In general, a minimax optimal solution consists of two ingredients: a least favorable scenario among all feasible ones and a procedure that is optimal under this scenario \cite{VerduPoor1984}. In robust detection, the scenario is determined by the distributions under each hypothesis, so that a minimax detector reduces to an optimal detector for the least favorable distributions. Compared to Bayesian or adaptive detectors, this has the advantage that a minimax detector is of the same form and complexity as a simple detector, only that it is designed under very specific worst-case assumptions. Owing to this property, minimax detectors can often be used as drop-in replacements for standard, non-robust methods.

Naturally, the minimax paradigm of tolerating uncertainty instead of reducing it is not always appropriate. In fact, the minimax design objective is frequently criticized for being ``too pessimistic'' or ``too conservative''. This criticism is sometimes justified, but just as often based on misconceptions or overgeneralizations. Whether or not a robust detector should be used mainly depends on the \emph{type of uncertainty}: if the unknown parameters can be estimated reliably and efficiently, the additional information used by adaptive procedures usually make them the method of choice. However, in cases where no efficient estimators exist, the environment changes too rapidly to estimate the unknown parameters, or entire distributions are subject to uncertainty---the latter being the focus of this paper---minimax optimal detectors are an attractive alternative that offers high reliability at comparatively low-complexity. Some common questions and reservations concerning the applicability of minimax robust detectors will also be addressed in Sec.~\ref{ssec:conclusions}, in the form of an FAQ, after having discussed the fundamentals of minimax robust detection for two hypotheses.

In general, the type of uncertainty that can be handled by robust detectors needs to be such that the set of feasible distributions is ``large'' enough to contain all or most distributions of interest, yet ``small'' enough to contain a meaningful worst-case scenario. If the uncertainty sets are too small, they do not guarantee robustness; if they are chosen too large, the worst-case is indeed too pessimistic and the minimax detector stops working well under regular conditions. The latter effect is known as \emph{over robustification} and will be further illustrated in Sec.~\ref{ssec:design_and_implementation} and Sec.~\ref{sec:application}. Three non-parametric uncertainty models that offer a good trade-off between robustness and nominal performance will be revised in Sec.~\ref{ssec:uncertainty_models}.

\emph{Defining} the least favorable distributions among the feasible ones is rather straightforward: they are those that minimize a suitably chosen performance metric. The main difficulty in designing minimax detectors lies in \emph{identifying and characterizing} these least favorable distributions. For this purpose, two qualitative, intuition based properties of the latter are useful to keep in mind throughout the remainder of the paper. First, least favorable distributions are such that the probability of confusing them for each other, which corresponds to the error probabilities of the underlying test, is maximized. Second, least favorable distributions are maximally similar in the sense that they minimize an appropriately chosen divergence or distance measure. It is not difficult to see that these two properties are closely related. Interestingly, while the characterization via maximum error probabilities might appear to be more natural at first, the characterization based on divergence measures turns out to provide some additional insights and to be more easily generalizable to sequential tests and tests for multiple hypotheses.

%%%%%%%%%%%%%%%%%%%%%%%%%%%%%%%%%%%%%%%%%%%%%%%%%%%%%%%%%%%%%%%%%%%%%%%%%%%%%%%%
\section{Optimal Tests for Two Hypotheses}
\label{sec:two_hypotheses_optimal}
%%%%%%%%%%%%%%%%%%%%%%%%%%%%%%%%%%%%%%%%%%%%%%%%%%%%%%%%%%%%%%%%%%%%%%%%%%%%%%%%

Before entering the subject of minimax detection, it is useful to briefly revise optimal non-robust detectors in order to introduce basic concepts as well as jargon and notation. Let $\bm{X}_N = (X_1, \ldots, X_N)$ be a sequence of random variables defined on a common sample space $(\mathcal{X}, \mathcal{F})$. In what follows, the joint distribution of $\bm{X}_N$ is dented by $\mathbb{P}_{\bm{X}_N}$ and the distribution of the individual $X_n$ by $P_{X_n}$. Assume for now that $X_1, \ldots, X_N$ are independent and identically distributed (i.i.d.) according to a distribution $P_X$, that is
\begin{equation}
  \mathbb{P}_{\bm{X}_N} = \prod_{n=1}^N P_{X_n} = P_{X}^N.
\end{equation}
The goal of a binary detector is to decide between the two hypotheses
\begin{equation}
  \begin{aligned}
    \mathcal{H}_0\colon & \quad P_X = P_0, \\
    \mathcal{H}_1\colon & \quad P_X = P_1,
  \end{aligned}
  \label{eq:simple_hypotheses}
\end{equation}
where $P_0$ and $P_1$ are two given distributions and $\mathcal{H}_0$ and $\mathcal{H}_1$ are referred to as the \emph{null} and \emph{alternative} hypothesis, respectively. Both $P_0$ and $P_1$ are assumed to admit probability density functions (PDFs) $p_0$ and $p_1$, respectively.%
\footnote{
  Note that this assumption is not restrictive since there always exists a reference measure $\mu$, for example $\mu = P_0 + P_1$, such that both $P_0$ and $P_1$ are absolutely continuous with respect to $\mu$. Usually, as here, the standard Lebesgue measure for continuous distributions is the reference measure of interest yielding PDFs. Or, in the case of discrete distributions, the counting measure is the reference measures of interest yielding probability mass function (PMFs) \cite{BauerBurckel2011}.
}
A statistical test for $\mathcal{H}_0$ against $\mathcal{H}_1$ is defined by a randomized decision rule
\begin{equation*}
  \delta\colon \mathcal{X}^N \to [0,1],
\end{equation*}
where $\delta = \delta(\bm{x}_N)$ denotes the conditional probability of deciding for the alternative hypothesis, given the observations $\bm{x}_N = (x_1, \ldots, x_N)$. 

In practice, the effect of randomization on the performance of the detector is often negligible. It is introduced here for two reasons. First, it is helpful from a technical point of view since the set of randomized decision rules can be shown to be convex, while the set of non-randomized decision rules is not. Second, more importantly, randomization can indeed play a crucial role for robust detectors, especially in the small sample size regime. This aspect is discussed in more detail in Section~\ref{ssec:design_and_implementation}.

Given a decision rule $\delta$, the corresponding type I and type II error probabilities are given by
\begin{align}
  \mathrm{Pr}(\text{``reject } \mathcal{H}_0 \text{ when it is true''}) &= \mathbb{E}_{\mathbb{P}_0}[\,\delta(\bm{X}_N)\,], \\
  \mathrm{Pr}(\text{``reject } \mathcal{H}_1 \text{ when it is true''}) &= \mathbb{E}_{\mathbb{P}_1}[1-\delta(\bm{X}_N)],
\end{align}
where $\mathbb{E}_{\mathbb{P}_0}$ and $\mathbb{E}_{\mathbb{P}_1}$ denote the expected value taken with respect to distributions $\mathbb{P}_0 = P_0^N$ and $\mathbb{P}_1 = P_1^N$, respectively. Based on the error probabilities, different cost functions can be formulated to quantify the performance of a detector. Three common examples are defined below.

\begin{definition}Common cost functions in (robust) detection: 
  \begin{enumerate}
    \item Weighted sum error cost
          \begin{multline}
            C_\text{WSE}(\delta; \mathbb{P}_0, \mathbb{P}_1) =  \mathbb{E}_{\mathbb{P}_0}[\,\delta(\bm{X}_N)\,] \\ + \eta \, \mathbb{E}_{\mathbb{P}_1}[1- \delta(\bm{X}_N)],
            \label{eq:weighted_sum_error}
          \end{multline}
          where $\eta$ denotes a positive cost coefficient;
    \item Bayes error cost
          \begin{multline}
            C_\text{BE}(\delta; \mathbb{P}_0, \mathbb{P}_1) = \mathrm{Pr}(\mathcal{H}_0) \mathbb{E}_{\mathbb{P}_0}[\,\delta(\bm{X}_N)\,] \\ + \mathrm{Pr}(\mathcal{H}_1) \mathbb{E}_{\mathbb{P}_1}[1- \delta(\bm{X}_N)],
          \end{multline}
          where $\mathrm{Pr}(\mathcal{H}_0)$ and $\mathrm{Pr}(\mathcal{H}_1)$ denote the prior probabilities of the hypotheses, and
    \item Neyman--Pearson cost
          \begin{equation}
            C_\text{NP}(\delta; \mathbb{P}_0, \mathbb{P}_1) = \mathbb{E}_{\mathbb{P}_1}[1- \delta(\bm{X}_N)],
          \end{equation}
          where $\delta$ is required to satisfy a constraint on the type I error probability
          \begin{equation}
            \mathbb{E}_{\mathbb{P}_0}[\,\delta(\bm{X}_N)\,] \leq \alpha_0,
            \label{eq:np_constraint}
          \end{equation}
          with $\alpha_0$ being a preset level.
  \end{enumerate}
  \label{def:costs}
\end{definition}

The optimal decision rule is then defined as the one that minimizes the cost $C$ for a given pair of distributions $(P_0, P_1)$, that is,
\begin{equation}
  \delta^* \in \argmin_\delta \; C(\delta; \mathbb{P}_0, \mathbb{P}_1).
  \label{eq:delta_opt}
\end{equation}
It is well known that the three cost functions introduced above all lead to the same optimal decision rule $\delta^*$, namely, the so-called \emph{likelihood ratio test} \cite{Poor1994, Kay1998, Lehmann2005, Levy2008}. 

\begin{theorem}
  The decsion rule
  \begin{equation}
    \delta^*(\bm{x}_N) \begin{cases}
                        = 1,        & z(\bm{x}_N) > \lambda \\
                        \in [0, 1], & z(\bm{x}_N) = \lambda \\
                        = 0,        & z(\bm{x}_N) < \lambda
                       \end{cases}
    \label{eq:lr_test}
  \end{equation}
  where $z \colon \mathcal{X}^N \to \mathbb{R}_+$ denotes the
  likelihood ratio
  \begin{equation}
    z(\bm{x}_N) = \frac{\mathrm{d} \mathbb{P}_1}{\mathrm{d} \mathbb{P}_0}(\bm{x}_N) = \prod_{n=1}^N\frac{p_1(x_n)}{p_0(x_n)}
  \end{equation}
  and $\lambda > 0$ denotes the detection threshold, is optimal in the sense of the weighted sum error, with $\lambda = 1/\eta$, the Bayes error, with $\lambda = \mathrm{Pr}(\mathcal{H}_0)/\mathrm{Pr}(\mathcal{H}_1)$, and the Neyman--Pearson error, with $\lambda$ chosen such that the constraint in \eqref{eq:np_constraint} is satisfied with equality.
  \label{th:optimal_test}
\end{theorem}

This strong optimality property of the likelihood ratio test extends to the minimax case in a natural manner, namely, by replacing the likelihood ratio of the nominal distributions with that of the least favorable distributions. However, the question how to define, characterize and calculate least favorable distributions is non-trivial. For the two-hypothesis case, it is answered in the following section.

%%%%%%%%%%%%%%%%%%%%%%%%%%%%%%%%%%%%%%%%%%%%%%%%%%%%%%%%%%%%%%%%%%%%%%%%%%%%%%%%
\section{Minimax Tests for Two Hypotheses}
\label{sec:two_hypotheses}
%%%%%%%%%%%%%%%%%%%%%%%%%%%%%%%%%%%%%%%%%%%%%%%%%%%%%%%%%%%%%%%%%%%%%%%%%%%%%%%%

The idea underlying robust detection is to relax the assumption that the distributions $P_0$ and $P_1$ in \eqref{eq:simple_hypotheses} are known exactly without giving up the benefits of model based techniques entirely. This is achieved by allowing the true distributions to lie within a neighborhood around the nominal distributions $P_0$ and $P_1$. Mathematically, this means replacing the simple hypotheses in \eqref{eq:simple_hypotheses} with composite hypotheses of the form
\begin{equation}
  \begin{aligned}
    \mathcal{H}_0 &\colon P_{X_n} \in \mathcal{P}_0, \\
    \mathcal{H}_1 &\colon P_{X_n} \in \mathcal{P}_1,
  \end{aligned}
  \label{eq:composite_hypotheses}
\end{equation}
for all $n = 1, \ldots, N$. Here $\mathcal{P}_0$ and $\mathcal{P}_1$ denote the sets of feasible distributions under the respective hypothesis and are usually referred to as \emph{uncertainty} or \emph{ambiguity} sets.\footnote{
  The vast majority of problems in minimax detection are of the form \eqref{eq:composite_hypotheses}, but sometimes additional complications are considered. For example, the uncertainty set can be defined \emph{jointly} for $P_0$ and $P_1$, that is, $(P_0, P_1) \in \mathcal{P}$. This means that the uncertainty under each hypothesis is allowed to depend on the true distribution under the other hypothesis. Another possible complication is to allow observations under one hypothesis to contain information about the distribution under the other hypothesis. However, both cases are non-standard and are not covered in this overview paper.
}  
The exact form of these sets is intentionally left unspecified at this point, but will be of importance later on. For now, it suffices to think of $\mathcal{P}_0$ and $\mathcal{P}_1$ as any two \emph{disjoint} sets of distributions, be it parametric or non-parametric. If $\mathcal{P}_0$ and $\mathcal{P}_1$ intersect, the two hypotheses in \eqref{eq:composite_hypotheses} become indistinguishable in the minimax sense since there exist distributions for which both hypotheses are true; this effect is comparable to the breakdown of a robust estimator and will be revisited in Sec.~\ref{ssec:calculating_lfds}. For the sake of a more compact notation, the hypotheses in \eqref{eq:composite_hypotheses} are also written as
\begin{equation}
  \begin{aligned}
    \mathcal{H}_0 &\colon \mathbb{P}_{\bm{X}_N} \in \mathcal{P}_0, \\
    \mathcal{H}_1 &\colon \mathbb{P}_{\bm{X}_N} \in \mathcal{P}_1.
  \end{aligned}
  \label{eq:composite_hypotheses_joint}
\end{equation}

Note that in \eqref{eq:composite_hypotheses} and \eqref{eq:composite_hypotheses_joint} the random variables $X_1, \ldots, X_N$ are no longer assumed to be identically distributed. In fact, they do not even need to be independent as long as the dependencies between them are such that their \emph{conditional} distributions remain within the uncertainty sets. In this manner, uncertainty about the dependencies between the random variables can be absorbed in the distributional uncertainty. However, for strong dependencies this approach is rarely practical, since it can inflate the uncertainty sets to the point where they are no longer useful. In such cases, it is better to directly formulate the composite hypotheses in terms of the conditional distributions. This problem will be picked up again in Section~\ref{sec:sequential_detection}, but a detailed technical discussion of minimax detection for dependent data is beyond the scope of this paper. 

The minimax optimal decision rule is defined as the one that minimizes the \emph{maximum} cost $C$, where the maximum is taken over all pairs of feasible distributions $(\mathbb{P}_0, \mathbb{P}_1)$, that is,
\begin{equation}
  \delta^* \in \argmin_\delta \; \max_{(\mathbb{P}_0, \mathbb{P}_1) \in \mathcal{P}_0 \times \mathcal{P}_1} \; C(\delta; \mathbb{P}_0, \mathbb{P}_1).
\end{equation}
A necessary and sufficient condition for minimax optimally is that $(\delta^*; \mathbb{Q}_0, \mathbb{Q}_1)$ satisfies the saddle point condition \cite{VerduPoor1984, Levy2008}, that is,
\begin{equation}
  C(\delta^*; \mathbb{P}_0, \mathbb{P}_0) \leq  C(\delta^*; \mathbb{Q}_0, \mathbb{Q}_1) \leq C(\delta; \mathbb{Q}_0, \mathbb{Q}_1)
\end{equation}
for all $(\mathbb{P}_0, \mathbb{P}_1) \in \mathcal{P}_0 \times \mathcal{P}_1$. From the previous section it is clear that once the least favorable distributions are fixed, the optimal decision rule is simply a likelihood ratio test of the form \eqref{eq:lr_test}, with $\mathbb{P}_0$ and $\mathbb{P}_1$ replaced by $\mathbb{Q}_0$ and $\mathbb{Q}_1$. The question of how to determine the latter will accompany us throughout the remainder of the paper.

%%%%%%%%%%%%%%%%%%%%%%%%%%%%%%%%%%%%%%%%%%%%%%%%%%%%%%%%%%%%%%%%%%%%%%%%%%%%%%%%
\subsection{Characterizing Least Favorable Distributions}
\label{ssec:characterizing_lfds}
%%%%%%%%%%%%%%%%%%%%%%%%%%%%%%%%%%%%%%%%%%%%%%%%%%%%%%%%%%%%%%%%%%%%%%%%%%%%%%%%

Knowing that the optimal decision rule is a likelihood ratio test between the least favorable distributions, the design of minimax optimal tests reduces to finding the latter. Hence, being able to identify least favorable distributions is crucial to robust detection.

For the two hypotheses case, a first criterion was given by Huber in his seminal paper \cite{Huber1965}. It is based on a property known as \emph{stochastic dominance} and fixed in the following definition. 

\begin{criterion}[Stochastic Dominance]
  If a pair of distributions $(Q_0,Q_1)$ satisfies
		\begin{align}
	Q_0 \biggl[ \frac{q_1(X)}{q_0(X)} >    \lambda \biggr] & \geq P_0\biggl[ \frac{q_1(X)}{q_0(X)} >    \lambda \biggr] 
	\label{eq:lfd_sd_0} \\
	Q_1 \biggl[ \frac{q_1(X)}{q_0(X)} \leq \lambda \biggr] & \geq P_1\biggl[ \frac{q_1(X)}{q_0(X)} \leq \lambda \biggr]
	\label{eq:lfd_sd_1}
	\end{align}
	for all $(P_0,P_1) \in \mathcal{P}_0 \times \mathcal{P}_1$ and all $\lambda \geq 0$, then the joint distributions $\mathbb{Q}_0 = Q_0^N$ and $\mathbb{Q}_1 = Q_1^N$ are least favorable for all cost functions in Definition~\ref{def:costs}, all thresholds $\lambda$, and all sample sizes $N$. 
	\label{crit:stochastic_dominance}
\end{criterion}

Stochastic dominance is based on the intuition that the error probabilities of a minimax test should be maximum under the least favorable distributions. Criterion~\ref{crit:stochastic_dominance} makes this notion formal and precise: by inspection, the probabilities in \eqref{eq:lfd_sd_0} and \eqref{eq:lfd_sd_1} correspond to the two error probabilities of a single-sample likelihood ratio test between the two least favorable distributions with threshold $\lambda$. The least favorable \emph{joint distributions} of $\bm{X}_N$ under each hypothesis are simply the corresponding product distributions, meaning that the i.i.d.~case is also the worst case. Note that this is indeed a property of the least favorable distributions and not an \emph{assumption} made beforehand. That is, within the given uncertainty sets, even allowing dependencies cannot further increase the error probabilities of the minimax test. 

There are two more properties of the stochastic dominance criterion that are worth highlighting. First, $Q_0$ and $Q_1$ play two different roles in \eqref{eq:lfd_sd_0} and \eqref{eq:lfd_sd_1}. On the one hand, they define the test statistic and in turn the events of interest, namely $\frac{q_1(X)}{q_0(X)} \lessgtr \lambda$. On the other hand, they define the distributions with respect to which the probabilities of these events are taken. This coupling is typical for objective functions in minimax robust detection, and it can be found in all three criteria given in this section.

Second, the stochastic dominance criterion requires the pair $(Q_0,Q_1)$ to maximize the error probabilities \emph{jointly} for both type I and type II errors and \emph{jointly} for all positive likelihood ratio thresholds $\lambda$. In other words, $Q_0$ and $Q_1$ need to be independent of $\lambda$ and independent of which hypothesis is associated with which distribution. Only under these conditions do the properties of the single-sample test with particular threshold $\lambda$ carry over to tests with arbitrary thresholds and sample sizes. This is a strong requirement and the existence of a pair of least favorable distributions that satisfy it is not guaranteed, but critically depends on the uncertainty sets $\mathcal{P}_0$ and $\mathcal{P}_1$. However, before going into the details of different uncertainty models, two alternative characterizations of least favorable distributions are introduced that shed some light on their relation to statistical similarity measures.

Huber and Strassen \cite{HuberStrassen1973} showed that least favorable distributions in the sense of Criterion~\ref{crit:stochastic_dominance} also satisfy the following criterion.

\begin{criterion}[Minimum $f$-Divergence]
	If a pair of distributions $(Q_0,Q_1)$ minimizes
	\begin{equation}
	  D_{f}(P_1 \Vert P_0) = \int_\mathcal{X} f\biggl( \frac{p_1(x)}{p_0(x)} \biggr) p_0(x) \, \mathrm{d}x 
	  \label{eq:lfd_f-div}
	\end{equation}
	over $(P_0, P_1) \in \mathcal{P}_0 \times \mathcal{P}_1$ for all twice differentiable convex functions $f \colon \mathbb{R}_+ \to \mathbb{R}$, then the joint distributions $\mathbb{Q}_0 = Q_0^N$ and $\mathbb{Q}_1 = Q_1^N$ are least favorable for all cost functions in Definition~\ref{def:costs}, all thresholds $\lambda$, and all sample sizes $N$.
	\label{crit:f-divergence}
\end{criterion}  
The quantity in \eqref{eq:lfd_f-div} is known as $f$-divergence (or $\phi$-divergence), where $f$ is a convex function satisfying $f(1) = 0$. The class of $f$-divergences was introduced independently and almost simultaneously by Csisz\'ar \cite{Csiszar1963}, Morimoto \cite{Morimoto1963} and Ali and Silvey \cite{AliSilvey1966}. It includes many frequently encountered distances and divergences, such as the Kullback--Leibler (KL) divergence (relative entropy), the $\alpha$-divergence (R\'enyi entropy), the $\chi^2$-divergence, the Hellinger distance, and the total variation distance. A comprehensive survey on $f$-divergences and related distance measures can be found in \cite{Cha2007}. Also, see the box on the right for a brief discussion of some properties of $f$-divergences that make them a natural similarity measure in a detection context.

\begin{figure}[tb]
  \begin{tcolorbox}[title=\textbf{$f$-Divergences in Robust Detection}]
    In (robust) detection, $f$-divergences emerge naturally as an appropriate class of measures for the distance or similarity of distributions. This is not a coincidence. By construction, $f$-divergences admit several properties that one would intuitively expect from a similarity measure in a detection context \cite{AliSilvey1966}:
    \begin{enumerate}
      \item Transformations can only decrease similarity. That is,
            \begin{equation}
              X' = T(X) \; \Rightarrow \; D(P' \Vert Q') \leq D(P \Vert Q),
            \end{equation}
            where $T \colon \mathcal{X} \to \mathcal{X}'$ is a measurable function and $P'$ and $Q'$ are the distributions on $\mathcal{X}'$ induced by applying $T$ to $X$. This is a variant of the data-processing inequality, implying that a detector based on processed observations $X'$ cannot perform better than a detector based on the raw observations $X$.
            
      \item $D(P \Vert Q)$ is invariant under permutations of $\mathcal{X}$. This property implies that $D(P \Vert Q)$ depends on the sample space only via $P$ and $Q$, but not via $x$. It again emphasizes the fact that in detection the \emph{probability} of an event is of importance, not the event itself; recall the discussion in Sec.~\ref{ssec:context}.
          
      \item $D(P \Vert Q)$ is jointly convex in $P$ and $Q$. This property is interesting from a robustness perspective, since it implies that a pair of distributions that is \emph{locally} least favorable is also \emph{globally} least favorable. In other words, there is no such thing as a locally minimax detector based on locally least favorable distributions.
    \end{enumerate}
    
    These properties, in combination with some other natural requirements, can be used to define $f$-divergences in an axiomatic manner \cite{Csiszr2008}, thus proving a strong, theoretical backing for using $f$-divergences instead of other, alternative distance measures in a detection context.
    Finally, it is interesting to note that the properties stated above only admit an operational interpretation for (minimax) \emph{optimal} detectors. When using sub-optimal detectors, preprocessing the data or redefining events can indeed affect the performance of the detector and, in turn, local similarity peaks and valleys can emerge. A nice illustration of this effect is known as \emph{stochastic resonance} \cite{Gammaitoni1998, Palonpon1998}, where additional noise, added in a smart way, can in fact improve the performance of a sub-optimal detector.
  \end{tcolorbox}
  \vspace{-2ex}
\end{figure}

The minimum $f$-divergence criterion is based on the intuition that, in order to maximize the error probabilities, the least favorable distributions should be \emph{maximally similar}. However, it is not sufficient for them to be most similar with respect to a single similarity measure, but they need to jointly minimize \emph{all} $f$-divergences whose defining functions are twice differentiable. This joint optimality can be seen as the ``divergence domain'' counterpart to the property that the least favorable distributions need to be independent of the threshold $\lambda$ in Criterion~\ref{crit:stochastic_dominance}. In \cite{HuberStrassen1973}, it is shown that distributions admitting this property are so-called 2-alternating capacities in the sense of Choquet \cite{Choquet1954}. However, for the purpose of this paper the definition as universal minimizers of $f$-divergences is sufficient and more transparent. 

At first glance, Criterion~\ref{crit:f-divergence} might seem stricter than Criterion~\ref{crit:stochastic_dominance} since it has to hold for a whole class of convex functions $f$, not just for all positive scalars $\lambda$. Nevertheless, both can be shown to be exactly equivalent. Some insight into why this is the case can be obtained by looking at yet another characterization of least favorable distributions. 

\begin{criterion}[Maximum Weighted Sum Error]
  If a pair of distributions $(Q_0,Q_1)$ maximizes
	\begin{equation}
	  L(\lambda P_1 \Vert P_0) = \int_\mathcal{X} \min\{ p_0(x), \, \lambda p_1(x) \} \, \mathrm{d}x
	  \label{eq:lfd_weighted_sum_error}
	\end{equation}
	over $(P_0,P_1) \in \mathcal{P}_0 \times \mathcal{P}_1$ for all $\lambda \geq 0$, then the joint distributions $\mathbb{Q}_0 = Q_0^N$ and $\mathbb{Q}_1 = Q_1^N$ are least favorable for all cost functions in Definition~\ref{def:costs}, all thresholds $\lambda$, and all sample sizes $N$.
	\label{crit:weighted_sum_error}
\end{criterion}  

This characterization of distributions that minimize $f$-divergences has been derived and used in various forms in the literature \cite{Mussmann1979, Torgersen1991, Liese2008}. In the context of robust detection, it was recently shown in \cite{Fauss2016_old_bands}. In a sense, Criterion~\ref{crit:weighted_sum_error} bridges the first two criteria. On the one hand, it is not hard to show that
\begin{multline}
\int_\mathcal{X} \min\{ p_0(x), \, \lambda p_1(x) \} \, \mathrm{d}x \\
	= P_0\biggl[ \frac{p_1(X)}{p_0(X)} > \frac{1}{\lambda} \biggr] + \lambda P_1\biggl[ \frac{p_1(X)}{p_0(X)} \leq \frac{1}{\lambda} \biggr],
	\label{eq:weighted_sum_error_alt}
\end{multline}
which means that when performing a single-sample likelihood ratio test, the pair that solves \eqref{eq:lfd_weighted_sum_error} admits the largest weighted sum error probabilities among all feasible pairs. This is clearly in close analogy to Criterion~\ref{crit:stochastic_dominance}. However, Criterion~\ref{crit:weighted_sum_error} exploits more properties of the minimax likelihood ratio test, such as the connection between the cost coefficient $\eta$ in \eqref{eq:weighted_sum_error} and the likelihood ratio threshold $\lambda$. This makes it possible to replace the separate constraints on the two error probabilities with a single constraint on their weighted sum. Moreover, it can be shown that it is not necessary to consider a missmatch between test statistic and true distributions, as is the case on the right-hand side of \eqref{eq:lfd_sd_0} and \eqref{eq:lfd_sd_1}. As a consequence of these simplifications, Criterion~\ref{crit:weighted_sum_error} is usually easier to evaluate in practice.

There also exists a strong connection between Criterion~\ref{crit:weighted_sum_error} and Criterion~\ref{crit:f-divergence}, although it is less obvious. It is based on the observation that any $f$-divergence can be decomposed into a superposition of weighted total variation distances. The total variation distance between $P_0$ and $P_1$ is defined as the largest possible difference when calculating the probability of any event $\mathcal{E}$ under one distribution instead of the other. That is,    
\begin{equation}
  D_\text{TV}(P_0, P_1) = \sup_{\mathcal{E} \in \mathcal{F}} \; \lvert P_1(\mathcal{E}) - P_0(\mathcal{E}) \rvert,
\end{equation}
where $\mathcal{F}$ denotes the $\sigma$-algebra of measurable events. If both $P_0$ and $P_1$ admit a probability density function, the total variation distance can be written as
\begin{align}
	D_\text{TV}(P_1 \Vert P_0) &= \frac{1}{2} \int_\mathcal{X} \lvert p_1(x) - p_0(x) \rvert \, \mathrm{d}x \\
	&= \frac{1}{2} \int_\mathcal{X} \left\lvert \frac{p_1(x)}{p_0(x)} - 1 \right\rvert p_0(x) \, \mathrm{d}x.
\end{align}
Hence, under these mild assumptions, total variation is the $f$-divergence induced by
\begin{equation}
  f_\text{TV}(t) = \frac{1}{2} \lvert t - 1 \rvert.
\end{equation}    
Now consider a version of $f_\text{TV}$, where $t$ is weighted by a nonnegative scalar $\lambda$, that is,
\begin{equation}
  f_{\lambda\text{TV}}(t) = \frac{1}{2} \bigl( \lvert \lambda t - 1 \rvert - \lvert \lambda - 1 \rvert \bigr).
  \label{eq:weighted_tv_func}
\end{equation} 
The second term in \eqref{eq:weighted_tv_func} is merely a re-normalization so that $f_{\lambda\text{TV}}(1) = 0$ for all $\lambda \geq 0$. In a slight abuse of notation, the divergence $D_{f_{\lambda\text{TV}}}$ is in the following written as 
\begin{equation}
  D_{f_{\lambda\text{TV}}}(P_1 \Vert P_0) = D_\text{TV}(\lambda P_1 \Vert P_0),  
\end{equation}
which emphasizes the interpretation of $\lambda$ as a weight and will generalize in a natural way to the multi-hypothesis case discussed in later sections. In \cite{Guntuboyina2014} it was shown that for every twice differentiable function $f$, the corresponding $f$-divergence can be written as
\begin{equation}
  D_f(P_1 \Vert P_0) = \int_0^{\infty} D_\text{TV}(\lambda P_1 \Vert P_0) \, f''(\lambda) \, \mathrm{d} \lambda,
  \label{eq:f-div_spectrum}
\end{equation}
where $f''$ denotes the second derivative of $f$.\footnote{This result can be generalized to non-differentiable $f$ by replacing $f''$ with an appropriate curvature measure. The technical details will not be entered here, but can be found, for example, in \cite{Liese2012}.} The right-hand side of \eqref{eq:f-div_spectrum} is known as the \emph{spectral representation} of $D_f$. It implies that every $f$-divergence can be composed by superimposing weighted elementary $f$-divergences of the total variation type. 

From \eqref{eq:f-div_spectrum} it follows that for a pair of distributions $(Q_0, Q_1)$ to minimize all $f$-divergences induced by twice differentiable functions $f$, it is sufficient that it minimizes $D_\text{TV}(\lambda P_1 \Vert P_0)$ for all $\lambda > 0$. The last step to arrive at Citerion~\ref{crit:weighted_sum_error} is to note that for two real scalars $a, b$ it holds that
\begin{equation}
	\lvert a - b \rvert = a + b - 2 \min\{ a \,,\, b\},
\end{equation}
so that 
\begin{equation}
  D_\text{TV}(\lambda P_1 \Vert P_0) = \min\{1, \lambda\} - L(\lambda P_1 \Vert P_0).
  \label{eq:weighted_tv_distance}
\end{equation}
Hence, $D_\text{TV}(\lambda P_1 \Vert P_0)$ is minimum if and only if $L(\lambda P_1 \Vert P_0)$ in \eqref{eq:lfd_weighted_sum_error} is maximum. 

Before proceeding further, it is important to note that all three criteria presented in this section are necessary for the least favorable distributions to factor ($\mathbb{Q} = Q^N$) and to be independent of both the sample size, $N$, and the detection threshold, $\lambda$. However, they are \emph{not} necessary for minimax optimality in general. That is, there can exist minimax optimal tests whose least favorable distributions do not satisfy Criteria~\ref{crit:stochastic_dominance}--\ref{crit:weighted_sum_error}. Such tests, however, are significantly less well-studied, significantly harder to design, and their usefulness in practice is limited. In fact, %with the exception of sequential detectors, which are discussed in Sec.~\ref{sec:sequential_detection}, 
we are not aware of any commonly used stricly minimax robust detector that does not satisfy Criteria~\ref{crit:stochastic_dominance}--\ref{crit:weighted_sum_error}. This problem and the question of how it can be overcome will be picked up again later in this section as well as in Sec.~\ref{sec:sequential_detection}, in the context of robust sequential detection.

At this point, it becomes hard to make more concrete statements about least favorable distributions, their existence, and their properties without fixing the uncertainty sets $\mathcal{P}_0$ and $\mathcal{P}_1$. Hence, in the next section, three commonly used uncertainty models are introduced and discussed in detail. These three models were chosen because they are flexible and useful in practice, yet admit tractable least favorable distributions. Moreover, they admit interesting theoretical properties that help to shed a light on certain fundamental properties of minimax robust detectors in general.

%------------------------------------------------------------------------------%
\subsection{Uncertainty Models}
\label{ssec:uncertainty_models}
%------------------------------------------------------------------------------%
In this section, three useful non-parametric uncertainty models are detailed, which capture different types of uncertainty corresponding to different effects in real-world applications. For all of them, the minimax detector is well-defined, simple to implement, and meaningful in the sense of the discussion in Section~\ref{sec:minimax_principle}. Moreover, they can be argued to form a single, larger class of uncertainty models. A more detailed discussion of this aspect is deferred to the next subsection.

%- - - - - - - - - - - - - - - - - - - - - - - - - - - - - - - - - - - - - - - %
\subsubsection[Epsilon-contamination uncertainty]{\underline{$\varepsilon$-contamination uncertainty}}
\begin{figure*}
	\centering
	\includegraphics{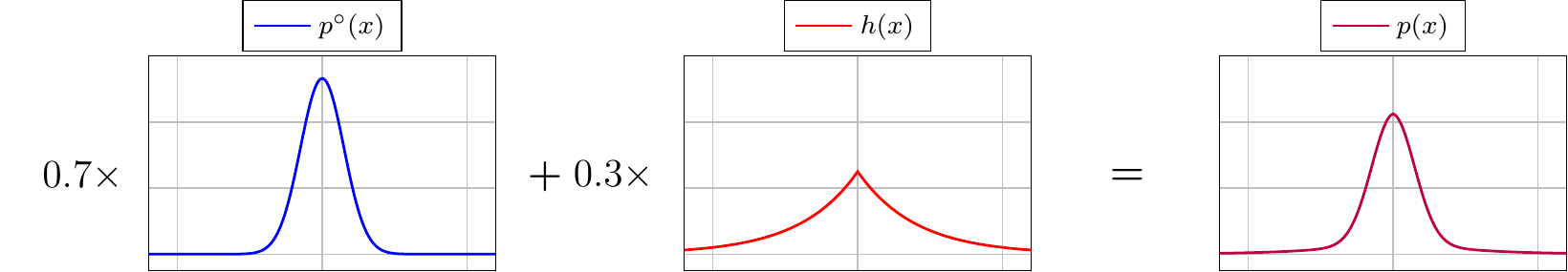}%
  \caption{Example of a distribution under $\varepsilon$-contamination with outlier ratio $\varepsilon = 0.3$. Here, clean samples from a normal distribution (left) are contaminated by outliers from a distribution with heavy tails (middle) causing the mixture distribution (right) to notably deviate from the nominal case.}
	\label{fig:outlier_model}
\end{figure*}

%- - - - - - - - - - - - - - - - - - - - - - - - - - - - - - - - - - - - - - - %
One of the oldest and most common uncertainty models in both detection and estimation is the $\varepsilon$-contamination model \cite{Huber1965}. It is based on the idea that the majority of the data indeed follow an ideal model (nominal model), whereas a fraction $\varepsilon < 0.5$ of the data can be outliers. Here the term outlier is used in the sense that a data point does not contain any useful information about the nominal model at all, but was drawn independently from a different distribution. Formally, the $\varepsilon$-contamination model is defined as
\begin{equation}
  \boxed{\mathcal{P}_\varepsilon(P^{\circ}) = \big\{\, P : P = (1-\varepsilon) P^{\circ} + \varepsilon H \,\big\} \Big. }
\end{equation} 
with $\varepsilon \in [0, 0.5)$, $P^{\circ}$ denoting a known nominal distribution, and $H$ denoting the distribution of the outliers, which can be any distribution defined on the given sample space. See Fig.~\ref{fig:outlier_model} for a graphical illustration.

The $\varepsilon$-contamination uncertainty model is particularly appropriate for scenarios in which a suitable model for the observed system exists, but individual measurements or data points can be severely corrupted. Typical examples for such scenarios are impulsive noise in radio systems \cite{Blackard1993, Fernandez2014}, motion artifacts in biomedical data \cite{Strasser2012, Schaeck2018}, or defective sensors in monitoring systems \cite{Rizzoni1991, Sharma2010}. However, the concept of outlies can also be applied in cases where the clean and corrupted data points are not entirely independent. For example, mislabeled entries in data sets \cite{Frenay2014, Fauss_2019_ICASSP} or corrupted bits in digitally stored data \cite{Bairavasundaram2008} can also be modeled as outliers. In such cases, $\varepsilon$-contamination is a pessimistic approximation of the true, more complex uncertainty model.

In general, the $\varepsilon$-contamination model is pessimistic in the sense that it requires the corresponding robust detector to be able to handle all possible outliers in the data, irrespective of how unlikely or nonsensical they are. In some scenarios, this requirement can be too strict. For example, in applications where the distributions generating the clean data are discrete or have a bounded support, gross outliers can easily be identified---one can think of counting the number of defective products in a lot or measuring the angle between two beams. In other applications, the data acquisition procedure has been perfected to a degree where it can safely be assumed not to produce severe outliers, for example, in laboratory experiments under highly controlled conditions. 

On the other hand, the assumption that only a fraction of the data points is subject to a model mismatch can be too optimistic. In most real-world scenarios, even the nominal model only holds approximately so that \emph{all} observations are subject to a \emph{moderate} model mismatch, instead of \emph{a few} observations being subject to a \emph{severe} model mismatch. In fact, this is the case for almost all procedures that are based on strong distributional assumptions, such as Gaussianity, uniformity, or independence, which are rarely satisfied exactly in practice. 
In such cases, instead of assuming outliers in the data, it is more natural to assume that the true distributions are not identical to the nominal distributions, but only \emph{similar}. 

%- - - - - - - - - - - - - - - - - - - - - - - - - - - - - - - - - - - - - - - %
\subsubsection[f-Divergence ball uncertainty]{\underline{$f$-Divergence ball uncertainty}}
\begin{figure}
  \centering
  \includegraphics[scale=0.85]{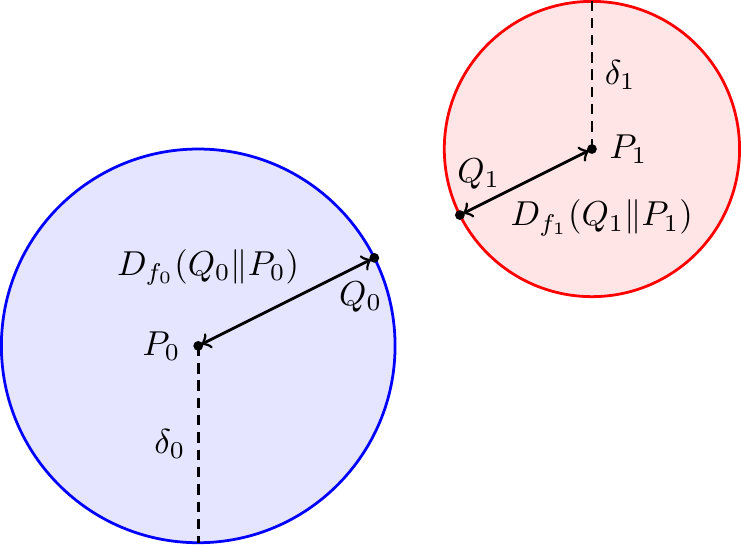}
  \caption{Illustration of the $f$-divergence ball uncertainty model, where each ball represents the feasible distributions under the corresponding hypothesis. The least favorable distributions correspond to the two closest points not being contained in the same ball.}
  \label{fig:f-div_ball_model}
\end{figure}

%- - - - - - - - - - - - - - - - - - - - - - - - - - - - - - - - - - - - - - - %
A common way of modeling this type of uncertainty is via divergence balls. More precisely, the true distribution is assumed to lie within a ball of radius $\zeta > 0$ around the nominal distribution $P^{\circ}$:
\begin{equation}
  \boxed{\mathcal{P}_{f, \zeta}(P^\circ) = \big\{\, P : D_f(P \Vert P^{\circ}) \leq \zeta \,\big\} \Big.}
  \label{eq:f-divergence_ball}
\end{equation} 
The exact distance $D$ defining this ball can be chosen depending on the application. As indicated by the notation in \eqref{eq:f-divergence_ball}, the class of $f$-divergences again arises as a natural choice that has been shown to offer a favorable compromise between being useful in practice and having favorable theoretical properties; compare the box on page 7. The $f$-divergence ball uncertainty model is illustrated in Fig.~\ref{fig:f-div_ball_model}.

Many special cases of the $f$-divergence ball uncertainty model have been studied in the literature, including KL divergence balls \cite{Levy2009, Gul2017_minimax_robust}, $\alpha$-divergence balls \cite{Gul2016_alpha_divergence}, Hellinger distance balls \cite{Gul2014_Hellinger_distance}, and combinations of the former \cite{Gul2015_composite_distances}. We are not yet in a position to discuss how the choice of $f$ affects the least favorable distributions and the properties of the robust detector, but we will return to this question later on. 

%- - - - - - - - - - - - - - - - - - - - - - - - - - - - - - - - - - - - - - - %
\subsubsection[Density band uncertainty]{\underline{Density band uncertainty}}
\begin{figure}
  \centering
  \includegraphics[scale=0.9]{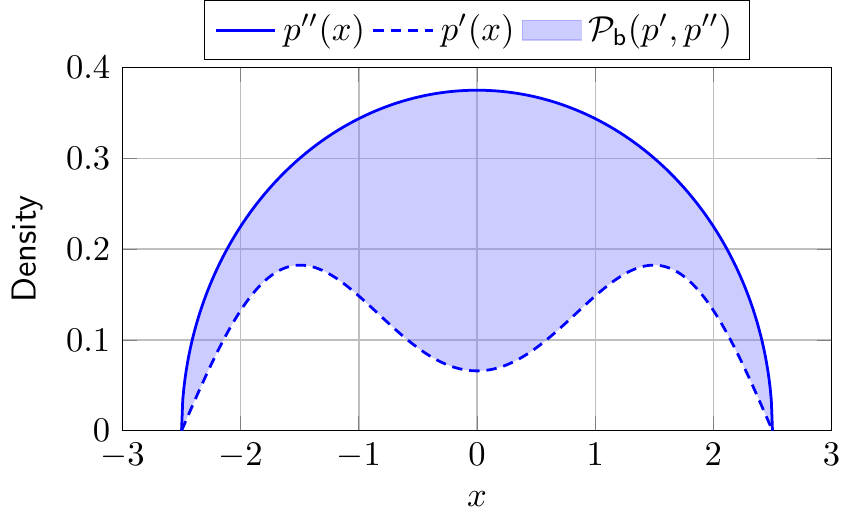}\\
  \includegraphics[scale=0.9]{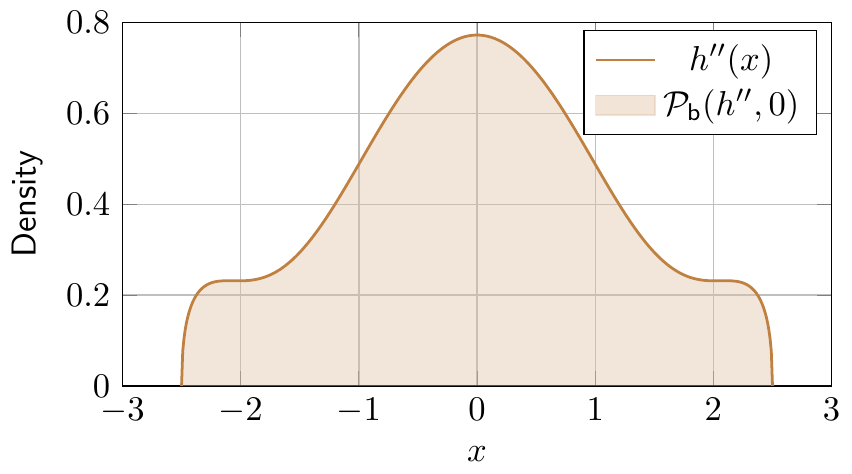}\\
  \caption{Illustration of a band uncertainty model for a density function (top) and the corresponding outlier density function (bottom). Here the outlier ratio in \eqref{eq:eps_band} is given by $\varepsilon = 0.4$. }
  \label{fig:density_band_model}
\end{figure}

%- - - - - - - - - - - - - - - - - - - - - - - - - - - - - - - - - - - - - - - %
An alternative to defining a neighborhood of similar distributions via a divergence ball around a nominal distribution is to allow the true density function to deviate from the nominal density function by a certain amount, more precisely,
\begin{equation}
  \boxed{\mathcal{P}_\text{b}(p', p'') = \big\{\, P : p'(x) \leq p(x) \leq p''(x) \,\big\} \Big.}
  \label{eq:density_band}
\end{equation} 
where $p'$ and $p''$ denote point-wise lower and upper bounds, respectively. This type of uncertainty model is known as density band model \cite{Kassam1981, Fauss2016_old_bands}; see Fig.~\ref{fig:density_band_model} for an illustration. Similar to $f$-divergence balls, it can be used to model uncertainties in the shape of distributions, but with the difference that neither a nominal distribution nor a divergence need to be introduced explicitly. This has the advantage of making it more transparent which distributions are included in an uncertainty set, even to non-experts. For example, it can be much easier for a practitioner to specify a band of typical densities for a certain random phenomenon than to specify a nominal distribution and a suitable divergence. Moreover, the amount of uncertainty in a density band model can be defined \emph{locally} for different regions of the sample space by choosing tighter or loser bounds, whereas it is controlled \emph{globally}, by a single parameter $\zeta$, for divergence balls. However, this additional flexibility can also be a disadvantage of the density band model. For example, when constructing uncertainty sets based on training data, estimating the radius of a divergence ball is typically easier and more accurate than estimating a confidence interval for a density function. 

Despite its conceptual connection to divergence balls, the density band model can also be interpreted as a variant of the $\varepsilon$-contamination model with additional constraints on the outlier distribution. More precisely, $\mathcal{P}_\text{b}$ in \eqref{eq:density_band} can equivalently be written as
\begin{equation}
  \mathcal{P}_\text{b}(p', p'') = \{\, P : p(x) = p'(x) + \varepsilon h(x), \; \varepsilon h(x) \leq \Delta_\text{b}(x) \,\},
\end{equation}
where $\Delta_\text{b}(x) = p''(x) - p'(x)$. This analogy can be made closer by defining the outlier ratio of a band model as
\begin{equation}
  \varepsilon = 1- \int p'(x) \, \mathrm{d}x
  \label{eq:eps_band}
\end{equation}
and its nominal distribution as that which admits the density
\begin{equation}
  p^\circ(x) = \frac{1}{1-\varepsilon} p'(x).
  \label{eq:band_nominal_density}
\end{equation}
Using these definitions, the band model in \eqref{eq:density_band} can be written as a constrained $\varepsilon$-contamination model \cite{Fauss2016_old_bands}
\begin{equation}
  \mathcal{P}_\text{b}(p', p'') = \{\, P : P = (1-\varepsilon) P^{\circ} + \varepsilon H, \; H \in \mathcal{P}_\text{b}(0, h'') \,\},
  \label{eq:band_outl}
\end{equation}
where the outlier density is bounded from above by
\begin{equation}
  h''(x) = \frac{p''(x) - p'(x)}{\varepsilon}.
\end{equation}

Both interpretations are useful to keep in mind. While the one in \eqref{eq:density_band} is typically more intuitive, the constrained $\varepsilon$-contamination model in \eqref{eq:band_outl} allows for an easier comparison to the other two uncertainty models since it admits an explicit nominal distribution and a global uncertainty parameter $\varepsilon$.

As a concluding remark, constructing uncertainty sets in practice always implies making a compromise between modeling the application-specific uncertainty as accurately as possible and keeping the model simple enough to be able to identify and calculate least favorable distributions. The three models presented above are by no means a comprehensive or exhaustive selection, and some applications might require an entirely different model. Nevertheless, they provide a good starting point in the sense that they are flexible enough to cover a wide range of distributional uncertainties, while at the same time allowing for an efficient calculation of their least favorable distributions. The latter is a topic in its own right and detailed in the next section.

%------------------------------------------------------------------------------%
\subsection{Calculating Least Favorable Distributions}
\label{ssec:calculating_lfds}
%------------------------------------------------------------------------------%
Having characterized the least favorable distributions implicitly, the question of how to calculate them explicitly arises. In this section, this question is answered for the two-hypothesis case and the three uncertainty models introduced in Sec.~\ref{ssec:uncertainty_models}. First, the least favorable distributions under uncertainty of the density band type are introduced, the corresponding results under uncertainty of the $f$-divergence ball and $\varepsilon$-contamination type are then obtained as special cases of the former. 

%- - - - - - - - - - - - - - - - - - - - - - - - - - - - - - - - - - - - - - - %
\subsubsection*{Density band uncertainty}
%- - - - - - - - - - - - - - - - - - - - - - - - - - - - - - - - - - - - - - - %
Let two uncertainty sets $\mathcal{P}_0$ and $\mathcal{P}_1$ of the form \eqref{eq:density_band} be given. In \cite{Fauss2016_old_bands} it is shown that in this case the pair $(Q_0,Q_1) \in \mathcal{P}_0 \times \mathcal{P}_1$ is least favorable if the densities $(q_0,q_1)$ satisfy
\begin{align}
  q_0(x) & = \min \{ p''_0(x) \,,\, \max \{ c_0 q_1(x) \,,\, p'_0(x) \} \},
  \label{eq:band_lfd_0} \\
  q_1(x) & = \min \{ p''_1(x) \,,\, \max \{ c_1 q_0(x) \,,\, p'_1(x) \} \},
  \label{eq:band_lfd_1}
\end{align}
for some $c_0, c_1 > 0$. The $\min$ and $\max$ operators on the right-hand side of \eqref{eq:band_lfd_0} and \eqref{eq:band_lfd_1} have to be read pointwise and guarantee that $q_0$ and $q_1$ are within the feasible bands. In words, $q_0$ is the \emph{projection} of $q_1$ onto the band of feasible densities $\mathcal{P}_0$, where the projection is performed by scaling $q_1$ such that the right-hand side of \eqref{eq:band_lfd_0} integrates to one. Analogously, $q_1$ is the projection of $q_0$ onto $\mathcal{P}_1$. Similar projection operators are commonly found in other areas of robustness \cite{Kassam1977, Poor1981, Poor1982_smoothing, Geraniotis1985, Geraniotis1987}, but typically without the coupling between the least favorable distributions, which is a characteristic of robust detection. 

\begin{figure*}
	\centering
	\includegraphics[scale=0.9]{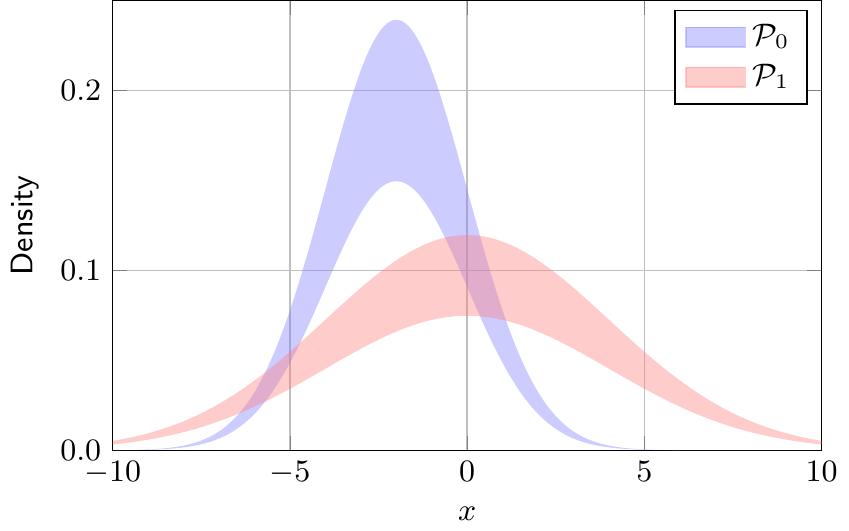}%
	\hspace{1em}
	\includegraphics[scale=0.9]{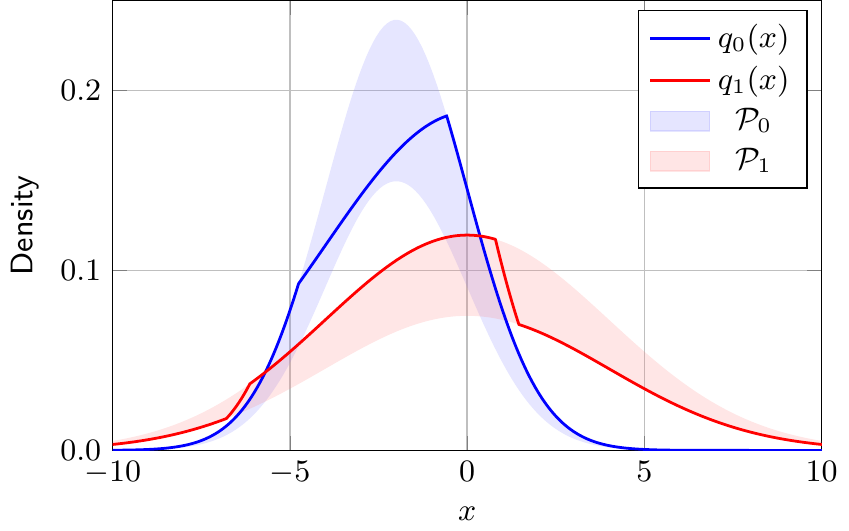}\\[1ex]
	\includegraphics[scale=0.9]{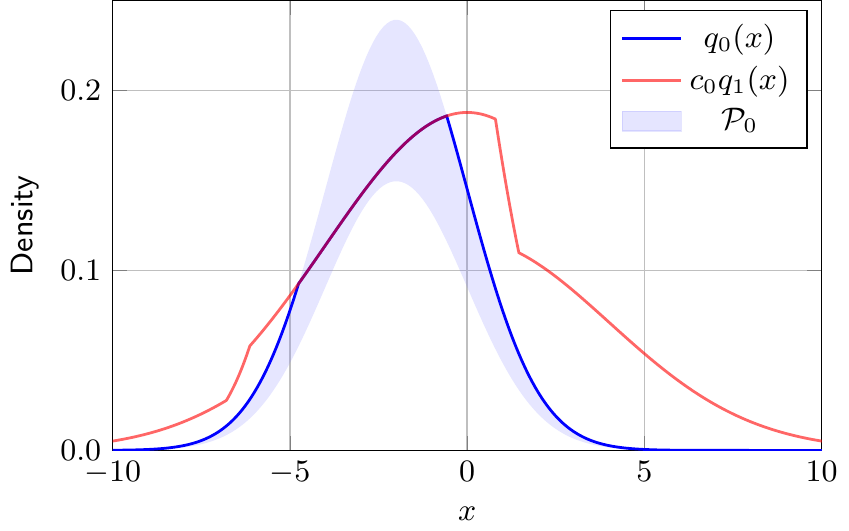}%
	\hspace{1em}
	\includegraphics[scale=0.9]{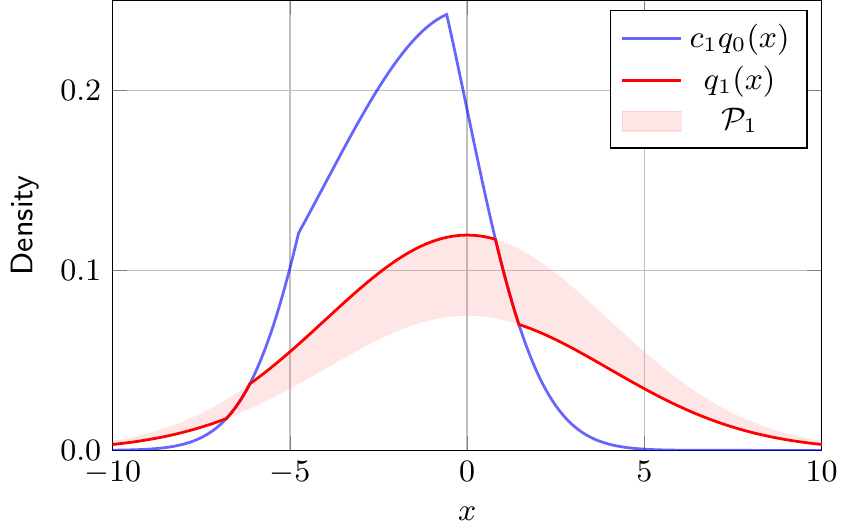}
	\caption{Example of a density band uncertainty set (top left) and the corresponding least favorable densities (top right). In the bottom plots it can be seen how the least favorable densities either coincide with one of their bounds or are scaled versions of their respective counterpart. The scaling factors $c_0$ and $c_1$ are determined by \eqref{eq:band_lfd_0} and \eqref{eq:band_lfd_1}.}
	\label{fig:lfds_band}
\end{figure*}

An example of a density band uncertainty model and its least favorable densities is given in Fig.~\ref{fig:lfds_band}. Here, the upper and lower bounds were chosen to be scaled Gaussian densities, more precisely,
\begin{equation}
  \begin{aligned}
    p'_i(x) = a \, p_\mathcal{N}(x; \mu_i, \sigma_i^2), \\
    p''_i(x) = b \, p_\mathcal{N}(x; \mu_i, \sigma_i^2),
  \end{aligned}
  \label{eq:density_bounds_example}
\end{equation}
where $p_\mathcal{N}(\bullet ; \mu, \sigma^2)$ denotes the density of a Gaussian distribution with mean $\mu$ and variance $\sigma^2$. For the uncertainty sets shown in Fig.~\ref{fig:lfds_band}, indicated by the shaded areas, the parameters are $a = 0.75$, $b = 1.2$, $\mu_0 = -2$, $\mu_1 = 0$, $\sigma_0^2 = 4$, and $\sigma_1^2 = 16$. It can be seen how the least favorable densities either coincide with one of the bounds or, on intervals where at least one of them lies in the interior of the band, are scaled versions of each other. Note that the least favorable densities are not guaranteed to be unique, that is, it can be possible to construct (infinitely) many density pairs satisfying \eqref{eq:band_lfd_0} and \eqref{eq:band_lfd_1}. However, the scaling factors $c_0$ and $c_1$ as well as the likelihood ratio $q_1/q_0$ can be shown to be unique. The properties of the latter will be investigated more closely in Sec.~\ref{ssec:design_and_implementation}.

A straightforward yet efficient way of calculating the least favorable densities is to iteratively solve \eqref{eq:band_lfd_0} and \eqref{eq:band_lfd_1} for $c_0$ and $c_1$. That is, starting from an initial guess $(q_0^{(0)}, q_1^{(0)})$, one constructs a sequence of pairs $(q_0^{(k)}, q_0^{(k)})$, $k = 1,2,\ldots$ via
\begin{align}
  q_0^{(k)}(x) & = \min \{ p''_0(x) \,,\, \max \{ c_0^{(k)} q_1^{(k-1)}(x) \,,\, p'_0(x) \} \},
  \label{eq:band_lfd_0_iterative} \\
  q_1^{(k)}(x) & = \min \{ p''_1(x) \,,\, \max \{ c_1^{(k)} q_0^{(k)}(x) \,,\, p'_1(x) \} \}.
  \label{eq:band_lfd_1_iterative}
\end{align}
Note that the only unknowns in \eqref{eq:band_lfd_0_iterative} and \eqref{eq:band_lfd_1_iterative} are the scalars $c_0^{(k)}$ and $c_1^{(k)}$ so that each update reduces to finding a scaling factor such that the projected density on the right-hand side integrates to one. Mathematically, this translates to finding the scalar root of a monotonic function, which is a well-known problem in numerics that can be solved by off-the-shelf algorithms. Moreover, this procedure is independent of the underlying sample space and its dimensions. As long as the right-hand sides of \eqref{eq:band_lfd_0_iterative} and \eqref{eq:band_lfd_1_iterative} can be integrated over the sample space, the least favorable densities can be computed iteratively. For high-dimensional problems, representing and integrating the least favorable distributions can become a non-trivial task. However, even in such cases, state-of-the-art approximation and integration techniques \cite{Dahlquist2008} in combination with modern hardware are usually powerful enough to obtain close approximations.

%- - - - - - - - - - - - - - - - - - - - - - - - - - - - - - - - - - - - - - - %
\subsubsection*{$\varepsilon$-contamination uncertainty} 
%- - - - - - - - - - - - - - - - - - - - - - - - - - - - - - - - - - - - - - - %
Being able to calculate least favorable densities for the band model also enables one to calculate least favorable densities for the $\varepsilon$-contamination model. As stated in \eqref{eq:band_outl}, the density band model can be interpreted as a constrained $\varepsilon$-contamination model, so that the latter can be recovered as a special case of the former. First, since the outlier distributions are unbounded under $\varepsilon$-contamination uncertainty, the upper bounds $p''_0$, $p''_1$ and in turn $h_0''$, $h_1''$ do not bind. Moreover, according to \eqref{eq:band_nominal_density}, the lower bounds $p'_0$ and $p'_1$ are scaled versions of the nominal densities. In combination, this yields least favorable densities that are of the form
\begin{align}
  q_0(x) & = \max \{ c_0 q_1(x) \,,\, (1-\varepsilon_0) p_0(x) \},
  \label{eq:outl_lfd_0_coupled} \\
  q_1(x) & = \max \{ c_1 q_0(x) \,,\, (1-\varepsilon_1) p_1(x) \}.
  \label{eq:outl_lfd_1_coupled}
\end{align}
Finally, it can be shown that $q_0, q_1$ on the right-hand side of the above equations can be replaced by the nominal densities $p_0, p_1$. In order to see this, note that $L_\lambda$ in Criterion~\ref{crit:weighted_sum_error} can be written as
\begin{equation*}
  \int_\mathcal{X} \min\{ q_0(x), \lambda q_1(x) \} \, \mathrm{d}x = Q_0(\mathcal{X}_{1,\lambda}) + \lambda Q_1(\mathcal{X}_{0,\lambda})
\end{equation*}
where
\begin{align}
  \mathcal{X}_{0,\lambda} = \{ x \in \mathcal{X} : q_0(x) > \lambda q_1(x) \}, \\
  \mathcal{X}_{1,\lambda} = \{ x \in \mathcal{X} : q_0(x) \leq \lambda q_1(x) \},
\end{align}
so that $\mathcal{X}_{0,\lambda} \cup \mathcal{X}_{1,\lambda} = \mathcal{X}$ and $\mathcal{X}_{0,\lambda} \cap \mathcal{X}_{1,\lambda} = \emptyset$. Under $\varepsilon$-contamination uncertainty, it holds that
\begin{align}
  Q_0(\mathcal{X}_{1,\lambda}) &= (1-\varepsilon_0) P_0(\mathcal{X}_{1,\lambda}) + \varepsilon_0 H_0(\mathcal{X}_{1,\lambda}) \\
  &\leq (1-\varepsilon_0) P_0(\mathcal{X}_\lambda) + \varepsilon_0
\end{align}
and
\begin{align}
  Q_1(\mathcal{X}_{0,\lambda}) &= (1-\varepsilon_1) P_1(\mathcal{X}_{0,\lambda}) + \varepsilon_1 H_1(\mathcal{X}_{0,\lambda}) \\
  &\leq (1-\varepsilon_1) P_1(\mathcal{X}_{0,\lambda}) + \varepsilon_1
\end{align}
The upper bounds are attained if $H_0(\mathcal{X}_{1,\lambda}) = H_1(\mathcal{X}_{0,\lambda}) = 1$, which implies that $H_0$ and $H_1$ need to be orthogonal in order to be least favorable. 

Now, consider the region on which $q_0$ does not attain its lower bounds. On this region it holds that  
\begin{equation}
  q_0(x) > (1-\varepsilon_0) p_0(x) \; \Rightarrow \; h_0(x) > 0.
\end{equation}
Hence, $h_1(x) = 0$ and $q_1(x) = (1-\varepsilon_1) p_1(x)$, so that $q_1$ in \eqref{eq:outl_lfd_0_coupled} can be replaced by $p_1$; the factor $(1-\varepsilon_1)$ can be absorbed in the free parameter $c_1$. The same line of arguments also applies to \eqref{eq:outl_lfd_1_coupled}. This yields the least favorable densities
\begin{align}
  q_0(x) & = \max \{\, c_0 p_1(x) \,,\, (1-\varepsilon_0) p_0(x) \,\}, \\
  q_1(x) & = \max \{\, c_1 p_0(x) \,,\, (1-\varepsilon_1) p_1(x) \,\}.
\end{align}
This result was obtained by Huber \cite{Huber1965} without recourse to the band model, but the proof via this connection is arguably more instructive. The fact that $q_0$ and $q_1$ are decoupled under $\varepsilon$-contamination means that there is no need for an iterative procedure to determine $c_0$ and $c_1$, which in turn simplifies their calculation. Under density band uncertainty, it is in general not possible to construct orthogonal outlier densities, which explains the coupling between $q_0$ and $q_1$ in the general equations \eqref{eq:band_lfd_0} and \eqref{eq:band_lfd_1}.

\begin{figure*}
	\centering
	\includegraphics[scale=0.9]{lfds.pdf}%
	\hspace{1em}
	\includegraphics[scale=0.9]{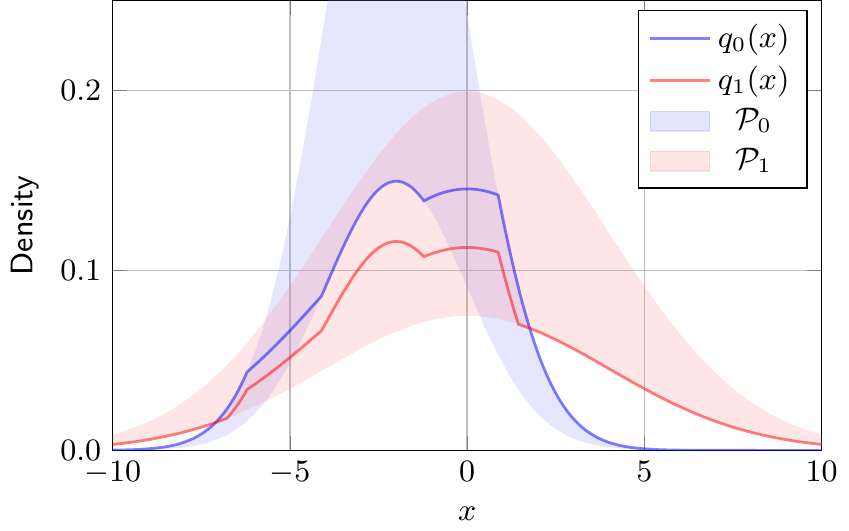}\\[1ex]
	\includegraphics[scale=0.9]{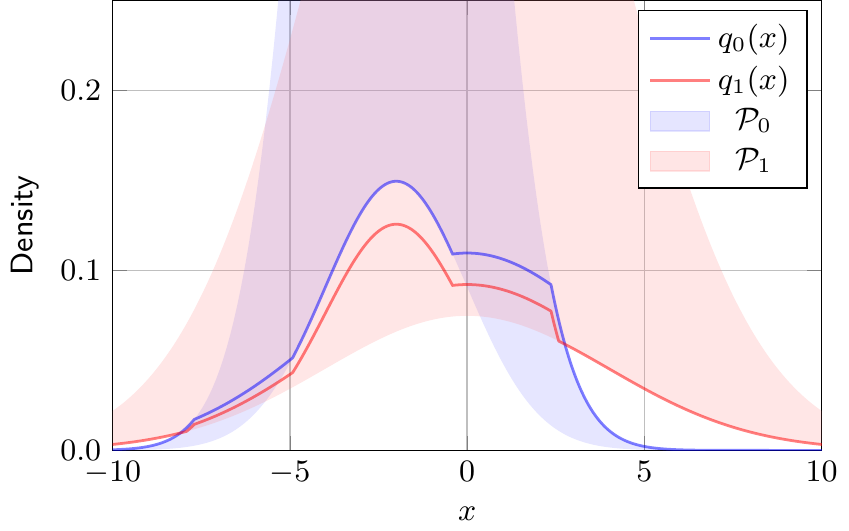}%
	\hspace{1em}
	\includegraphics[scale=0.9]{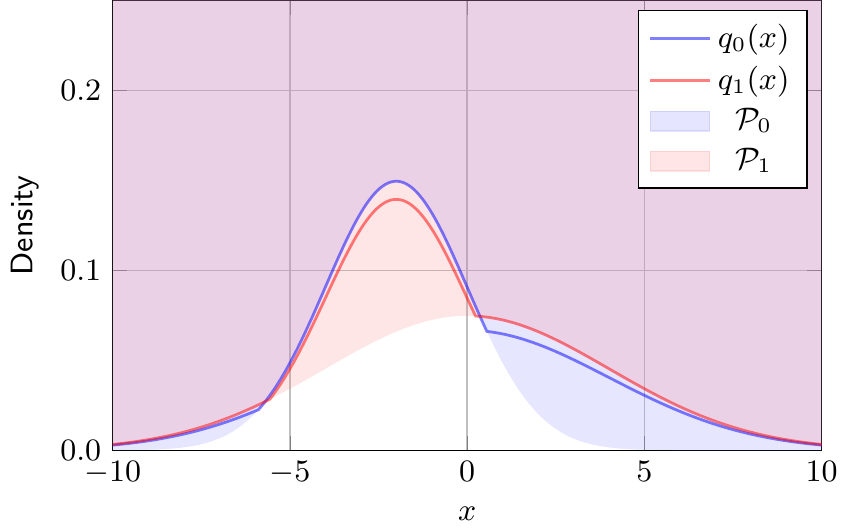}
	\caption{Example of how an $\varepsilon$-contamination model can be obtained by relaxing a density band model. The upper bounds are of the form \eqref{eq:density_bounds_example}, with $c'' = 1.2$ (top left), $c'' = 2$ (top right), $c'' = 5$ (bottom left), and $c'' = \infty$ (bottom right), the latter corresponding to unconstrained $\varepsilon$-contamination. Purple shaded areas indicate overlapping density bands.}
	\label{fig:lfds_transition}
\end{figure*}

The transition from a density band model to an $\varepsilon$-contamination model can be illustrated by successively relaxing the upper bounds of the density band model in Fig.~\ref{fig:lfds_band}. The effect of this relaxation on the least favorable densities is shown in Fig.~\ref{fig:lfds_transition}. As can be seen, the more the upper bounds are relaxed, the more freely the probability mass can be moved and the more overlap there is between the distributions. Finally, under unconstrained $\varepsilon$-contamination, the least favorable distributions become similar enough to almost overlap completely. In fact, if the outlier ratio, which is \SI{25}{\percent} in this example, is further increased, the least favorable distributions in the lower right plot of Fig.~\ref{fig:lfds_transition} become indeed identical, meaning that in the worst case the two hypotheses become indistinguishable in the minimax sense. 

This effect is the closest equivalent to what is know as a \emph{breakdown point} in robust estimation. The breakdown point of an estimator determines how large a ratio of outliers in the data an estimator can tolerate without deviating arbitrarily far from the true value of the parameter. Hence, for outlier ratios above its breakdown point, an estimator is no longer guaranteed to perform better than randomly guessing the unknown parameter. In analogy, the breakdown point of a detector can be defined as the ratio of outliers it can tolerate while still performing better than random guessing. This breakdown happens exactly when the two least favorable distributions become identical and, in turn, the likelihood ratio becomes a constant. It is important to note that while in estimation the breakdown point is a property of the \emph{estimator}, in detection it is a property of the \emph{uncertainty sets}, more precisely, of the nominal distributions under each hypothesis. In the example shown in Fig.~\ref{fig:lfds_transition}, the outlier ratio is \SI{25}{\percent}, which is well below the theoretical limit of \SI{50}{\percent}. However, for the given nominal distributions, this outlier ratio is already close to the breakdown point, which is at approximately \SI{28}{\percent}. Choosing the nominal distributions to be more or less similar, decreases or increases the breakdown point accordingly.

\begin{figure*}
	\centering
	\includegraphics[scale=0.9]{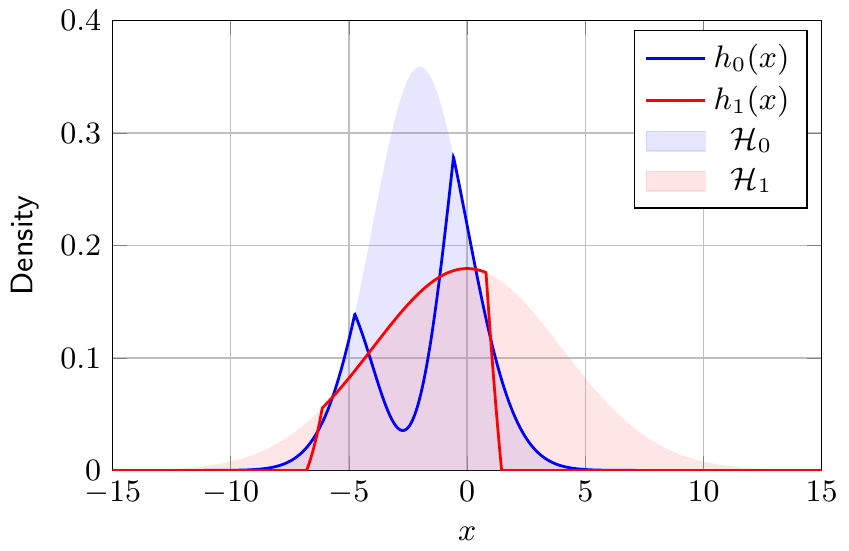}%
	\hspace{1em}
	\includegraphics[scale=0.9]{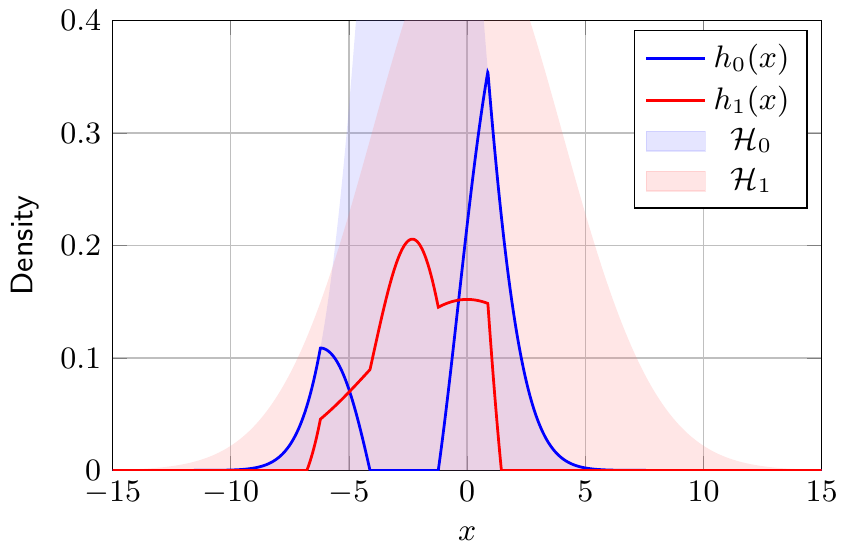}\\[1ex]
  \includegraphics[scale=0.9]{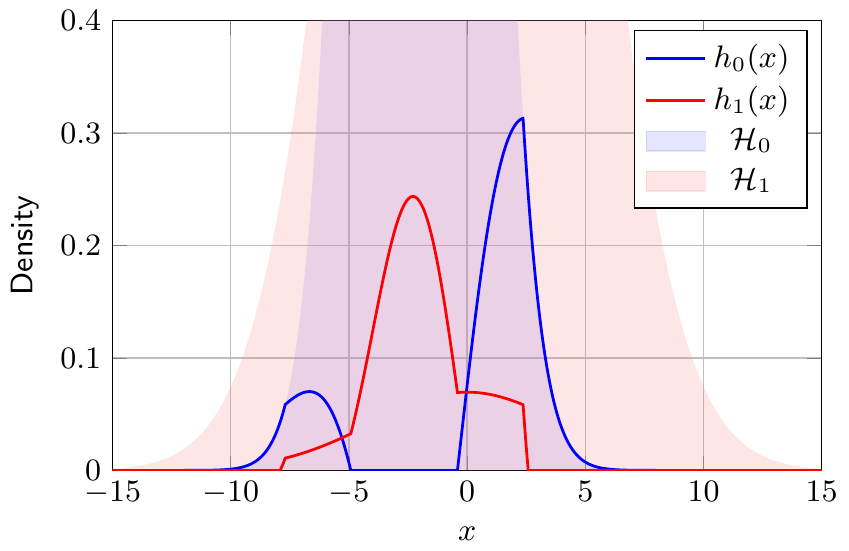}%
  \hspace{1em}	
  \includegraphics[scale=0.9]{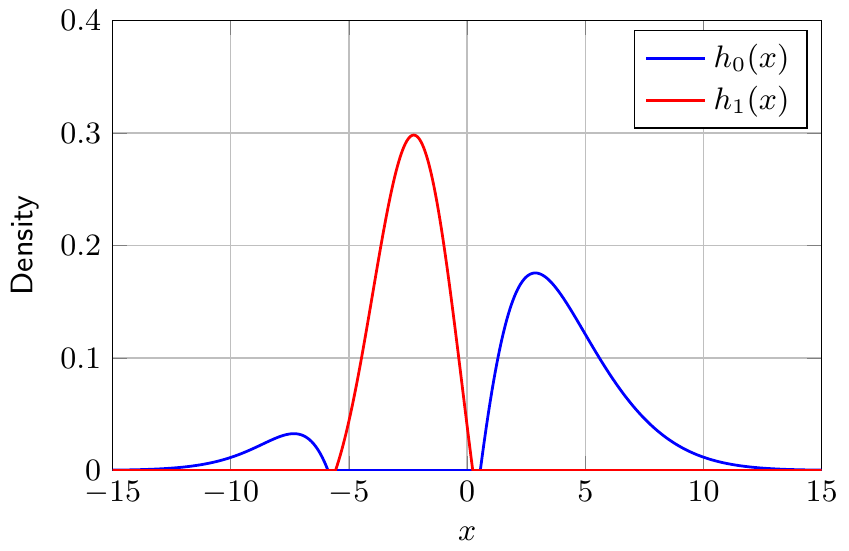}\\
	\caption{Outlier densities corresponding to the least favorable distributions in Fig.~\ref{fig:lfds_transition}. While the outlier densities are orthogonal under $\varepsilon$-contamination (lower right), the coupling in \eqref{eq:band_lfd_0} and \eqref{eq:band_lfd_1} forces them to overlap under density band uncertainty. Purple shaded areas indicate overlapping density bands.}
	\label{fig:lfds_outliers}
\end{figure*}

The outlier densities corresponding to the least favorable distributions in Fig.~\ref{fig:lfds_transition} are plotted in Fig.~\ref{fig:lfds_outliers}. Here, ``outlier density'' refers to the two densities $h_0$, $h_1$ which satisfy $q_0 = p'_0 + \varepsilon_0 h_0(x)$ and $q_1 = p'_1 + \varepsilon_1 h_1(x)$, where $\varepsilon_0$ and $\varepsilon_1$ are as in \eqref{eq:eps_band}. For the band model, the sets of feasible outlier distributions in the sense of \eqref{eq:band_outl} are indicated by the shaded areas. It can clearly be seen how the outlier densities start to overlap less as their constraints are relaxed. Interestingly, in the upper right and lower left plot, $h_1$ does not ``fill the gap'' left by $h_0$, although the constraints would allow for it. However, merely concentrating probability mass in order to reduce the spread of $q_1$ is not minimax optimal, since it ignores the coupling between $q_0$ and $q_1$. Figuratively speaking, the ``gap'' left by $h_0$ needs to be large enough to fit $h_1$ without violating the constraints on its shape in \eqref{eq:band_lfd_0} and \eqref{eq:band_lfd_1}.

Another intersting observation is that, although the outlier densities admit very particular shapes with sharp transitions and large intervals of probability zero, their probability mass is concentrated on a finite interval. That is, in contrast to robust estimation, the worst-case outliers are not extreme values generated by heavy-tailed distributions. Instead, they are in the same range as the clean data and are generated by slightly shifted and distorted versions of the nominal densities. An intuitive explanation for this observation is that in detection the ``most misleading'' samples are not those which are extremely large or small, but those that appear as if they had been generated under the hypothesis that should be rejected; recall the discussion in the introduction. Consequently, the least favorable distributions generate outliers that mimic the clean data under the opposite hypothesis. From a practitioner's point of view, this phenomenon highlights the fact that, even in the absence of impulsive noise and for seemingly clean data, deviations from the nominal model can still be present and can have a detrimental effect on detection performance. The outlier density $h_1$ shown in the lower right plot of Fig.~\ref{fig:lfds_outliers}, for example, could very well correspond to a defective sensor randomly oscillating around a small negative value. In summary, this example indicates that encountering a close-to-worst-case scenario is not just a mere theoretical possibility, but it is a real danger that should be taken seriously when designing detectors for critical applications. 

%- - - - - - - - - - - - - - - - - - - - - - - - - - - - - - - - - - - - - - - %
\subsubsection*{$f$-divergence ball uncertainty}
%- - - - - - - - - - - - - - - - - - - - - - - - - - - - - - - - - - - - - - - %
For the $f$-divergence ball uncertainty model, the situation is more complex than for the previous two uncertainty models. In general, least favorable distributions in the sense of the three criteria given in Sec.~\ref{ssec:characterizing_lfds} do not exist. This is the case for the KL divergence, the $\chi^2$-divergence, the Hellinger divergence, the R\'enyi divergence, and many more. As mentioned in Sec.~\ref{ssec:characterizing_lfds}, this does not imply that no minimax optimal tests exists. But it means that the least favorable distributions do not factor and depend on the sample size and the detection threshold. Moreover, least favorable distributions of this kind are significantly harder to characterize since no clear criteria comparable to those in the previous section exist. Naturally, this is a major obstacle for the use of $f$-divergence ball uncertainty models in practice. Although this obstacle cannot be overcome completely, it can be worked around by exploiting a connection between density bands and $f$-divergence balls, which makes it possible to design robust tests with least favorable distributions that admit the strong properties enforced by Criteria~\ref{crit:stochastic_dominance}-\ref{crit:weighted_sum_error} without sacrificing strict minimax optimality.

This connection is based on \emph{single-sample tests}, whose least favorable distributions exist under much milder conditions. For any given $\lambda > 0$, the cost function $L_\lambda$ in Criterion~\ref{crit:weighted_sum_error} can be maximized under $f$-divergence ball uncertainty. This is the case because $L_\lambda$ is jointly concave in the pair $(P_0, P_1)$ and $f$-divergence balls are convex sets of distributions. Let these minimizers of $L_\lambda$ be denoted by $Q_{0,\lambda}$ and $Q_{1,\lambda}$, where the subscript indicates the dependence on $\lambda$. $Q_{0,\lambda}$ and $Q_{1,\lambda}$ are indeed least favorable, but only for a single-sample test $(N = 1)$ with fixed likelihood ratio threshold $\lambda$. If the threshold changes, $Q_{0,\lambda}$ and $Q_{1,\lambda}$ need to be recalculated. Clearly, this weaker minimax property is not very useful in practice, where redesigning tests is costly and having multiple, independently drawn observations $(N > 1)$ is by far the most common scenario. 

In \cite{Fauss2018_icassp}, it is shown that for every $f$-divergence ball model there exists an equivalent density band model that admits the same single-sample least favorable distributions $Q_{0,\lambda}$ and $Q_{1,\lambda}$, but for which the latter are in fact least favorable in the strong sense of Criteria~\ref{crit:stochastic_dominance}-\ref{crit:weighted_sum_error}. In other words, given an uncertainty set of the $f$-divergence ball type and a threshold parameter $\lambda$, an equivalent density band model can be constructed such that the least favorable distributions of the latter coincide with the single-sample least favorable distributions of the former. Moreover, the bounds of the equivalent density band model can be shown to be simply scaled versions of the nominal densities. Schematically, this connection can be depicted as follows:
\begin{align}
  \begin{pmatrix} \lambda \\ \mathcal{P}_{f_0}(P_0, \zeta_0) \\ \mathcal{P}_{f_1}(P_1, \zeta_1) \end{pmatrix} \quad &\Rightarrow \quad \begin{pmatrix} \mathcal{P}_\text{band}(a_0 p_0, b_0 p_0) \\[0.5ex] \ \mathcal{P}_\text{band}(a_1 p_1, b_1 p_1) \end{pmatrix} 
  \label{eq:f-div_band_equivalence} \\
  \Downarrow \hspace{1.2cm} & \hspace{1.2cm} \Downarrow \notag \\
  (\tilde{Q}_{\lambda, 0}, \tilde{Q}_{\lambda, 1}) \quad &= \quad (Q_0, Q_1)
\end{align}
The scaling factors $a_0$, $b_0$ and $a_1$, $ b_1$ depend on $\zeta_0$, $\zeta_1$ and $f$ and can be obtained from the (generalized) inverses of the derivatives of the function $f$. Alternatively, the four constants can be determined numerically, by successively constructing least favorable densities until the $f$-divergence ball constraints of the original model are satisfied with equality. See \cite{Fauss2018_icassp} for more details on both methods. 

From \eqref{eq:f-div_band_equivalence}, it is clear that the least favorable distributions of every $f$-divergence ball uncertainty model are again of the form in \eqref{eq:band_lfd_0} and \eqref{eq:band_lfd_1}, with the bounds being scaled versions of the nominal densities. In fact, it can be shown that the band model in Fig.~\ref{fig:lfds_band} is equivalent to a KL divergence ball model, $f(t) = t \log(t)$, with radii $\zeta_0 \approx 0.0136$ and $\zeta_1 \approx 0.0242$.

As mentioned in Section~\ref{ssec:uncertainty_models}, for special cases of $f$-divergences, more explicit expressions for (single-sample) least favorable distributions can be found in the literature. Evaluating these directly is usually easier than first constructing the equivalent band model and then solving \eqref{eq:band_lfd_0} and \eqref{eq:band_lfd_1}. Also note that, having calculated the least favorable densities in one way or another, it is not difficult to construct the equivalent band model. As discussed in Sec.~\ref{ssec:uncertainty_models}, this can be helpful to translate $f$-divergence ball uncertainty into a form that is easier to interpret. Moreover, the fact that the upper bounds of the equivalent band model are scaled versions of the nominal densities provides a nice illustration for why $f$-divergence balls cannot contain distributions that are significantly more heavy-tailed than the nominal distributions: increasing the $f$-divergence ball radius increases this scaling factor, but does not affect the type of decay. However, one should always keep in mind that, although their least favorable densities are identical, the sets of feasible distributions on the left and right-hand side of \eqref{eq:f-div_band_equivalence} are different in general. 

We conclude this section with a historical remark. The least favorable densities for both the band model and the $\varepsilon$-contamination model had both been derived long ago; the latter by Huber \cite{Huber1965}, the former by Kassam \cite{Kassam1981}. However, the form in which the least favorable densities of the band model were stated made it hard to work with them in practice. The respective theorem in \cite{Kassam1981} distinguishes between four special cases, each involving a piecewise definition of the densities. In order to know which case holds, one has to check the existence or non-existence of in total six constants that have to be chosen such that the solutions are valid densities. This might partially explain why the band model never became as popular as the $\varepsilon$-contamination model in robust statistics. In fact, it can be argued that its popularity has been decreasing. While it used to be one of the standard models in robust testing and filtering in the 1980s \cite{KassamPoor1985, Poor1983}, today many signal processing practitioners do not seem to be aware of its existence and recent books on robust statistics ignore it entirely \cite{Maronna2006, Levy2008}. Based on the discussion above, we strongly encourage practitioners and researchers to have a second, closer look at the density band model: in practice, it provides a more flexible alternative to the common outlier model with little increase in complexity, and in theory, it provides useful insights into fundamental connections between uncertainty sets based on outliers and $f$-divergence balls.

%------------------------------------------------------------------------------%
\subsection{Detector Design and Implementation}
\label{ssec:design_and_implementation}
%------------------------------------------------------------------------------%
Assume that two sets $\mathcal{P}_0$ and $\mathcal{P}_1$ have been determined that adequately describe the model uncertainty and a pair of least favorable distributions $(Q_0, Q_1)$ has been calculated. According to Theorem~\ref{th:optimal_test}, the minimax optimal test is then a likelihood ratio test with threshold $\lambda > 0$ and test statistic
\begin{equation}
  z^N(\bm{x}_N) = \prod_{n=1}^N\frac{q_1(x_n)}{q_0(x_n)}.
  \label{eq:likelihood_ratio_product}
\end{equation}
This test statistic is referred to as \emph{minimax likelihood ratio} in what follows. A closer look at it provides some insight into the kind of counter measures that are called for by the three types of uncertainty introduced above.

\begin{figure*}
	\centering
	\includegraphics[scale=0.9]{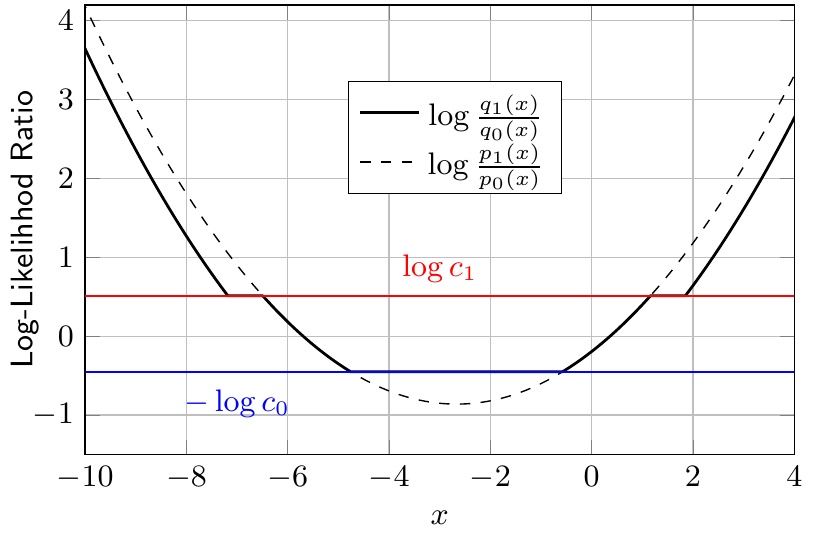}%
	\hspace{1em}
	\includegraphics[scale=0.9]{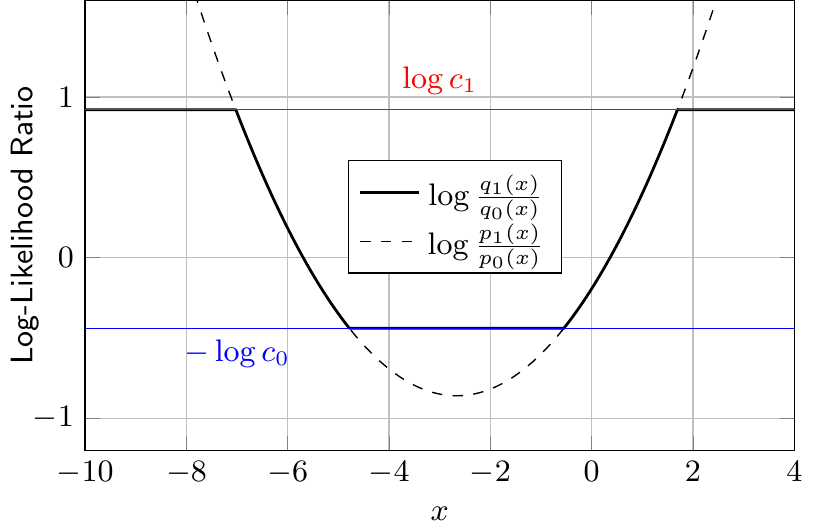}\\
	\caption{Examples of minimax log-likelihood ratios under density band uncertainty (left) and $\varepsilon$-contamination (right). The uncertainly model is the same as in Fig.~\ref{fig:lfds_band}. The $\varepsilon$-contamination model is obtained by letting $p_0'', p_1'' \to \infty$, which corresponds to an outlier ratio of $\varepsilon = 0.25$.}
	\label{fig:llr}
\end{figure*}

From the general form of the least favorable densities in \eqref{eq:band_lfd_0} and \eqref{eq:band_lfd_1}, it follows that for a density band uncertainty model the minimax likelihood ratio $\frac{q_1(x)}{q_0(x)}$ can take on six different values at any point $x$, namely,
\begin{equation}
  \frac{q_1(x)}{q_0(x)} \in \left\{ \frac{p'_1(x)}{p'_0(x)} \,,\, \frac{p''_1(x)}{p'_0(x)} \,,\, \frac{p'_1(x)}{p''_0(x)} \,,\, \frac{p''_1(x)}{p''_0(x)} \,,\, \frac{1}{c_0} \,,\, c_1 \right\}. 
  \label{eq:likelihood_cases_band}
\end{equation}
That is, over the entire sample space, the minimax likelihood ratio is determined by the density bounds and the two constants $c_0$, $c_1$. Consequently, there are two, not necessarily connected, regions on which the likelihood ratio is constant. These regions can clearly be identified in Fig.~\ref{fig:llr}, where, on the left-hand side, the minimax likelihood ratio is plotted for the band model in Fig.~\ref{fig:lfds_band}. 

For the $\varepsilon$-contamination model, the set of possible minimax likelihood ratio values reduces to
\begin{equation}
  \frac{q_1(x)}{q_0(x)} \in \left\{ \frac{1-\varepsilon_1}{1-\varepsilon_0} \frac{p_1(x)}{p_0(x)} \,,\,   \frac{1}{c_0} \,,\, c_1 \right\}. 
  \label{eq:likelihood_cases_outl}
\end{equation}
That is, the minimax likelihood ratio is either a scaled version of the nominal likelihood ratio or a constant, which is illustrated on the right-hand side of Fig.~\ref{fig:llr}. Note that in Fig.~\ref{fig:llr} $\varepsilon_0 = \varepsilon_1$ so that the minimax likelihood ratio coincides with the nominal one on the respective region of the sample space. 

Finally, for uncertainty sets of the $f$-divergence ball type, the minimax likelihood ratio can again take on six values, namely,
\begin{equation}
	\left\{ \frac{a_1}{a_0} \frac{p_1(x)}{p_0(x)} \,,\, \frac{b_1}{a_0} \frac{p_1(x)}{p_0(x)} \,,\, \frac{a_1}{b_0} \frac{p_1(x)}{p_0(x)} \,,\, \frac{b_1}{b_0} \frac{p_1(x)}{p_0(x)} \,,\, \frac{1}{c_0} \,,\, c_1 \right\}.
  \label{eq:likelihood_cases_fdiv}
\end{equation}
This results in minimax likelihood ratios of the same type as for the density band model, with the ratios of the upper and lower bounds being replaced by scaled versions of the nominal likelihood ratio.

While all three uncertainty models admit two regions of constant likelihood ratio, what distinguishes them and the corresponding robust tests is the shape and location of these regions. For $\varepsilon$-contamination models, the minimax likelihood ratio is a \emph{clipped} version of the nominal one, as depicted on the right-hand side of Fig.~\ref{fig:llr}. This seminal result was first proved by Huber in \cite{Huber1965}. A clipped likelihood ratio limits the influence of any single observation on the outcome of the test so that even extreme outliers cannot overwrite the evidence provided by the majority of the data. This renders the detector robust against noise from heavy-tailed distributions. However, as shown in the previous section, the least favorable distributions themselves are not necessarily heavy-tailed. 

The density band model, being a generalization of the $\varepsilon$-contamination model, provides more subtle means of robustification than just clipping. An example of a minimax likelihood ratio based on a band model can be seen on the left-hand side of Fig.~\ref{fig:llr}. While the negative part of the likelihood ratio is clipped, just as it is for the $\varepsilon$-contamination model, the positive part is merely ``dampened'', meaning that the influence of highly indicative observations is reduced, but not bounded. This is a consequence of the assumption that the outlier distribution itself is constrained, compare \eqref{eq:band_outl}, so that extreme outliers are too rare to justify the performance loss incurred by clipping. Likelihood ratios of this form are sometimes referred to as \emph{compressed} \cite{Fauss2016_old_bands}, in the sense that on any given interval the absolute value of the minimax likelihood ratio is bounded by the absolute value of the nominal likelihood ratio.

\begin{figure*}
	\centering
	\includegraphics[scale=0.9]{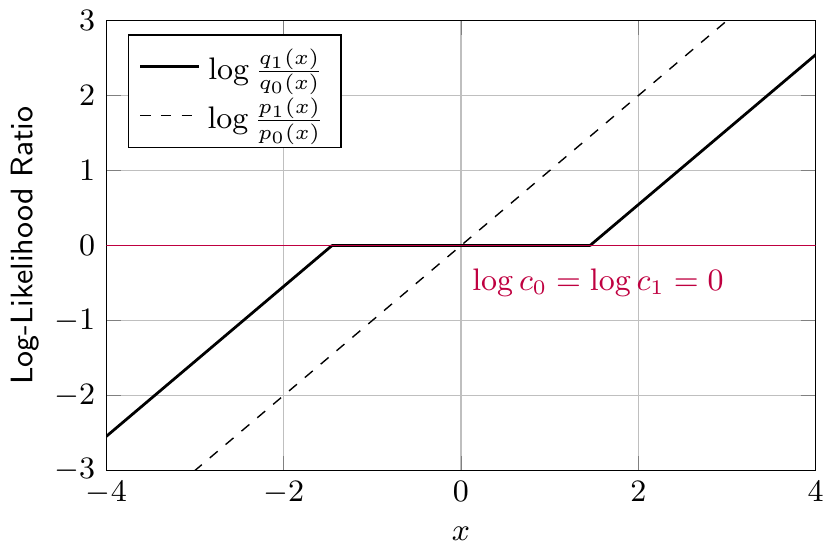}%
	\hspace{1em}
	\includegraphics[scale=0.9]{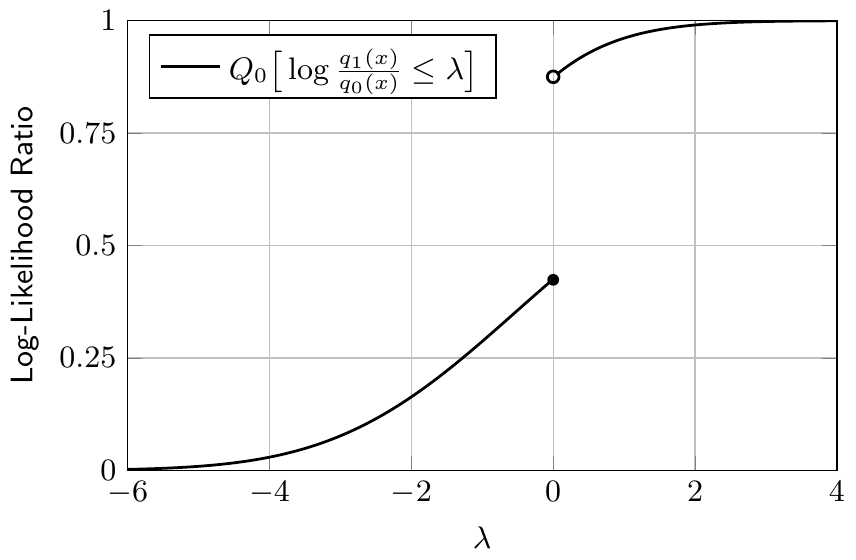}
	\caption{Example of a minimax log-likelihood ratio under density band or $f$-divergence ball uncertainty (left), which results in censored observations on an interval around $x = 0$. The distribution function of the log-likelihood ratio under $Q_0$ is depicted on the right.}
	\label{fig:llr_cens}
\end{figure*}

Finally, there exists a third type of minimax likelihood ratio that can emerge from both the band model and the $f$-divergence ball model; it is illustrated by the example in Fig.~\ref{fig:llr_cens}, which is obtained form a density band model of the form \eqref{eq:density_bounds_example} with $a = 0.7$, $b = 3$, $\mu_1 = -\mu_0 = 2$, and $\sigma_0^2 = \sigma_1^2 = 4$. It can be shown that for this particular uncertainty set, the two constants $c_0$ and $c_1$ coincide, so that there is a single region where the likelihood ratio is constant. Moreover, in this example, the constant equals one, meaning that the observations falling in this region do not contribute to the test statistic, but are entirely ignored. Hence, this scheme is usually referred to as \emph{censoring} \cite{Rago1996}. The intuition underlying censoring is that samples providing little evidence for either hypothesis should not be trusted, since this evidence is likely to be exclusively due to random deviations from the nominal model. Hence, censoring can be interpreted as a data cleaning procedure, where the cleaning happens implicitly when evaluating the test statistic. 

Censoring or rejecting data points is an old, well-known technique for dealing with contaminated data sets \cite{Pearson1936, Grubbs1950} and still plays an important role in modern robust statistics \cite{Zhang2010, Gupta2014, Wang2019}. In robust estimation, outlier rejection arises naturally as a consequence of so-called redescending weight functions, which assign a vanishingly small or even zero weight to large observations \cite{Shevlyakov2008, Zoubir2012}. Interestingly, in robust detection, censoring has almost the opposite effect. A minimax optimal procedure never rejects observations with extremely large nominal likelihood ratio, but can reject observations with small nominal likelihood ratio. In other words, while censoring in estimation prevents rare, grossly corrupted observations from overruling the majority of the data, censoring in robust detection prevents the errors introduced by small but frequent model deviations from accumulating in the test statistic. This example highlights the fact that there are indeed connections between robust estimation detection, but that results from one are not guaranteed to carry over to the other in a straightforward manner.

Another important aspect of minimax optimal detectors is that they may require randomized decision rules in order to perform well in practice. This problem was briefly touched upon in Sec.~\ref{sec:two_hypotheses_optimal}, however, we are now in a better position to explain it. Using the censored likelihood ratio depicted in Fig.~\ref{fig:llr_cens} as an example, it is clear that when evaluating this test statistic a value of \emph{exactly} one occurs with non-zero probability, that is, the distribution of the likelihood ratio contains a point mass. This can clearly be seen in the plot on the right-hand side of Fig.~\ref{fig:llr_cens}, where the cumulative distribution function of the censored log-likelihood ratio is plotted under the corresponding least favorable distribution $Q_0$. By inspection, the probability of erroneously rejecting $\mathcal{H}_0$ jumps from approximately \SI{12.5}{\percent} to approximately \SI{50.7}{\percent} when $\lambda$ changes from $+0$ to $-0$. Consqeuently, a detector with a reasonable false alarm rate can be reduced to a random coin flip, depending on which decision rule is applied when the likelihood ratio evaluates to one. The same considerations apply to the probability of erroneously rejecting $\mathcal{H}_1$. 

\begin{figure}
	\centering
	\includegraphics[scale=0.9]{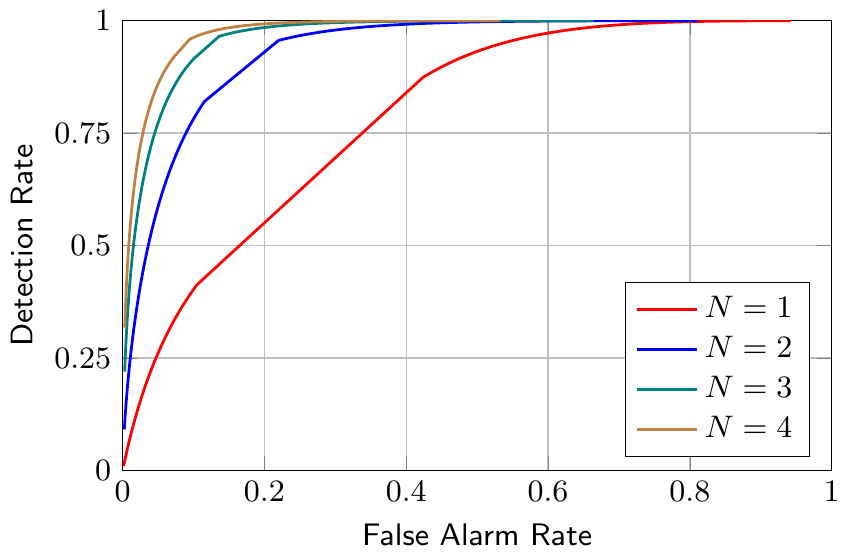}\\
	\includegraphics[scale=0.9]{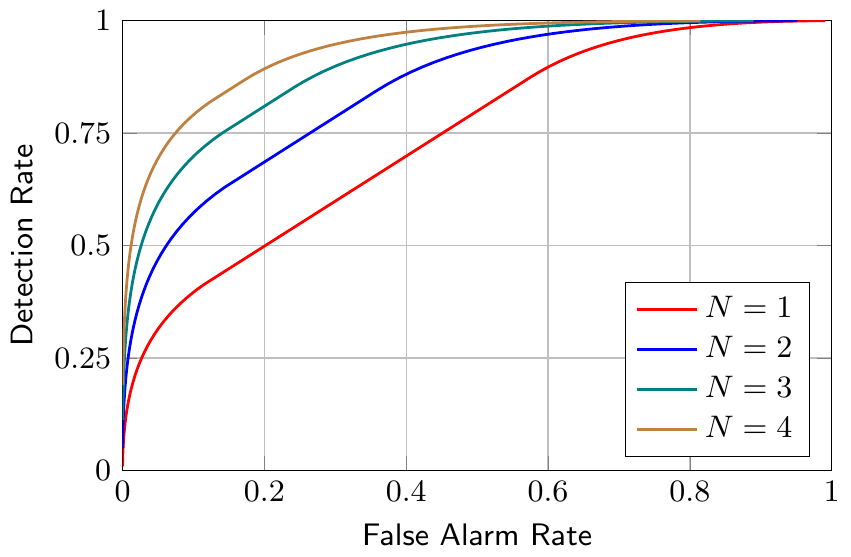}
	\caption{Receiver operating characteristics (ROC) of the minimax detector with test statistic shown on the left-hand side of Fig.~\ref{fig:llr_cens} under the nominal distributions (top) and the least favorable distributions (bottom).}
	\label{fig:llr_roc}
\end{figure}

The effect of having a point mass in the distribution of the test statistic can better be illustrated by inspecting the receiver operating characteristic (ROC) of the corresponding detector. The ROC of the minimax detector with the censored test statistic shown in Fig.~\ref{fig:llr_cens} is depicted in Fig.~\ref{fig:llr_roc}. The upper plot shows the ROC under the nominal, Gaussian distributions ($a = b = 1$ in \eqref{eq:density_bounds_example}), and the lower plot shows the ROC under the least favorable distributions. In both cases, it can be seen that on regions where the error probabilities are approximately balanced, the ROC is linear, which implies that the corresponding points of operation require a randomized decision rule. Especially under the least favorable distributions, all useful points of operation fall in this category. 

Large linear segments in the ROC are a strong indicator that the uncertainty sets have been chosen too large and the detector has been ``over robustified''. The larger the uncertainty sets, the more similar are the least favorable distributions, and the more is the distribution of the minimax likelihood ratio concentrated around one. In the extreme case that the uncertainty sets intersect each other, the test reduces to random guessing and the ROC becomes entirely linear. In practice, this collapse can easily be avoided by making sure that the uncertainty sets are disjoint. However, observing a highly concentrated test statistic or a strong influence of the randomization rule on the outcome should generally ring an alarm bell. 

The effect of point masses is most detrimental for censored test statistics. It can also occur for clipped and compressed likelihood ratios. However, for these, the point masses are usually smaller and, in particular for the clipped test, occur at values that are too large or too small to be useful thresholds in practice. The effect also becomes less critical for large sample sizes, since summing the individual log-likelihood ratios spreads the distribution and smoothes out the point masses. This effect can be observed in Fig.~\ref{fig:llr_roc}, where the ROCs are plotted for different sample sizes. Nevertheless, point masses in the distributions of the test statistic cannot be avoided entirely, so the potential need for randomization should be kept in mind when designing robust detectors, the more so the smaller the sample size and the larger the uncertainty sets. A good example for the importance of randomized decision rules and how to design them optimally can be found in \cite{Gul2016_alpha_divergence}, where the technical aspects are discussed in more detail.

Randomization issues aside, the performance of a minimax robust detector can be analyzed in analogy to the performance of a regular likelihood ratio detector. From the Chernoff–Stein Lemma \cite[Sec.~11.8]{CoverThomas1991} it follows that the error probabilities of a likelihood ratio test for two hypotheses of the form \eqref{eq:simple_hypotheses} decrease exponentially. More precisely, keeping one error probability fixed and letting the sample size grow, the other error probability decays exponentially with exponents $D_\text{KL}(Q_0 \Vert Q_1)$ and $D_\text{KL}(Q_1 \Vert Q_0)$, that is,
\begin{align}
  \mathbb{E}_{\mathbb{P}_0}[\,\delta^*(\bm{X}_N)\,] &\approx 2^{-N D_\text{KL}(P_0 \Vert P_1)}
  \label{eq:err_exp_0}
  \intertext{and}  
  \mathbb{E}_{\mathbb{P}_1}[1-\delta^*(\bm{X}_N)\,] &\approx 2^{-N D_\text{KL}(P_1 \Vert P_0)}
  \label{eq:err_exp_1}
\end{align}  
for large $N$. In the context of robustness, however, the performance under distribution mismatch is more interesting. In \cite{BoroumandFabregas2020}, it is shown that the error exponents of a likelihood ratio test that is designed for the hypotheses in \eqref{eq:simple_hypotheses}, but is evaluated under different (not necessarily least favorable) distributions $Q_0$, $Q_1$, are given by 
\begin{align}
  \mathbb{E}_{\mathbb{Q}_0}[\,\delta^*(\bm{X}_N)\,] &\approx 2^{-N ( D_\text{KL}(P_0 \Vert Q_1) - D_\text{KL}(P_0 \Vert Q_0))}
  \label{eq:err_exp_mismatch_0}
  \intertext{and}
  \mathbb{E}_{\mathbb{Q}_1}[1-\delta^*(\bm{X}_N)\,] &\approx 2^{-N ( D_\text{KL}(P_1 \Vert Q_0) - D_\text{KL}(P_1 \Vert Q_1) )}.
  \label{eq:err_exp_mismatch_1}
\end{align}
Note that for \eqref{eq:err_exp_mismatch_0} and \eqref{eq:err_exp_mismatch_1} to hold, the exponents on the right-hand side need to be positive, meaning that $Q_0$ is more similar to $P_0$ than to $P_1$ and $Q_1$ is more similar to $P_1$ than to $P_0$. Using \eqref{eq:err_exp_mismatch_0} and \eqref{eq:err_exp_mismatch_1}, the performance of optimal and minimax optimal likelihood ratio tests can be approximated for large sample sizes. Of course, in practice, Monte Carlo simulations are often used as an alternative or additional way of evaluating the performance of a detector under more complex distributions and for small sample sizes.

Before concluding this section, two remarks about designing and analyzing robust detectors are in place. First, the form of the error exponents under mismatch suggests that the KL divergence is the ``most natural'' uncertainty measure. This conception can be misleading. In \eqref{eq:err_exp_mismatch_0} and \eqref{eq:err_exp_mismatch_1}, the KL divergence quantifies the \emph{asymptotic performance} of a detector. The $f$-divergence in the definition of the uncertainty model in \eqref{eq:f-divergence_ball} specifies the \emph{type of uncertainty}. The latter should always be chosen according to an honest assessment of the model at hand, not based on mathematical convenience. The performance evaluation is a separate step. In other words, the error exponents of a robust detector should be a consequence of the uncertainty model and not the other way around.

A second important aspect when designing robust detectors is that one should not focus exclusively on the performance under one particular pair of distributions---be it the nominal distributions or the least favorable distributions. After all, the very idea underlying robustness is that it is more important to perform well under \emph{all} feasible distributions than to perform optimally under any specific distribution. Ultimately, the minimax design approach is merely a means to this end.

%------------------------------------------------------------------------------%
\subsection{Conclusions}
\label{ssec:conclusions}
%------------------------------------------------------------------------------%
The discussion of the two-hypothesis case so far has revealed that 
\begin{itemize}
  \item minimax optimal tests are standard likelihood ratio tests based on least favorable distributions;
  \item the latter can be characterized by their property to maximize error probabilities and minimize $f$-divergences;
  \item there are useful uncertainty models for which least favorable distributions exist and can be calculated efficiently in practice;
  \item robust detectors for these uncertainty models admit clipped, censored, or compressed test statistics.
\end{itemize}
All in all, minimax robust tests for two-hypothesis are well-understood and, except for some caveats, such as the potential need for randomized decision rules, they can be designed and operated in a relatively straightforward manner. In most cases, transitioning from a non-robust test to a robust one boils down to a simple transformation of the test statistic. Yet, despite their drop-in nature and potential benefits, robust detectors are still not widespread in practice, in particular for uncertainty models that are not of the $\varepsilon$-contamination type. In light of this, we highly recommend practitioners in signal processing, data science, and related areas to try robust detectors for themselves. A toolbox for the calculation of least favorable distributions in MATLAB and Python is available online \cite{GitHubRepo}. For answers to some common questions and resevation concerning the use of robust detectors in practice, also see the FAQ box on the right.

While tests for two hypotheses are arguably most common in practice, there are also many applications in which a decision for one out of \emph{multiple} options needs to be made. The state-of-the-art in robust testing is less complete for this problem, however, some interesting findings have been obtained recently, which are detailed in the next two sections.

%%%%%%%%%%%%%%%%%%%%%%%%%%%%%%%%%%%%%%%%%%%%%%%%%%%%%%%%%%%%%%%%%%%%%%%%%%%%%%%%
\section{Minimax Detection for Multiple Hypotheses}
\label{sec:multiple_hypotheses}
%%%%%%%%%%%%%%%%%%%%%%%%%%%%%%%%%%%%%%%%%%%%%%%%%%%%%%%%%%%%%%%%%%%%%%%%%%%%%%%%

The problem of deciding for one out of multiple possible hypotheses, also known as \emph{statistical classification} in the literature, is a natural generalization of binary hypothesis testing and has applications in signal processing \cite{Miao2002}, communications \cite{Wei2000}, and other areas \cite{Bou2016}. 

In the previous sections it has become clear that the existence of least favorable distributions that are independent of the sample size and the threshold parameter is crucial for the existence of minimax optimal tests. Unfortunately, the existence of distributions that are least favorable in this strong sense can only be guaranteed for tests for two hypotheses. As a consequence, the design of minimax optimal tests for multiple hypotheses becomes significantly harder. In this section, it is discussed why this is the case, why robust detection for multiple hypotheses is a fundamentally different problem, and how \emph{sequential hypothesis testing} can help in solving this problem.

\begin{tcolorbox}[title=\textbf{Robust Detection in Practice: FAQ}]    
  \begin{description}
    \item[Q:] Robustness seems like a waste of resources. Why not aim for optimal performance, if the conditions allow for it?
          
    \item[A:] If the goal is to achieve optimal performance under ideal conditions, then robust detectors are indeed not a good choice. However, in practice it can be better to guarantee ``good enough'' performance under a variety of conditions. For example, the nominal sensitivity of a smoke detector under lab conditions is irrelevant if it is supposed to work reliably for many years in a dusty basement or a poorly insulated attic.
    
    \item[Q:] Why should a detector be designed such that it performs optimally only in pathological worst-case scenarios? Those will never arise in practice!
    
    \item[A:] Many worst-case scenarios, particularly in detection, are not pathological at all; recall the least favorable distributions in Fig.~\ref{fig:lfds_outliers}. Moreover, and more importantly, the point of robust detectors is not that they perform optimally under a \emph{single} peculiar distribution, but that they perform well under \emph{all} distributions in the uncertainty set.%\\
          
    \item[Q:] Robust detectors seem to be very inflexible. Would it not be better to adapt to the true distribution?
    
    \item[A:] Whether or not the true distribution should be estimated critically depends on the type of uncertainty; compare the discussion in Sec.~\ref{sec:minimax_principle}. As a rule of thumb, whenever the uncertainty is such that resolving it seems to require non-parametric estimators, robust detectors are likely to be a better performing alternative.
    
    \item[Q:] There are detectors that have been shown to be asymptotically optimal under mild assumptions. Does this not make robust detectors obsolete?
    
    \item[A:] Asymptotically, distributional uncertainty becomes an almost negligible issue. As the sample size goes to infinity, any unknown property of the true distribution can be inferred with arbitrary precision from the data. In other words, having access to an unlimited amount of data is as good as having perfect knowledge of their distribution. In practice, however, the sample sizes are typically nowhere near the asymptotic regime. In fact, interpreting ``asymptotically optimal'' as ``guaranteed to be close-to optimal'' is a misconception that can lead to a false and potentially dangerous sense of security. Robust detectors, on the other hand, not only provide actual performance guarantees, but also force the designer to explicitly take the uncertainty into account.
          
    \item[Q:] Robust detection seems to be rather niche, is it not better to use well-established techniques?
    
    \item[A:] Robust detection is undeniably niche; but it does not have to stay niche. The theory is well understood and software packages are readily available. 
  \end{description}
\end{tcolorbox}

%------------------------------------------------------------------------------%
\subsection{Optimal Tests for Multiple Hypotheses}
\label{ssec:multi_hypotheses_optimal}
%------------------------------------------------------------------------------%
In analogy to the two-hypothesis case, it is useful to briefly revise optimal non-robust test for multiple hypotheses before entering the discussion of their robust counterparts. Again, assume for now that $\bm{X}_N = (X_1, \ldots, X_N)$ are i.i.d.~with common distribution $P_X$ such that
\begin{equation}
  \mathbb{P}_{\bm{X}_N} = \prod_{n=1}^N P_{X_n} = P_X^N.
\end{equation}
The goal is now to decide between the hypotheses
\begin{equation}
  \mathcal{H}_k\colon \; P_X = P_k,
  \label{eq:multiple_simple_hypotheses}
\end{equation}
where $k = 0, \ldots, K$ and all $P_k$ are known exactly. The decision rule of the test is defined as
\begin{equation}
  \bm{\delta}\colon \mathcal{X}^N \to [0,1]^{K+1},
  \label{eq:decision_rule_multi_hyp}
\end{equation}
where $\bm{\delta} = (\delta_0, \ldots, \delta_K)$ is a $K+1$-dimensional vector whose $k$th entry denotes the conditional probability of deciding for the $k$th hypothesis given the observations $(x_1,\ldots,x_n)$. 

Throughout this section, the focus will be on the weighted sum error cost
\begin{equation}
  C_\text{WSE}(\bm{\delta}; \pmb{\mathbb{P}}) =  \sum_{k=0}^K \lambda_k \mathbb{E}_{\mathbb{P}_k}[1- \delta_k(\bm{X})],
  \label{eq:wse_multiple_hyp}
\end{equation}
where $\pmb{\mathbb{P}} = (\mathbb{P}_0, \ldots, \mathbb{P}_K)$ denotes a vector of distributions, $\lambda_k$, $k = 1, \ldots, K$, denote positive cost coefficients and it is assumed without loss of generality that $\lambda_0 = 1$. Note that the Bayes error cost function can be obtained form \eqref{eq:wse_multiple_hyp} by setting $\lambda_k = \frac{\text{Pr}(\mathcal{H}_k)}{\text{Pr}(\mathcal{H}_0)}$. 

The optimal decision rule is again defined as the one that minimizes a cost function for a given vector of distributions $\pmb{\mathbb{P}}$. For the cost function in \eqref{eq:wse_multiple_hyp}, it can be shown that the optimal decision rule is a \emph{weighted maximum likelihood test} \cite{DeGroot1970, Domingos1997}.

\begin{theorem}
  The optimal decsion rule for the cost function in \eqref{eq:wse_multiple_hyp} is given by
  \begin{equation}
    \delta_k^*(\bm{x}) \begin{cases}
                        \leq 1, & \lambda_k z_k(\bm{x}_N) \geq \lambda_l z_l(\bm{x}_N) \; \forall \, l \neq k \\
                        = 0, & \text{otherwise}
                       \end{cases} 
    \label{eq:lr_test_multi_hyp}
  \end{equation}
  where $k = 0, \ldots, K$, and $z_k \colon \mathcal{X}^N \to \mathbb{R}_+$ is defined as
  \begin{equation}
    z_k(\bm{x}_N) = \frac{\mathrm{d} \mathbb{P}_k}{\mathrm{d} \mathbb{P}_0}(\bm{x}_N) = \prod_{n=1}^n\frac{p_k(x_n)}{p_0(x_n)}.
    \label{eq:lhr_multi_hyp}
  \end{equation}
  \label{th:optimal_test_multi_hyp}
\end{theorem}
The first case on the right-hand side of \eqref{eq:lr_test_multi_hyp} is defined as an inequality in order to allow for randomization.

From the discussion in the previous section it is clear that any minimax optimal robust test needs to be based on the decision rule in \eqref{eq:lr_test_multi_hyp} in combination with a vector $\pmb{\mathbb{Q}}$ of least favorable distributions. However, characterizing or even defining the latter in a meaningful manner is a much more delicate problem for multiple hypotheses.

%------------------------------------------------------------------------------%
\subsection{Characterizing Least Favorable Distributions}
\label{ssec:characterizing_lfds_multi}
%------------------------------------------------------------------------------%
Just as in the two-hypothesis case, uncertainty under multiple hypotheses is introduced by allowing the true distributions to lie in a set of feasible distributions, that is,
\begin{equation}
  \mathcal{H}_k \colon P_{X_n} \in \mathcal{P}_k,
  \label{eq:composite_hypotheses_multi_hyp}
\end{equation}
for all $n = 1, \ldots, N$, where $\mathcal{P}_k$ denotes the uncertainty set under $\mathcal{H}_k$. In analogy to \eqref{eq:composite_hypotheses_joint}, \eqref{eq:composite_hypotheses_multi_hyp} is written more compactly as
\begin{equation}
  \mathcal{H}_k \colon \mathbb{P}_{\bm{X}_N} \in \mathcal{P}_k.
  \label{eq:composite_hypotheses_joint_multi_hyp}
\end{equation}

In Section~\ref{ssec:characterizing_lfds}, three equivalent criteria for the characterization of least favorable distributions of binary detection problems were given. All three criteria can be extended to the multi-hypothesis case. However, here only the generalized version of Criterion~\ref{crit:f-divergence} is stated since it is the most instructive.

\begin{criterion}[Minimum $f$-Dissimilarity]
	If a vector of distributions $(Q_0, \ldots, Q_K)$ minimizes
	\begin{equation}
	  D_{f}(P_1, \ldots, P_K \Vert P_0) = \int_\mathcal{X} f\biggl( \frac{p_1(x)}{p_0(x)}, \ldots, \frac{p_K(x)}{p_0(x)} \biggr) p_0(x) \, \mathrm{d}x 
	  \label{eq:lfd_f-div_multi_hyp}
	\end{equation}
	over $(P_0, \ldots, P_K) \in \mathcal{P}_0 \times \dots \mathcal{P}_K$ for all twice differentiable convex functions $f \colon \mathbb{R}_+^K \to \mathbb{R}$, then the joint distributions $\mathbb{Q}_k = Q_k^N$ are least favorable w.r.t.~the cost function in \eqref{eq:wse_multiple_hyp} for all thresholds $\lambda$ and all sample sizes $N$.
	\label{crit:f-divergence_multi_hyp}
\end{criterion}  

In a nutshell, Criterion~\ref{crit:f-divergence} is obtained from Criterion~\ref{crit:f-divergence_multi_hyp} by replacing the class of $f$-divergences by that of $f$-\emph{dissimilarities}. The latter are  a natural extension of $f$-diverences to multiple distributions and inherit all relevant properties. They were first proposed and studied by Gy\"orfi and Nemetz in the 1970s \cite{GyorfiNemetz1975, GyorfiNemetz1977, GyorfiNemetz1978} and play an important role in the general theory of statistical decision making. In particular, the connection between $f$-dissimilarities and Bayesian risks has been a topic of high interest in statistics \cite{Nguyen2009}, signal processing \cite{Varshney2011} and machine learning \cite{Reid2011}. 

Although Criterion~\ref{crit:f-divergence_multi_hyp} is a neat extension, its significance is of a purely negative nature. In a nutshell, there exist no distributions for $K > 1$ that satisfy Criterion~\ref{crit:f-divergence_multi_hyp} under useful (nontrivial) uncertainty models. As a consequence, no multi-hypothesis equivalent to the minimax optimal tests for two hypotheses exists. 

\begin{figure*}
	\centering
	\includegraphics[scale=0.9]{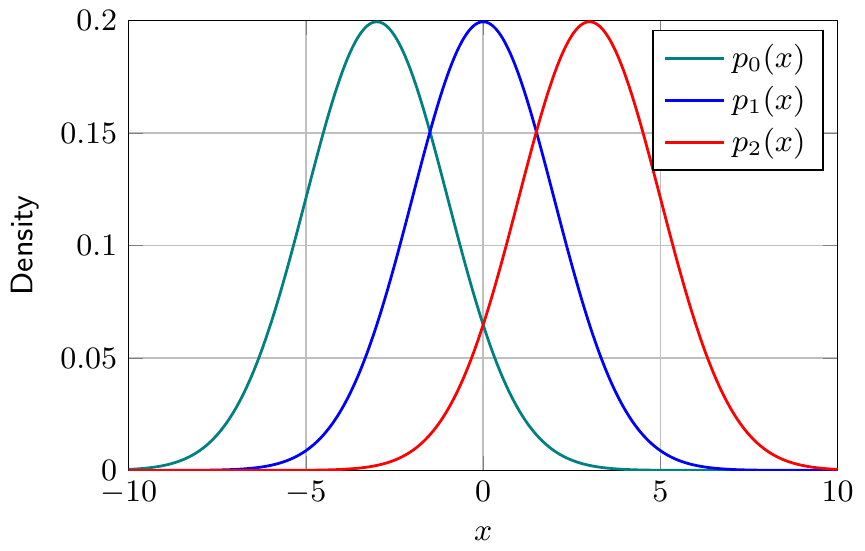}%
	\hspace{1em}
	\includegraphics[scale=0.9]{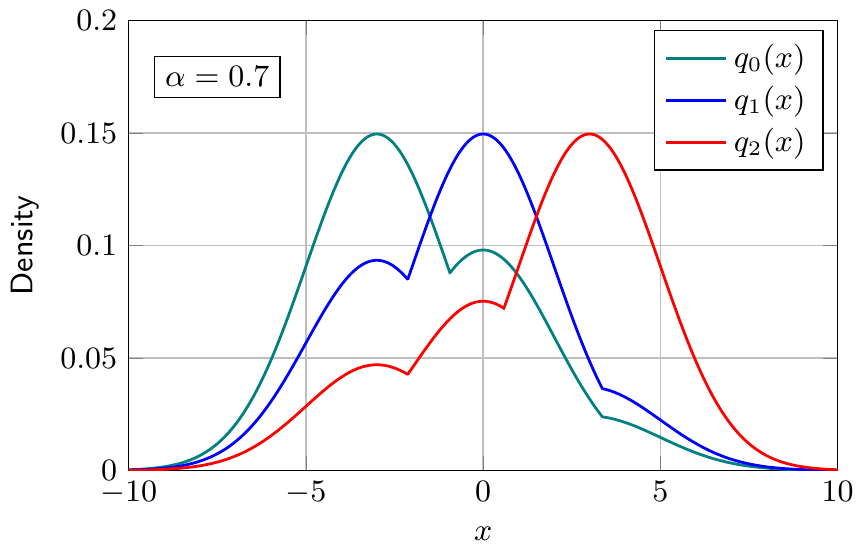}\\[1ex]
	\includegraphics[scale=0.9]{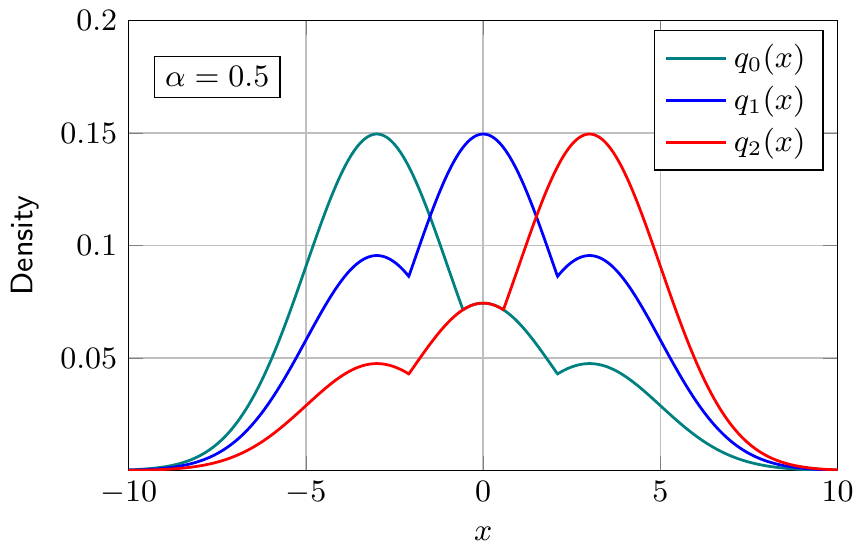}%
	\hspace{1em}
	\includegraphics[scale=0.9]{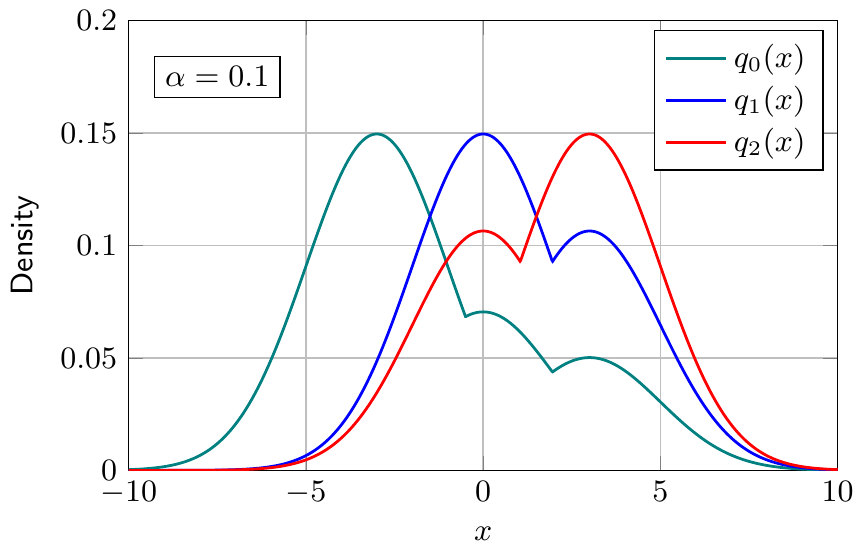}
	\caption{Examples of different least favorable distributions for the $f$-dissimilarity in \eqref{eq:convex_KL} under \SI{25}{\percent} contamination and with nominal densities shown in the top left plot. By inspection, different weights $\alpha$ lead to different minimizers.}
	\label{fig:dissimilarity_illustration}
\end{figure*}

Before discussing a possible way of working around this problem, it is instructive to look at an example that illustrates why least favorable distributions for multiple hypotheses cannot be defined in analogy to those for two hypotheses. To this end, consider the following $f$-dissimilarity, which is simply a convex sum of KL divergences,
\begin{multline}
  D_f(P_1, P_2 \Vert P_0) = \alpha D_\text{KL}(P_1 \Vert P_0) \\ + (1-\alpha) D_\text{KL}(P_2 \Vert P_0), \quad \alpha \in [0, 1].
  \label{eq:convex_KL}
\end{multline}
It is not hard to verify that the right-hand side of \eqref{eq:convex_KL} is a valid $f$-dissimilarity. Now, consider three uncertainty sets $\mathcal{P}_0$, $\mathcal{P}_1$, $\mathcal{P}_2$ of the $\varepsilon$-contamination type with \SI{25}{\percent} contamination ratio and nominal distributions that are shifted Gaussians with identical variance $\sigma^2 = 4$ and shift $\Delta \mu = 3$. The nominal densities are shown in the top left plot of Fig.~\ref{fig:dissimilarity_illustration}. The other three plots of Fig.~\ref{fig:dissimilarity_illustration} depict the densities of the distributions that minimize \eqref{eq:convex_KL} under this uncertainty model for different values of $\alpha$. It can clearly be seen how different combination weights affect the shape of the least favorable distributions and cause probability mass to be shifted towards $P_0$, in the upper right plot, $P_1$, in the lower left plot, and $P_2$, in the lower right plot. The effect illustrated with this example holds true in general: for more than two distributions, different similarity measures lead to different weights for the pairwise similarities and, as a consequence, to least favorable distributions that depend on $f$.

The same idea can be expressed in terms of error probabilities. In a test for two hypotheses, there is only one possible source of errors: confusing $P_0$ and $P_1$. In a test for multiple hypotheses, there are also multiple sources of error, namely any pair of distributions $P_k$ and $P_{j \neq k}$ that can be confused. However, making $P_0$ and $P_1$ as similar as possible, thus maximizing this particular probability of confusion, might decrease the similarity between, say, $P_0$ and $P_2$---compare the different least favorable distributions in Fig.~\ref{fig:dissimilarity_illustration}. Hence, there exist in general no distributions that jointly maximize \emph{all} pairwise error probabilities. 

These considerations already hint at a common approach to overcoming the problems with the design of minimax optimal detectors for multiple hypotheses. Namely, instead of \emph{jointly} testing for $K+1$ hypotheses, one can test each pair of hypotheses $\mathcal{H}_k$, $\mathcal{H}_{j \neq k}$ \emph{individually}. This kind of procedure is known as \emph{mulitple testing} or \emph{mulitple comparison} and has been studied extensively in the literature; see \cite{Hochberg1987, Tukey1991, BenjaminiHochberg1995, Shaffer1995, Zoubir2002_Sensors, Benjamini2010, Dutta2012, AriasCastro2017, Rabinovich2017} and the references therein. A particularly attractive implementation of a detector based on this approach is the sequential \emph{matrix} likelihood/probability ratio  test \cite{Tartakovsky2014}, whose robust version is based on separate pairs of least favorable distributions for each pairwise test \cite{Leonard2018b, Leonard2018a}. Sequential tests will be revisited in more detail shortly. Detectors based on robust versions of multiple comparison procedures are indeed robust in the sense that they do not breakdown under model uncertainties. However, they are \emph{not} minimax robust and do not necessarily provide performance guarantees. The difficulties in obtaining such guarantees for multiple pairwise tests stems from the fact that some of the pairwise tests are performed under neither the null hypothesis nor the alternative hypothesis. That is, while the test is supposed to decide for either $P_0$ or $P_1$, in reality the data might have been drawn from a third distribution, say, $P_2$. Unless $P_2$ is by chance included in $\mathcal{P}_0$ or $\mathcal{P}_1$, the behavior of the minimax tests under $P_2$ is not considered in its design.

We have only scratched the surface, and we will not enter a more detailed discussion of robust pairwise tests at this point. Since multiple testing is a distinct topic with its very own characteristics and terminology, an in-depth discussion is beyond the scope of this paper. Moreover, it would lead away from the central theme of strictly minimax detection.

Returning to the latter, it can be summarized that minimax robust detectors for multiple hypotheses are still in their infancy and are held back not just by implementation issues, but by fundamental theoretical limitations. One possible way forward, which is implicitly already followed by methods using multiple pairwise tests, is to give up on strict minimax optimality and to focus on defining and analyzing robustness in detection in closer analogy to robustness in estimation. This avenue is discussed in more detail in Section~\ref{sec:beyond_minimax}. Alternatively, instead of relaxing the definition of robustness, the problem formulation of the underlying test can be relaxed by not assuming the sample size $N$ to be fixed beforehand. This leads into the territory of  \emph{sequential} tests, which adapt the sample size to the observed data. The existence of strictly minimax optimal sequential tests has recently been shown in \cite{Fauss2019_aos}. Combining the benefits of adaptive procedures with strict robustness guarantees makes them a promising choice for future applications that require both efficiency and reliability.

%%%%%%%%%%%%%%%%%%%%%%%%%%%%%%%%%%%%%%%%%%%%%%%%%%%%%%%%%%%%%%%%%%%%%%%%%%%%%%
\section{Minimax Sequential Detection}
\label{sec:sequential_detection}
%%%%%%%%%%%%%%%%%%%%%%%%%%%%%%%%%%%%%%%%%%%%%%%%%%%%%%%%%%%%%%%%%%%%%%%%%%%%%%

The idea underlying sequential detection is the following: instead of fixing the number of samples, $N$, in advance, samples are collected one at a time until the probability of making a correct decision based on the current collection is sufficiently large. The advantage of this approach is that it can adapt the sample size to the quality of the observations at hand. That is, the sample size can be reduced if the first $n < N$ observations are already highly significant, or it can be increased if $N$ observations are not yet significant enough. In practice, the latter case occurs less frequently so that sequential tests typically admit an \emph{expected} sample size considerably below that of fixed sample size tests with identical error probabilities \cite{Wald1947, Tartakovsky2014}. 

The price one has to pay for this increase in efficiency is that the number of samples becomes a function of the realization of the random sequence $(X_1, X_2, \ldots)$. Consequently, the sample size, or run-length, of a sequential test is itself a random variable whose properties have to be taken into account in the test design. A typical cost function in sequential detection is the weighted sum of the expected run-length and the error probabilities, that is
\begin{equation}
  C_\text{SEQ}(\bm{\delta}; \pmb{\mathbb{P}}) = \mathbb{E}_{\mathbb{P}_0}\bigl[ \tau \bigr] + \sum_{k=1}^K \lambda_k \mathbb{E}_{\mathbb{P}_k}\bigl[ 1-\delta_k(\bm{X}_\tau) \bigr],
  \label{eq:sequential_test_cost}
\end{equation}
where $\tau$ denotes the random stopping time of the test and $\bm{\delta}$ is a decision rule of the form \eqref{eq:decision_rule_multi_hyp}. Note that \eqref{eq:sequential_test_cost} corresponds to a test for $K$ hypotheses $\mathcal{H}_1$ to $\mathcal{H}_K$, that is, there is no null hypothesis. Instead, in what follows $\mathbb{P}_0$ refers to the distribution under which the expected run-length of the test is supposed to be minimum. In practice, $\mathbb{P}_0$ is often chosen as one of the hypothesized distributions $\mathbb{P}_k$, $k \geq 1$. However, in particular in the context of robust detection, other choices can also be useful. 

The reason for not including a null hypothesis in \eqref{eq:sequential_test_cost} is two-fold. On the one hand, this formulation leads to a notation that is more in line with the one used in the previous sections. On the other hand, it reflects the fact that in sequential hypothesis testing all error probabilities can be made arbitrarily small by increasing the expected sample size. Consequently, at least from a mathematical point of view, it becomes unnecessary to single out one hypothesis as a \emph{null} hypothesis.

\begin{figure}
	\centering
	\includegraphics[scale=0.9]{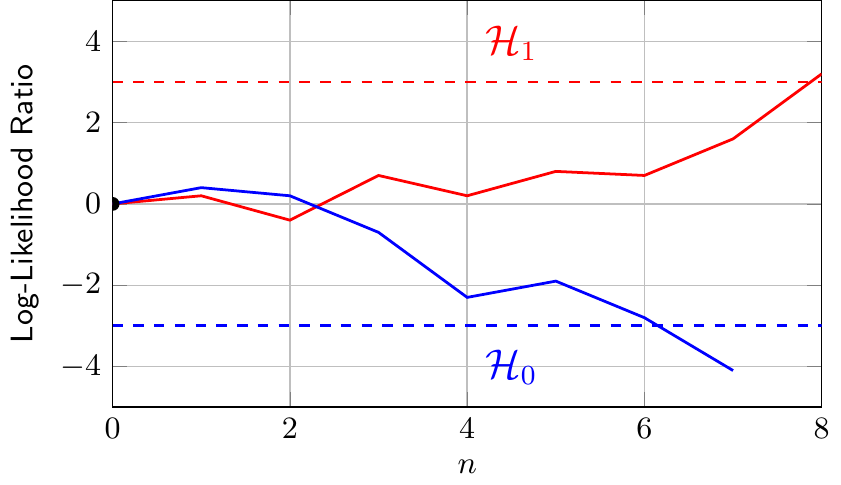}%
	\caption{Examples of log-likelihood ratio trajectories as functions of the sample number $n$. If the lower threshold ($\log B = -3$) is crossed, the test decides for $\mathcal{H}_0$, if the upper threshold ($\log A = 3$) is crossed, the test decides for $\mathcal{H}_1$. }
	\label{fig:sequential_test_illustration}
\end{figure}

In his seminal book on sequential detection \cite{Wald1947}, Wald showed that the optimal sequential test for two simple hypotheses $\mathcal{H}_1$ and $\mathcal{H}_2$ is a likelihood ratio test of the form \eqref{eq:lr_test}, but with two thresholds instead of one. More precisely, after the sample $x_n$ was observed, the likelihood ratio $z$ is updated according to
\begin{equation}
  z(\bm{x}_n) = z(\bm{x}_{n-1}) \frac{p_2(x_n)}{p_1(x_n)}
\end{equation}
and a decision is made according to the rule
\begin{equation}
  \delta^*(\bm{x}_n) = \begin{cases}
              1,      & z(\bm{x}_n) \geq A \\
              0,      & z(\bm{x}_n) \leq B, \\
              \text{continue sampling}, & \text{otherwise}
            \end{cases}
\end{equation}
where $A$ and $B$ are the threshold parameters. This so-called sequential probability ratio test (SPRT) is illustrated in Fig.~\ref{fig:sequential_test_illustration}. Wald showed that for correctly chosen thresholds $A$ and $B$, the SPRT is optimal in the sense that it satisfies the constraints on the error probabilities with equality and at the same time admits minimum expected run-length under both hypothesized distributions. Similar optimality properties of the SPRT also hold for Bayesian cost functions \cite{Tartakovsky2014}. 

Sequential tests for multiple hypotheses are covered in \cite{Novikov2009_multiple_hypotheses} and \cite{Fauss2019_aos}. While an in-depth discussion is beyond the scope of this paper, the underlying concepts can be understood without going into technical details. Essentially, a sequential test for multiple hypotheses compares the cost for stopping the test at the current sample to the expected cost for continuing the test using an optimal decision rule. This leads to a cost function of the form
\begin{equation}
  \rho(\bm{z}) = \min\{\, g(\bm{z}) \,,\, 1 + D(\bm{z}) \,\}
  \label{eq:optimal_cost_sequential}
\end{equation}
where $\rho$ denotes the optimal cost, which is given by the minimum of the cost for stopping at the current sample, $g$, and the cost for continuing the test, $1 + D$. Note that the additional cost of $1$ in the latter term arises from the extra sample that is taken when deciding to continue the test. The vector $\bm{z} = (z_1, \ldots, z_K)$ denotes the likelihood ratios defined in \eqref{eq:lhr_multi_hyp}, which can be shown to be a sufficient statistic for the cost function in \eqref{eq:sequential_test_cost}. The exact form of $\rho$ and $g$ depends on the distributions $\pmb{\mathbb{P}}$ as well as the weights $\bm{\lambda}$; see \cite{Novikov2009_multiple_hypotheses} for more details. In any case, both $\rho$ and $g$ are guaranteed to be concave in $\bm{z}$, a property that will become important shortly.

When running a sequential test for multiple hypotheses, each likelihood ratio $z_k$ is updated according to
\begin{equation}
  z_k(\bm{x}_n) = z_k(\bm{x}_{n-1}) \frac{p_k(x_n)}{p_0(x_n)},
  \label{eq:lr_update}
\end{equation}
$k = 0, \ldots, K$, and the test stops as soon the cost for continuing exceeds the cost for stopping, that is, 
\begin{equation}
  \tau = \min\bigl\{\, n \geq 0 : g(\bm{z}(\bm{x}_n)) \leq 1 + D(\bm{z}(\bm{x}_n)) \,\bigr\}
\end{equation}
Once the test stops, a decision is made according to Theorem~\ref{th:optimal_test_multi_hyp} with $N = \tau$. 

How does robustness enter in all of this? In analogy to the fixed sample size case, the minimax robust sequential test is formulated in terms of uncertainty sets of feasible distributions. In general, it makes sense to assume that the \emph{conditional} distribution $X_n|\bm{X}_{n-1}$ is subject to uncertainty. For random processes with Markovian representations this problem has been studied in \cite{Fauss2019_aos}. However, in order to reduce the technicalities, $(X_1, X_2, \ldots)$ are assumed to be independent in what follows. In this case, uncertainty is introduced as in \eqref{eq:composite_hypotheses_multi_hyp}, by assuming each $P_{X_n}$ to be within an uncertainty set $\mathcal{P}_k$ under the corresponding hypothesis $\mathcal{H}_k$. A special case is the uncertainty set $\mathcal{P}_0$, which determines the set of distributions over which the worst-case expected run-length is minimized. Natural choices for this set are $\mathcal{P}_0 = \mathcal{P}_k$ for some $k > 0$, or $\mathcal{P}_0 = \bigcup_{k \in \mathcal{K}} \mathcal{P}_k$, where $\mathcal{K} \subset \{1, \ldots, K\}$, that is, worst-case expected run-length under one or more hypotheses is considered. However, in principle, $\mathcal{P}_0$ can be chosen independently. It can even be chosen to be the set of \emph{all} distributions on the sample space, so that the run-length is guaranteed to remain bounded for all possible realizations of the underlying random process.\footnote{This particular problem, which is a nonparametric variant of the Kiefer--Weiss problem, is studied in \cite{FaussPoor2020}, and can be shown to be solved by a non-standard sequential test that admits randomized stopping rules and other counterintuitive properties. A detailed discussion of this corner case is beyond the scope of this paper.}

Once the uncertainty sets are fixed, the least favorable distributions need to be determined. While for fixed-sample-size tests this turned out to be an all but impossible task, it becomes feasible in the sequential case. In order to establish a connection to Criterion~\ref{crit:f-divergence_multi_hyp}, recall that $D$ on the right-hand side of \eqref{eq:optimal_cost_sequential} is the expected cost of continuing the test with an optimal decision rule. Hence, it can be expressed recursively as a function of the optimal cost $\rho$, namely,
\begin{equation}
  D(\bm{z}) = \mathbb{E}_{P_0}\biggl[ \rho\biggl(z_1 \frac{p_1(X)}{p_0(X)}, \ldots, z_K \frac{p_K(X)}{p_0(X)} \biggr) \biggr], 
  \label{eq:cost_continue}
\end{equation}
where the expectation is taken over the likelihood ratios after the next sample has been observed, that is, after an update of the form \eqref{eq:lr_update}. Since $\rho$ is a concave function, it holds that
\begin{align}
  - D(\bm{z}) &= \int_\mathcal{X} -\rho\biggl(z_1 \frac{p_1(x)}{p_0(x)}, \ldots, z_K \frac{p_K(x)}{p_0(x)} \biggr) p_0(x) \, \mathrm{d}x \\[0.5ex]
  &= D_{-\rho}\bigl(z_1 P_1, \ldots, z_K P_K \Vert P_0\bigr)
  \label{eq:rho-dissimilarity}
\end{align}
is a weighted $f$-dissimilarity, whose weights correspond to the current likelihood ratios. Alternatively, $D_{\rho}$, being a negative $f$-dissimilarity, can be considered an $f$-\emph{similarity}. It can be shown that given $\bm{X}_n = \bm{x}_n$, the least favorable distributions of the next sample, $X_{n+1}$, are those that minimize the $f$-dissimilarity in \eqref{eq:rho-dissimilarity} over the uncertainty sets. More formally,
\begin{equation}
  \bm{Q}_{X_{n+1}} \in \argmax_{P_k \in \mathcal{P}_k} \; D_{\rho}\bigl(z_1 P_1, \ldots, z_K P_K \Vert P_0\bigr),
  \label{eq:multi_lfds}
\end{equation}
where $z_k = z_k(\bm{x}_n)$. 

In a nutshell, the minimax sequential test works as follows: There exists a convex function $\rho$ that determines an appropriate $f$-similarity $D_\rho$. This $f$-similarity, weighted by the likelihood ratios $z_1, \ldots, z_K$, determines the least favorable distributions of the next observation. The latter are in turn used to update the likelihood ratios. The test stops once the similarity measure $D_\rho$ exceeds a certain threshold, determined by the function $g$. This stopping rule can be interpreted as the distributions becoming \emph{more similar}, that is, \emph{more costly to separate}, when continuing the test compared to stopping it and deciding based on the observations at hand. The  procedure of the robust sequential test is summarized in Table~\ref{alg:minimax_sequential}.

\begin{table}[tb]
  \begin{algorithmic}[1]
    \STATE \textbf{input} \par
    \hskip\algorithmicindent Cost functions $g$ and $\rho$ (implicitly dependent on $\lambda_1, \ldots, \lambda_K$)\par    
    \hskip\algorithmicindent Uncertainty sets $\mathcal{P}_0, \ldots, \mathcal{P}_K$
    \STATE \textbf{initialize} \par
    \hskip\algorithmicindent Set $n = 0$ \par 
    \hskip\algorithmicindent Set $z_k = 1$ for $k = 1, \ldots, K$.  
    \LOOP
    \STATE Calculate least favorable distributions
           \begin{equation*}
            \bm{Q} \in \argmax_{P_k \in \mathcal{P}_k} \; D_{\rho}\bigl(z_1 P_1, \ldots, z_K P_K \Vert P_0\bigr),
            \end{equation*}
    \IF{$g(\bm{z}) \leq 1 + D_{\rho}\bigl(z_1 Q_1, \ldots, z_K Q_K \Vert Q_0\bigr)$}
        \STATE Stop test with a decision according to \eqref{eq:lr_test_multi_hyp}
    \ELSE
        \STATE Set $n \leftarrow n + 1$ and take next observation $x_n$ 
        \STATE Update likelihood ratios
               \begin{equation*}
                  z_k \leftarrow z_k \frac{q_k(x_n)}{q_0(x_n)},
              \end{equation*}
    \ENDIF
    \ENDLOOP
  \end{algorithmic}
  \caption{Pseudo code of a minimax robust sequential test for $K$ hypotheses}
  \label{alg:minimax_sequential}
\end{table}

At this point, it is instructive to take a step back and return to the question of how the sequential test is able to solve the problems arising with minimax fixed-sample-size tests for multiple hypothesis that were discussed in the previous section. Recall that the fundamental obstacle turned out to be that for $K > 1$ no distributions exist that jointly minimize all $f$-dissimilarities. Instead, different choices of $f$ give rise to different definitions of joint similarity. Hence, it followed that minimax optimal tests for multiple hypotheses only exist for the single-sample case. This is where the sequential test comes in: \emph{By introducing the option to take another sample, it allows one to break a test based on $N$ samples down into $N$ consecutive single-sample tests, which in turn admit well defined least favorable distributions.} 

The price one has to pay for this is that the least favorable distributions themselves need to be updated with every new observation, that is, they become \emph{data dependent}. In this sense, it can be argued that the minimax robust sequential test is no longer true to the spirit of minimax robustness, whose characteristic feature, as discussed in Sec.~\ref{sec:minimax_principle}, is that the procedure can be determined entirely offline and does \emph{not} depend on the data. However, this does not mean that minimax sequential tests should in turn be classified as purely adaptive; they rather combine aspects of both robust and adaptive procedures:
The robust or \emph{static} aspect is that the function $\rho$ and the induced similarity measure $D_\rho$ are fixed throughout the test and can indeed be calculated offline. The \emph{dynamic} aspect is reflected in the likelihood ratios $z_1, \ldots, z_K$, which are updated online. Large likelihood ratio values imply that a decision for the corresponding hypothesis is likely, small likelihood ratio values imply that a decision is unlikely. Hence, distributions with large weights have a strong influence on the similarity measure, distributions with small weights have little influence. In other words, the similarity measure automatically adapts to the data---not by changing the similarity measure itself, but by changing the weights of the individual distributions. 

Since, by design, the fixed-sample-size minimax test needs to make a decision in a ``single shot'', it lacks this ability to successively adapt the similarity measure and weights to the data. In the two-hypothesis case, it was possible to turn this limitation into a strength, by exploiting the fact that a ``one fits all'' solution existed that is optimal with respect to all $f$-divergences at the same time. In the multi-hypothesis case, however, the class of similarity measures that need to be considered becomes too large to be tackled in this manner. By relaxing the problem formulation and allowing data depended stopping rules, it can be narrowed down again, at the expense of an increase in complexity.

This increase in complexity is in fact significant. While the underlying theory is well-understood by now, designing and implementing minimax sequential tests still requires a considerable computational effort. The procedure proposed in \cite{Fauss2019_aos} designs a minimax sequential test by alternatingly updating $\rho$ and $(Q_0, \ldots, Q_K)$. The former can be done by solving a system of coupled integral equations, the latter by solving the minimization problem in \eqref{eq:multi_lfds}. Numerical algorithms for both problems can be found in the general optimization literature. However, more efficient methods have been proposed recently that exploit the additional structure of both problems in the context of robust detection. In \cite{Fauss2015}, it has been shown that the function $\rho$ can be determined by solving a linear program, which is a standard problem in numerical optimization and can be solved efficiently even for large problem sizes. The minimization problem in \eqref{eq:multi_lfds} is studied in detail in \cite{Fauss2018_tsp}, where an iterative fixed point algorithm is proposed that works by repeatedly finding the scalar root of a monotonic function, which again is a well-known problem in numerics. Moreover, the proposed algorithm is highly parallelizable and has low memory requirements. In combination, both algorithms provide a powerful numerical tool that is tailored for the design of robust tests. Implementations in Python, MATLAB, and C are available online \cite{GitHubRepo}.

In order to make the results and concepts discussed so far more tangible, a numerical example of a minimax sequential test is presented next.

%------------------------------------------------------------------------------%
\subsection{A Numerical Example}
\label{ssec:numerical_example}
%------------------------------------------------------------------------------%
The example presented here is taken from \cite{Fauss2019_aos}, and the reader is referred to this reference for a more detailed discussion. In addition, \cite{Fauss2019_aos} also presents an example with dependent observations, which is beyond the scope of this overview. 

For this example, all $X_n$, $n \geq 1$, are assumed to be independent and distributed on the interval $\mathcal{X} = [-1,1]$. The task is to decide between the following three hypotheses:
\begin{equation}
  \begin{aligned}
    \mathcal{H}_1 &\colon P_{X_n} \in \mathcal{P}_1, \\
    \mathcal{H}_2 &\colon P_{X_n} \in \mathcal{P}_2, \\
    \mathcal{H}_3 &\colon P_{X_n}  = \mathcal{U}_{[-1,1]},
  \end{aligned}
  \label{eq:sequential_example_hyp}
\end{equation}
where $\mathcal{U}_{[a,b]}$ denotes the continuous uniform distribution on the interval $[a, b]$ and the uncertainty sets $\mathcal{P}_1$, $\mathcal{P}_2$ are of the density band form with
\begin{align}
  p_1'(x) &= a e^{-2x} + 0.1, & p_1''(x) &= a e^{-2x} + 0.3, 
  \label{eq:seq_unctny_set_1} \\
  p_2'(x) &= a e^{2x} + 0.1,  & p_2''(x) &= a e^{2x} + 0.3,
  \label{eq:seq_unctny_set_2}
\end{align}
where $a \approx 0.1907$ is chosen such that $P_1'(\mathcal{X}) = P_2'(\mathcal{X}) = 0.9$ and $P_1''(\mathcal{X}) = P_2''(\mathcal{X}) = 1.1$. The expected run-length is minimized under $\mathcal{H}_3$, that is, $P_0 = \mathcal{U}_{[-1,1]}$, so that $z_3 = 1$ and $\rho$ in \eqref{eq:optimal_cost_sequential} can be written as a function of $(z_1, z_2)$ only. 

The scenario in this example can arise, for instance, in monitoring applications, where $\mathcal{H}_3$ corresponds to an ``in control'' state in which the distribution of the data is known almost exactly, while $\mathcal{H}_1$ and $\mathcal{H}_2$ correspond to two different ``out of control'' states, with only partially known distributions. If it needs to be established that the system is ``in control'' before a certain procedure starts, it is reasonable to minimize the expected run-length of the test under the ``in control'' distribution, while still requiring it to be insensitive to distributional uncertainties in the ``out of control'' states.

\begin{figure}
	\centering
	\includegraphics[scale=0.9]{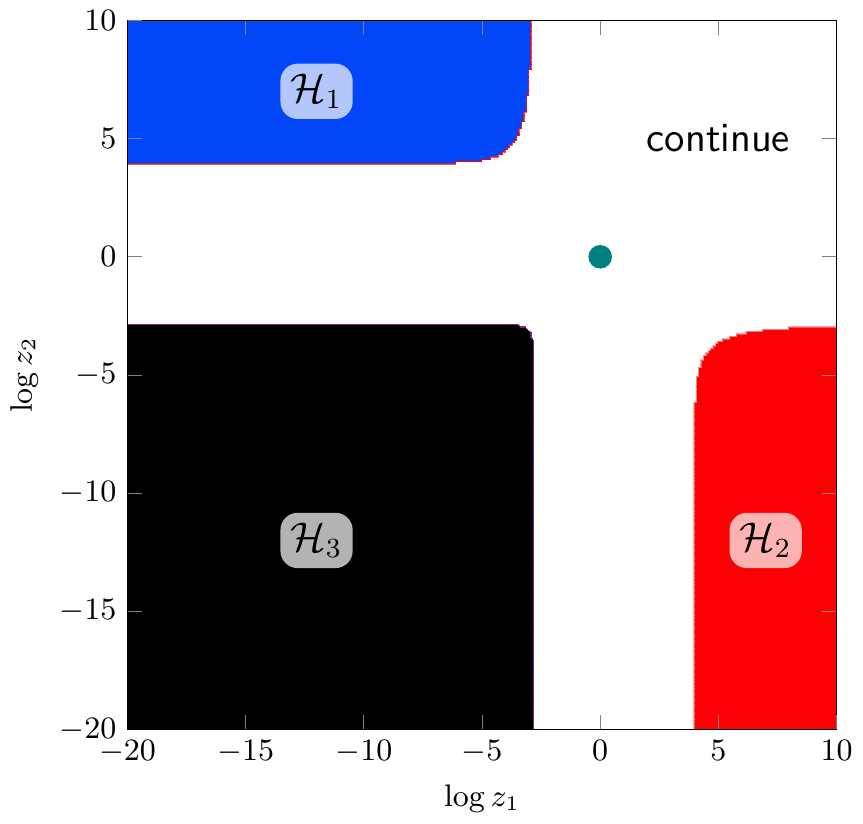}%
	\caption{Minimax optimal decision regions in the log-likelihood ratio plane when sequentially testing the hypotheses in \eqref{eq:sequential_example_hyp}.}
	\label{fig:policy}
\end{figure}

The example was solved numerically on a discretized likelihood ratio plane. The cost coefficients were chosen to be $\bm{\lambda}^* \approx (133.41, 133.41, 45.41)$, resulting in error probabilities of $\approx$\SI{1}{\percent}; see \cite{Fauss2019_aos} for how to set $\bm{\lambda}$ such that the test meets certain error probability constraints. The resulting testing policy is depicted in Fig.~\ref{fig:policy}. In analogy to the regular SPRT, the minimax optimal test consists of two corridors that correspond to a binary test between $\mathcal{H}_{\{1,2\}}$ and $\mathcal{H}_3$, respectively. Interestingly, there is a rather sharp intersection of the two corridors so that the test quickly reduces to a quasi-binary scenario. 

\begin{figure*}
	\centering
	\includegraphics[scale=0.9]{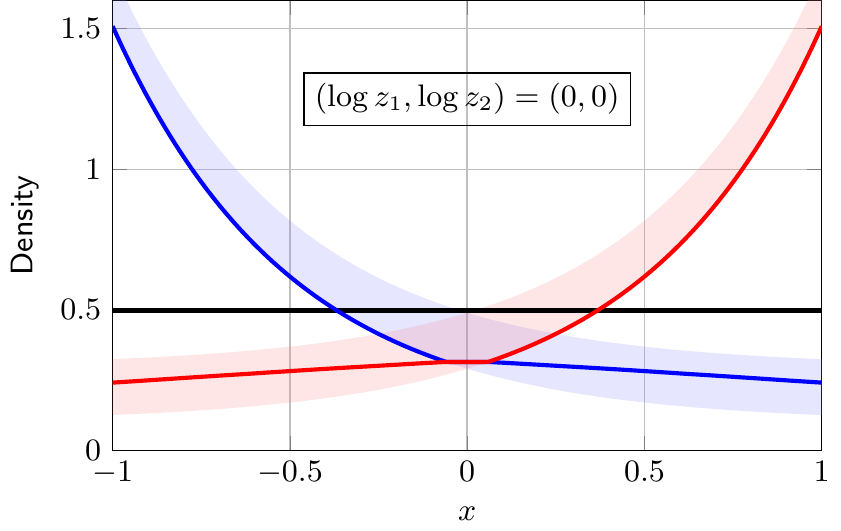}%
	\hspace{1em}
	\includegraphics[scale=0.9]{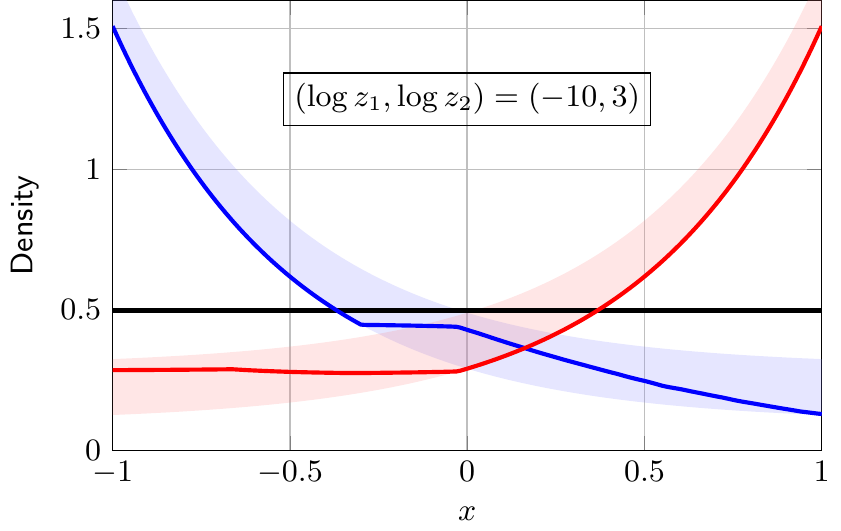}\\[1ex]
	\includegraphics[scale=0.9]{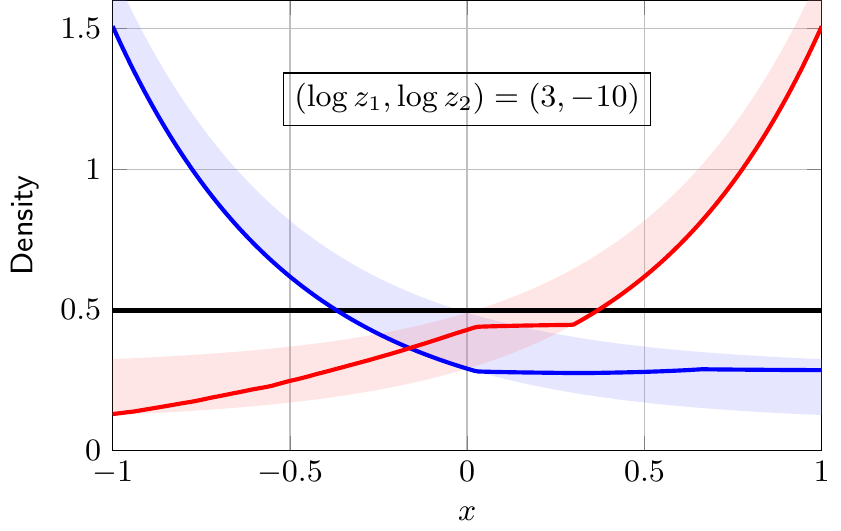}%
	\hspace{1em}
	\includegraphics[scale=0.9]{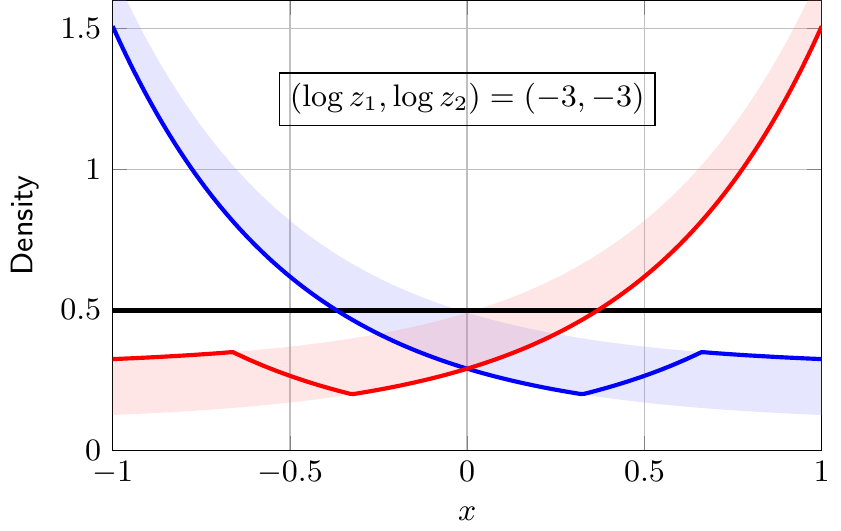}\\
	\includegraphics{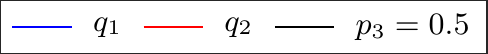}
	\caption{Examples of least favorable distributions for different values of the test statistic when sequentially testing the hypotheses in \eqref{eq:sequential_example_hyp}. The uncertainty sets in \eqref{eq:seq_unctny_set_1} and \eqref{eq:seq_unctny_set_2} are indicated by the shaded areas.}
	\label{fig:sequential_lfds}
\end{figure*}

Four examples of the corresponding least favorable densities are depicted in Fig.~\ref{fig:sequential_lfds}. In line with the previous discussion, the densities change significantly, depending on the state of the test statistic. In the top left plot, the test statistic is in its initial state, meaning that there is no preference for either hypothesis. Consequently, the least favorable densities are chosen such that all three distributions are equally similar to each other, which in this case implies that they are symmetric around the $y$-axis and that $q_1$ and $q_2$ jointly mimic the uniform density $p_3$. Also note that $q_1$ and $q_2$ overlap on an interval around $x = 0$, which can be considered the multi-hypothesis equivalent of the censored likelihood ratio discussed in Sec.~\ref{ssec:design_and_implementation}. As the test statistic is updated, the least favorable distributions change. In the upper right and the lower left plot of Fig.~\ref{fig:sequential_lfds} two cases are depicted where the test has a strong preference for $\mathcal{H}_1$ or $\mathcal{H}_2$, respectively; compare the decision regions in Fig.~\ref{fig:policy}. In both cases, the least favorable densities are no longer symmetric, but their probability masses are shifted, their tail-behavior is noticeably different, and the interval of overlap can no longer be observed. Finally, in the lower right plot, there is a strong preference for $\mathcal{H}_3$, which leads to $q_1$ and $q_2$ both shifting as much probability mass as possible to their tails in order to reduce the significance of the corresponding observations. It is interesting to observe that an imminent decision for $\mathcal{H}_3$ casues $q_1$ and $q_2$ to become less similar to each other in order to increase their joint similarity to $p_3$. This is in contrast to the initial state depicted in the upper left plot, where $q_1$ and $q_2$ are also such that they approximate $p_3$, but at the same time they need to be similar to each other.

Although it is not obvious from mere inspection of the least favorable densities, the idea of robustifying a test by setting the test statistic to a constant on certain regions of the sample space carries over to the multi-hypothesis case as well. However, it does so in a modified form. Namely, for each $k = 0, \ldots, K$, there exists a region of the sample space on which
\begin{equation}
  \rho_{z_k}'\biggl(z_1 \frac{q_1(x)}{q_0(x)}, \ldots, z_K \frac{q_K(x)}{q_0(x)} \biggr) = \text{const},
  \label{eq:derivative_contour}
\end{equation}
where $\rho_{z_k}'$ denotes the partial derivative of $\rho$ w.r.t.~$z_k$. In the one-dimensional case, \eqref{eq:derivative_contour} reduces to 
\begin{equation}
  \rho_{z_1}'\biggl( z_1 \frac{q_1(x)}{q_0(x)} \biggr) = \text{const.} \quad \Rightarrow \quad \frac{q_1(x)}{q_0(x)} = \text{const},
\end{equation}
which recovers the characteristic constant likelihood ratios of the two-hypothesis case. In the multi-hypothesis case, however, this simplification no longer holds. Instead, the likelihood ratios of the least favorable distributions, as functions of $x$, now follow the \emph{contour lines of the partial derivatives of $\rho$}. The latter can be shown to be closely related to the error probabilities and the expected run-length of the test. Hence, the constant likelihood ratios in the two-hypothesis case and the contour-line-tracing likelihood ratios in the multi-hypothesis case, both imply a constant \emph{performance} over the corresponding regions of the sample space. This is in line with the general property of minimax procedures to admit flat performance profiles over the uncertainty sets, recall the discussion in Sec.~\ref{sec:minimax_principle}. A more detailed discussion of this connection requires technical results that are beyond the scope of this paper, but can be found in \cite{Fauss2016_thesis, Fauss2017_ISI, Fauss2018_tsp}.

In summary, this example shows that robust sequential tests provide a viable option for enabling the use of strictly minimax tests for multiple hypotheses. The seamless extension to multiple hypotheses is a great advantage and opens up numerous potential applications in practice. However, at this point in time, one quickly runs into complexity constraints when trying to extend robust sequential tests beyond simple toy examples. Hence, more research is required before they will become useful in real-world scenarios. For now, the significant additional complexity will only be warranted in special corner cases. 

One promising application, namely, forward looking ground penetrating radar, is presented in the next section, where it is shown how robust detectors can already help to improve the state-of-the-art in some areas.

%%%%%%%%%%%%%%%%%%%%%%%%%%%%%%%%%%%%%%%%%%%%%%%%%%%%%%%%%%%%%%%%%%%%%%%%%%%%%%%%
\section{Applications of Robust Detectors}
\label{sec:application}
%%%%%%%%%%%%%%%%%%%%%%%%%%%%%%%%%%%%%%%%%%%%%%%%%%%%%%%%%%%%%%%%%%%%%%%%%%%%%%%%

In this section, examples of current and possible future applications of robust detectors are presented. In the first part, it is shown how robust techniques can lead to noteworthy performance improvements for the detection of buried objects, such as unexploded ordnance (UXO), via ground penetrating radar (GPR) \cite{Pambudi2020, Pambudi2019_EUSIPCO}. This application will be used as a realistic example to illustrate that robust detectors are able to work reliably under uncertainty and forgo the need to accurately estimate parameters in environments with impulsive noise and non-stationary clutter. In the second part, a brief outlook on promising future applications is given.

%------------------------------------------------------------------------------%
\subsection[Current Application: Ground Penetrating Radar]{Current Application: Ground Penetrating Radar\footnote{This is joint work with Fauzia Ahmad (Temple University, PA, USA).}}
\label{ssec:gpr}
%------------------------------------------------------------------------------%
GPR is a nondestructive method that uses electromagnetic radiation to detect buried targets \cite{Daniels1996, Zoubir2002_Sensors}. It can detect metallic as well as non-metallic targets having dielectric properties different from the background medium. A forward-looking GPR (FL-GPR) offers the advantage of a reduced risk to the operator and reduced risk of damaging the target \cite{Dogaru2012, Phelan2013, Comite2017, Comite2018}. However, a challenge of using FL-GPR is that the illuminating signals and the reflected signals experience substantial attenuation owing to the near cancellation of the direct and ground-reflected waves. Furthermore, the interface roughness and subsurface clutter, which are usually highly non-stationary, have a strong impact on FL-GPR performance. In order to reduce detection errors, these effects need to be compensated with an appropriate signal processing method.

\begin{figure}[tb]
	\centering
	\includegraphics[width=\columnwidth]{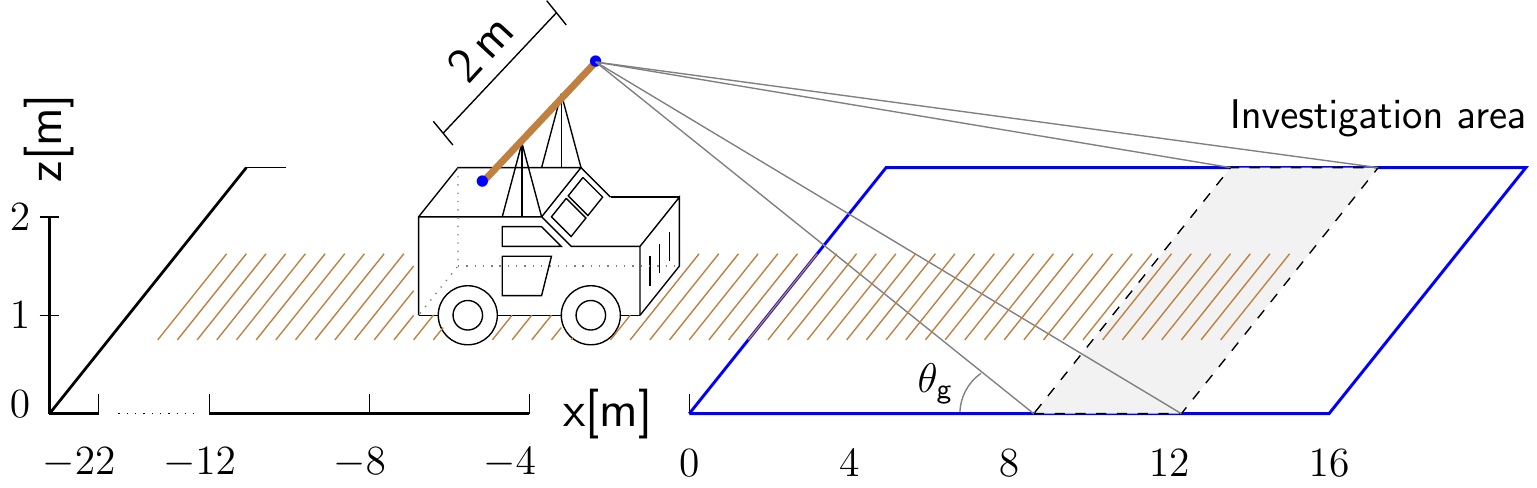}
	\caption{FL-GPR vehicle-based system with antenna array mounted on top. Brown lines on the ground indicate sensor array positions.}
	\label{fig:FLGPR}
\end{figure}

The numerical data used in this example were simulated using the Near-Field Finite-Difference Time-Domain software package, NAFDTD, developed by the U.S.~Army Research Laboratory (ARL) \cite{Dogaru2012}. Fig.~\ref{fig:FLGPR} illustrates the FL-GPR system with an antenna array mounted on top of a vehicle. In order to sense the investigation area, indicated by the blue rectangle in Fig.~\ref{fig:FLGPR}, the radar platform successively moves along the x-direction, starting at \SI{-22}{\meter}.

\begin{figure*}[tb]
	\centering
	\includegraphics[width=0.8\textwidth]{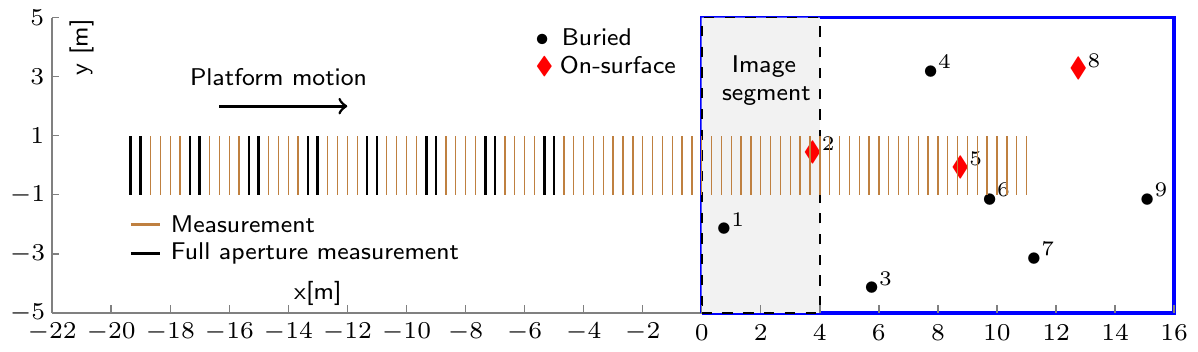}
	\caption{Top view of the FL-GPR measurement. The blue rectangle indicates the investigation area containing nine targets, either buried or placed on the ground surface.}
	\label{fig:measurement}
\end{figure*}

Fig.~\ref{fig:measurement} depicts the measurement configuration considered in the simulation. The investigation area is a rough surface environment containing nine targets. Six targets are buried at a depth of \SI{3}{\centi\meter}, five of which are metallic \{1, 3, 4, 6, 7\} and one is made of plastic \{9\}. The remaining targets, two plastic ones \{2, 8\} and a metallic one \{5\}, are placed on the surface. The ground is modeled as a dielectric medium which is non-dispersive, non-magnetic, and homogeneous. The surface roughness is described as a two-dimensional Gaussian random process; see \cite{Pambudi2020} for more details on the sensing setup.

A total of 90 array positions spaced $\Delta$x = \SI{0.33}{\meter} apart are considered, whose projections on the x-y plane are depicted as parallel brown lines in Fig.~\ref{fig:measurement}. An image segment is obtained by integrating eight full aperture measurements. Fig.~\ref{fig:measurement} indicates the eight measurements between \SI{-19.33}{} and \SI{-5}{\meter} used to image the first segment (dashed rectangle in Fig.~\ref{fig:measurement}). The full tomographic image is constructed by combining \num{4} segments, integrating \num{32} full aperture measurements in total. In order to improve the detection performance, ten additional images are generated by successively moving the radar closer to the target area. For more details on the multiview tomographic imaging approach, the reader is refered to \cite{Comite2017}.

\begin{figure}[tb]
	\centering
	\includegraphics[width=0.95\columnwidth]{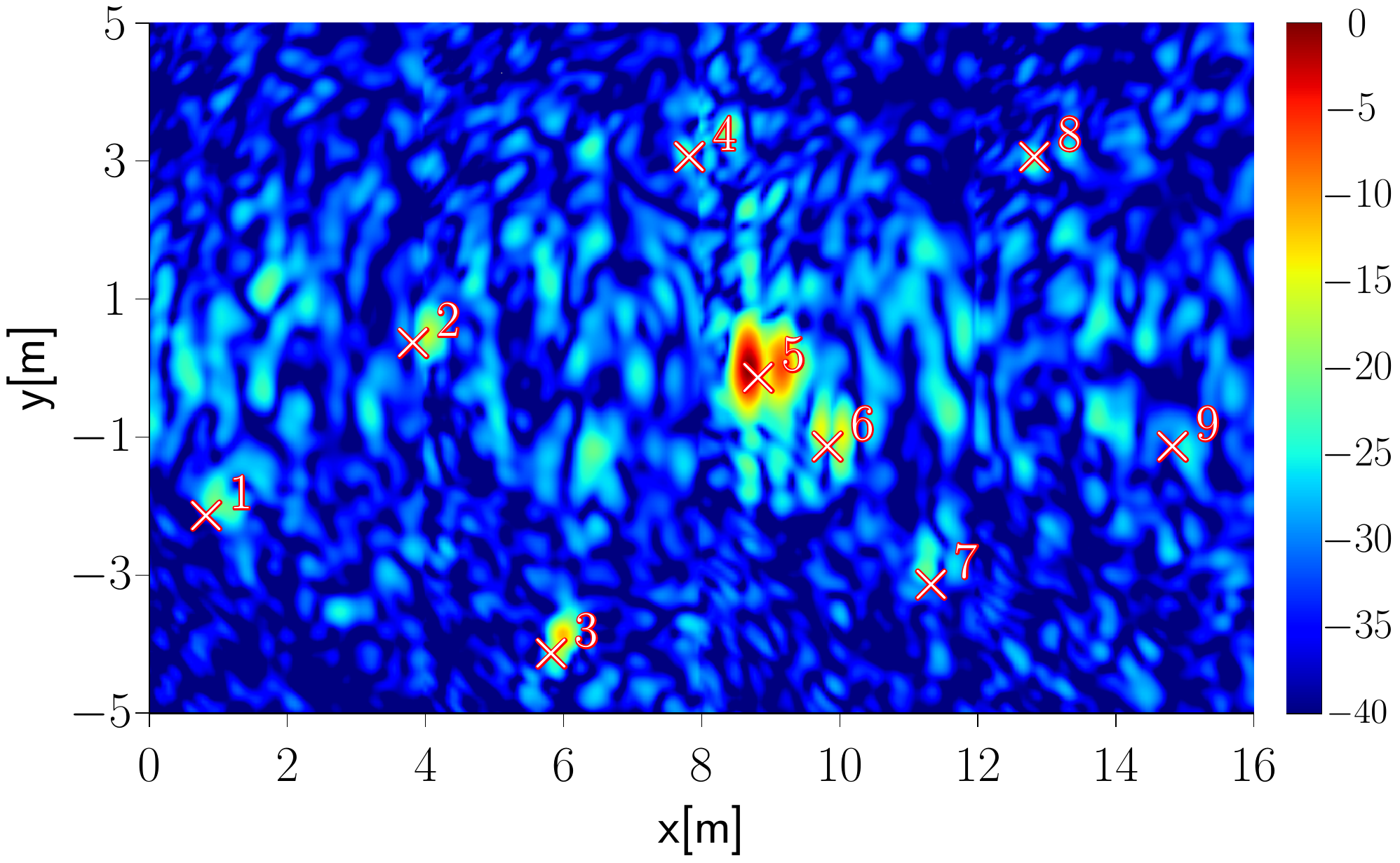}
	\caption{Normalized tomographic reconstruction of the scene in Fig.~\ref{fig:measurement} in dB scale.}
	\label{fig:tomographic}
\end{figure}

An example for a tomographic image obtained this way is shown in Fig.~\ref{fig:tomographic}. It is normalized to \SI{40}{\dB} dynamic range and consists of $1153$ pixels in downrange and $721$ pixels in crossrange with a resolution of \SI{5}{\centi\meter}. The targets located near the boundaries of the investigation area are less visible since they are outside of the mainlobe of the antenna array, and, compared to the targets placed on the surface, the buried targets are more challenging to detect due to the clutter caused by the radar back-scatter from the rough ground surface. 

\begin{figure}[tb]
	\centering
	\includegraphics[scale=0.9]{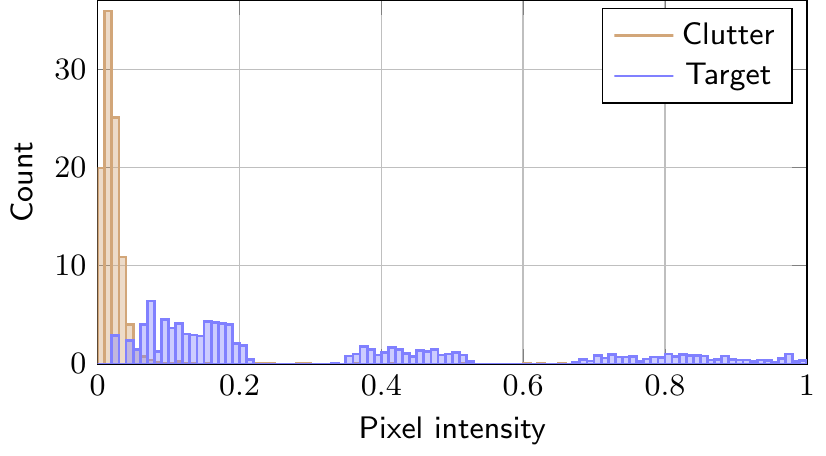}\\
	\includegraphics[scale=0.9]{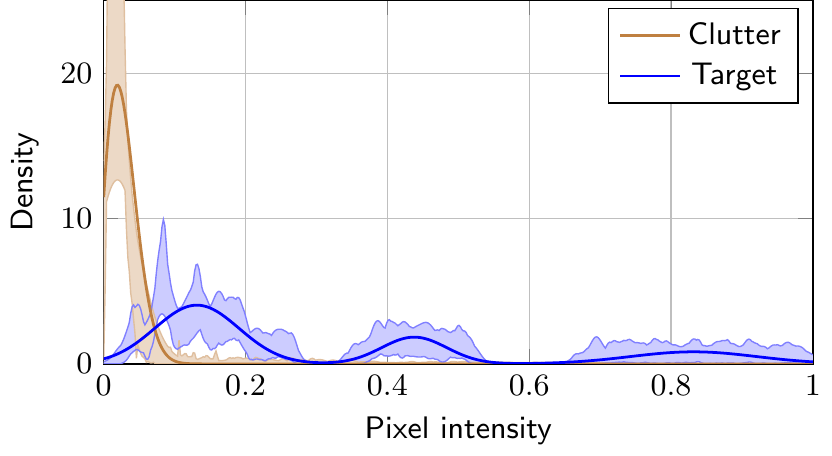}\\
	\caption{Histograms (top) and parametric density estimates with confidence intervals (bottom) of the return intensity of clutter and noise pixels.}
	\label{fig:histogram}
\end{figure}

Several methods to overcome the challenges of FL-GPR have been proposed \cite{Sun2003, Sun2005, Wang2007, Ton2010, Jin2012, Liao2014}. The approach followed here is based on a pixel-wise likelihood-ratio test to detect targets in the image domain \cite{Comite2017, Comite2018}. In order to model the distribution of the return intensity of targets and clutter, two training images were generated using the NAFDTD software. The first image is clutter-free, meaning it only contains target pixels; the second image is composed of only clutter pixels. In this way, two sets of training data are obtained that can be used to construct a probabilistic model under each hypothesis. The histograms of the training data sets are shown in the top plot of Fig.~\ref{fig:histogram}.

A common way of obtaining a nominal model is to fit a parametric family of distributions to the training data. The authors of \cite{Comite2017, Debes2009, Liao2012} showed that the clutter pixels are approximately Rayleigh distributed, while the distribution of the target pixels can be approximated by a Gaussian mixture distribution. The fitted densities corresponding to this nominal model are indicated by the solid lines in the bottom plot of Fig.~\ref{fig:histogram}.

While the nominal model appears to be a good fit in this case and has the advantage of being clean and simple, it is highly idealized and does not guard against deviations from the fitted distributions. Minimax robust detection, in combination with the uncertainty models introduced in Sec.~\ref{ssec:uncertainty_models}, provides a way of keeping the model complexity low, while still taking the inherent uncertainty in the fitted distributions into account. 

For this example, two uncertainty models were considered, namely, the density band model and an $\varepsilon$-contamination model. The density band model was obtained by bootstrapping the target and clutter densities from the training data using a Gaussian kernel density estimator and taking the lower and upper envelope of the bootstrap estimates to be the upper and lower density bounds; see \cite{Pambudi2019_EUSIPCO} for more details. The resulting uncertainty band is depicted in the bottom plot of Fig.~\ref{fig:histogram}. One of the advantages of the band model becomes apparent here, namely, that the amount of uncertainty can vary locally. For example, while the uncertainty band of the clutter intensity distribution is wide around its peak at small intensities, it becomes vanishingly narrow towards the tail of the distribution. This means that there is uncertainty about the true height and shape of the peak, but there is little uncertainty about gross outliers in the clutter data. 

However, since the density bands were bootstrapped from the training data, the question arises whether this uncertainty model really reflects reality or just reproduces characteristics of the particular realizations used for training. In order to see in how far the density model might be biased towards the training data, a simple $\varepsilon$-contamination model with an outlier rate of $\varepsilon = 0.4$ was considered as well. Both minimax detectors were then evaluated under a mixed scenario, with some segments of the ground admitting the same surface roughness as the training data and other areas admitting different surface roughness values. This is in line with the non-stationarity of the clutter returns that can be observed in practice. 

\begin{figure}[tb]
	\centering
	\includegraphics[width=0.9\columnwidth]{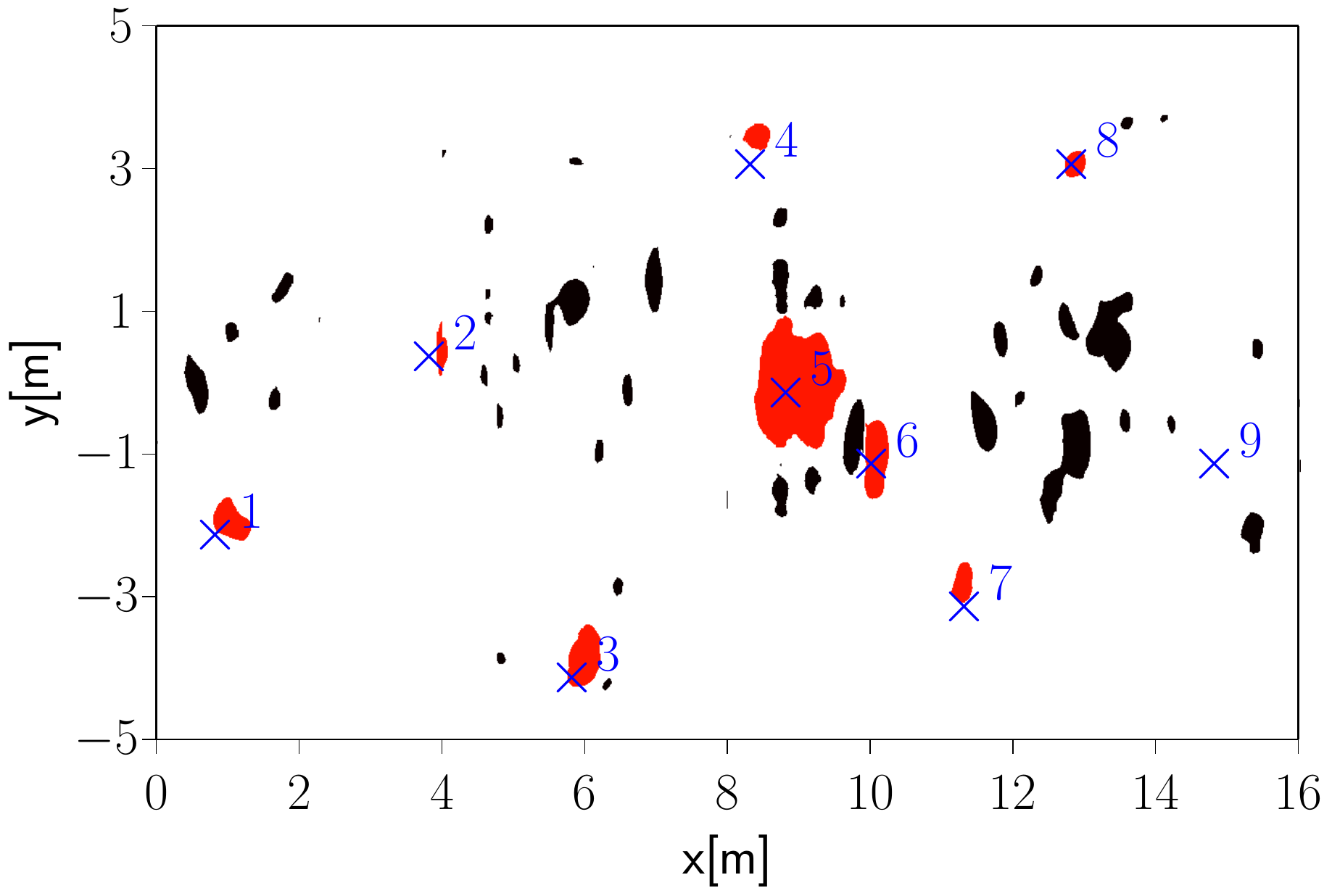}\\
	\includegraphics[width=0.9\columnwidth]{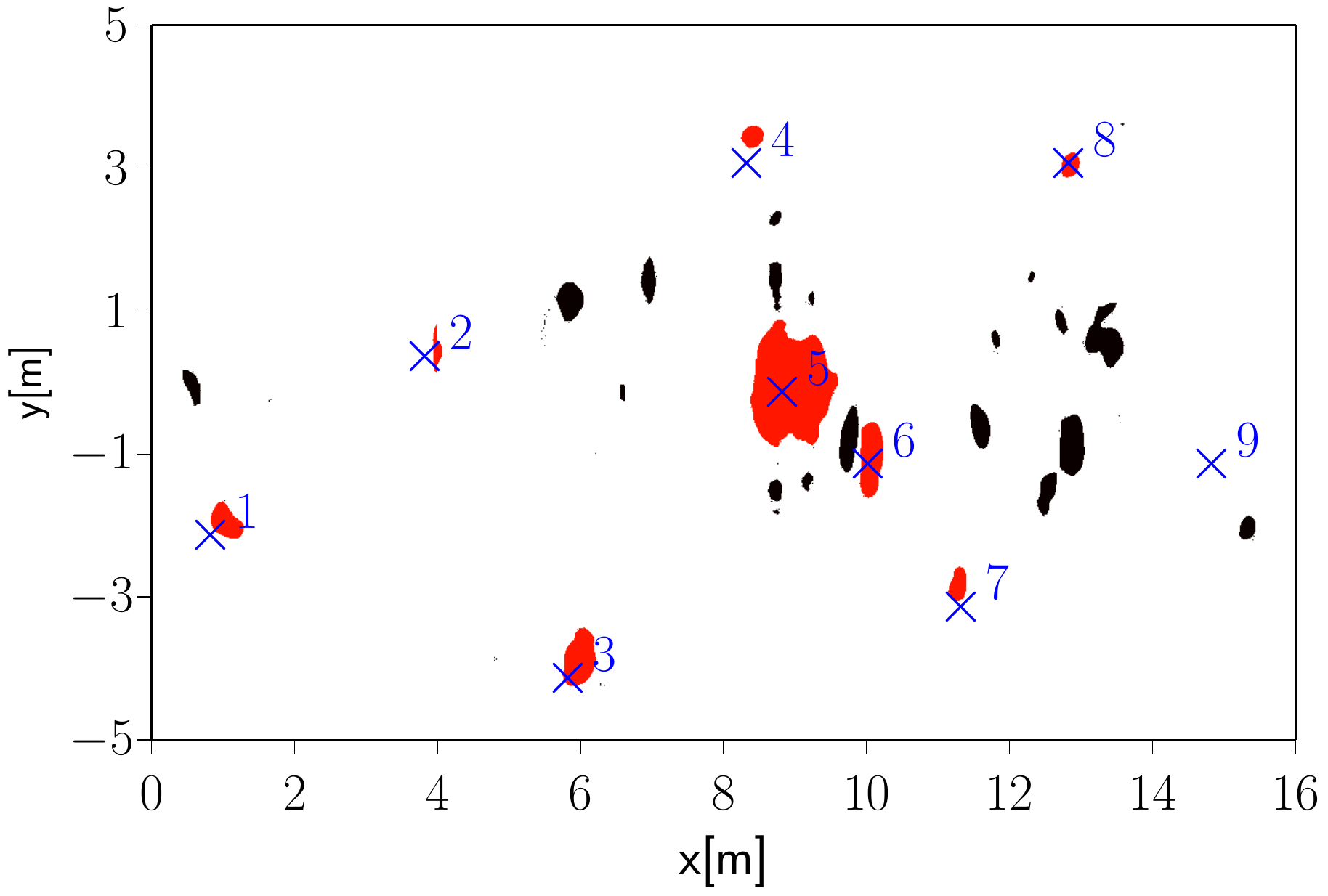}\\
	\includegraphics[width=0.9\columnwidth]{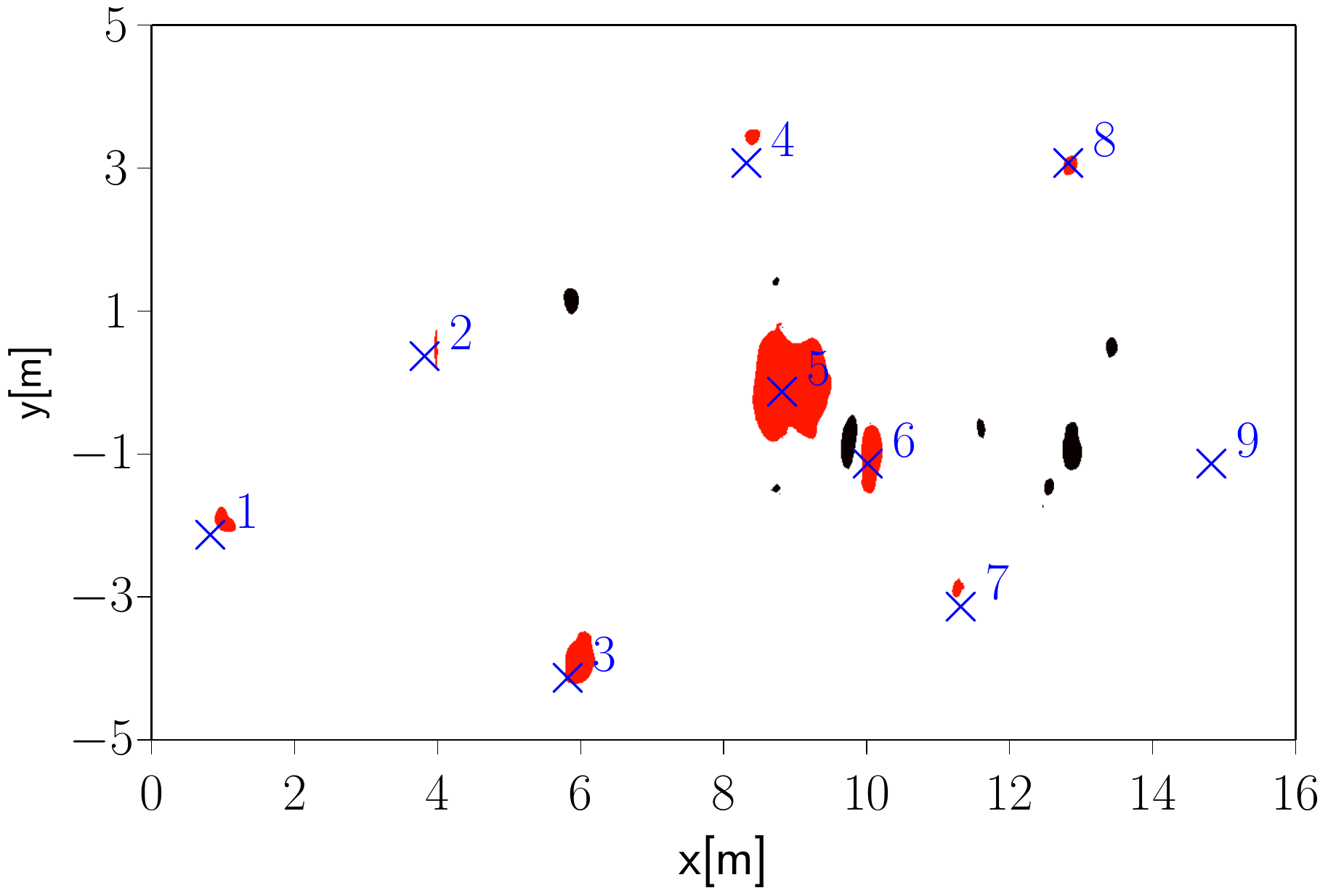}
	\caption{Detection results based on the tomographic reconstruction depicted in Fig.~\ref{fig:measurement}. By inspection, there is a notable performance improvement when replacing the standard likelihood ratio detector (top) with a minimax robust detector based on a density band (middle plot) or an $\varepsilon$-contamination (bottom plot) uncertainty model.}
	\label{fig:gpr_detection_results}
\end{figure}

The detection results of the standard likelihood ratio test based on the nominal model as well as the minimax tests under the density band and $\varepsilon$-contamination uncertainty are depicted in Fig.~\ref{fig:gpr_detection_results}. It can clearly be seen how incorporating uncertainty in the model reduces the number of false alarms, here indicated in black. The nominal test is too quick to declare a pixel with even medium return intensity a target, while the two robust tests require stronger evidence. Note that all tests have been designed such that they admit roughly the same detection rate in order to make for a fair comparison. Of course, the false alarm rate of the nominal test could be decreased by increasing the decision threshold, however, this would come at the expense of a decreased detection rate. Overall, the minimax test under $\varepsilon$-contamination performs best in this example, admitting the lowest false alarm rate. However, it can be seen that the outlier ratio of \SI{40}{\percent}, which was deliberately chosen large, is starting to affect the sensitivity of the test. The red regions around targets 4 and 7, for example, are visibly smaller than they are for the other two tests, meaning that these targets are likely to be missed under slightly less favorable conditions. This effect is to be expected. On the one hand, the outlier model allows for larger deviations from the nominal model, which makes it less sensitive to the choice of training data and in turn enables it to deal with the mixed surface areas better than the fitted density band model. On the other hand, assuming a large share of the data to be potential outliers causes the test to be exceedingly cautious. While the test has not been over robustified yet, recall the discussion in Sec.~\ref{ssec:design_and_implementation}, one would be well advised not to increase $\varepsilon$ any further.

In general, the question if and how training data should be used to fit uncertainty models has not been answered satisfactorily yet. On the one hand, training data provide useful information that should be taken into account, on the other hand, the goal of robust statistics is to work well even under conditions that are not typical, have not been observed before, or change too frequently to produce reliable training data. An interesting approach proposed in \cite{Gul2017_minimax_robust} is to combine two uncertainty models. For example, an $f$-divergence ball could be used to determine the least favorable distribution \emph{according to the training data}, which is then used as the nominal distribution in an outer $\varepsilon$-contamination model. In any case, further exploring and developing approaches to a data-driven design of uncertainty sets is a highly relevant topic for future research, which is likely to have direct implications on how robust detectors are and will be used in practice.

%------------------------------------------------------------------------------%
\subsection{Future Applications: Biomedical Engineering and Adversarial Machine Learning}
\label{ssec:future_applications}
%------------------------------------------------------------------------------%
GPR is a good example for an area where robust detectors can be applied today, as plug-in replacements for standard detectors. Beyond these areas, there are applications that are too complex for this simple plug-in method to work, yet are likely to benefit from robust detectors and the techniques underlying their design. Two of those, namely biomedical engineering and adversarial machine learning, are briefly introduced in this section.

Detection and classification problems in \emph{biomedical engineering} \cite{Basu2007, Han2013, Schaeck2018, Zoubir2018} typically show two characteristics that make them natural candidates for the application of minimax techniques: small sample sizes, with generating new samples often being either impossible or requiring elaborate experiments, and uncertainty about both the underlying biological processes as well as the quality of the measurements \cite{Basu2007, Muma2019}. Moreover, the observations often cannot be assumed to be independent and abnormalities have to be detected both reliably and quickly, which motivates the use of robust sequential detectors. Research on robust and efficient methods in this area is ongoing.

In \emph{adversarial machine learning} \cite{Tygar2011, Schaefer2015, McDaniel2016, Garnaev2016}, an adversary is assumed to generate samples with the aim of deceiving an algorithm or system that takes these samples as an input. In the context of detection and classification, the aim of the adversary is to be able to control the output of the detector as much as possible, irrespective of the underlying ground truth. This corresponds to a classic minimax scenario, the only difference being that the uncertainty set is interpreted as the set of actions available to the adversary. Interestingly, in several recent works on learning under adversarial settings \cite{Dziugaite2015, Nowozin2016, Ghasemipour2019}, $f$-divergences and other similarity measures are used as surrogate objective functions, meaning that the adversary's aim of mimicking a certain behavior is formulated as the problem of minimizing a certain $f$-divergence. The advantage of this approach is that no strong assumptions about the underlying detector need to be made. The disadvantage is that there are no rigorous guidelines which $f$-divergence should be used in the design/training process. We conjecture that the lessons learned in minimax detection can help to overcome this problem in the sense that, given the hypotheses and the action/uncertainty sets, the ``correct'' $f$-divergence can be identified, that is, the one corresponding to the true minimax optimal detector. In this way, the $f$-divergence loses its character of being merely a surrogate objective, while at the same time existing training procedures can be used. In other words, an optimal detector can be used for the training, only by carefully choosing the divergence measure and without actually implementing the detector. 

Another potentially fruitful area is to explore the connection between robust detectors and machine learning procedures in the other direction, that is, to answer the question of how machine learning techniques can be leveraged for the design of robust hypothesis tests. In particular, the implicit definition of the optimal cost function $\rho$ in \eqref{eq:optimal_cost_sequential} and \eqref{eq:cost_continue}, can be shown to correspond to a Bellman equation with state vector $\bm{z}$, a discount factor of $1$ (no discount), and a piecewise constant reward function. This correspondence suggests that reinforcement learning techniques \cite{Kar2013, Sutton2018, Mohri2018} can be used to learn the optimal dissimilarity measure $\rho$ from given sets of training data or even in an online manner. Since $\rho$ completely characterizes the minimax optimal test, this would be sufficient for an implementation in practice. Moreover, the spectral representation of $f$-divergences discussed in Sec.~\ref{ssec:characterizing_lfds} shows that the latter can be decomposed into superpositions of functions of the form $f(t) = \max\{0, \lambda t\}$. Functions of this type are known as \emph{rectifiers} in the machine learning literature and are among the most commonly used activation functions in (deep) neural networks. This suggests neural networks with rectifier activation functions as a natural way of implementing and learning complex divergence measures. However, as of now, it is not clear if and how this approach would extend to multidimensional divergences, such as $f$-dissimilarities. Nevertheless, we conjecture that robust tests based on a machine learning aided design could provide highly useful detectors that combine the power of modern learning techniques with the strict robustness properties of classic minimax procedures.

These considerations already point towards a fundamental \emph{limit} of minimax robust detectors in practice: the more complex problem formulations, cost functions, and uncertainty models are required, the less likely it becomes that the minimax approach can be followed in an uncompromising manner. While advances in machine learning and approximation theory will continue to push this limit, there certainly is a need for robust detection that goes beyond the minimax principle. A brief outlook on this topic is given in the next section.

%%%%%%%%%%%%%%%%%%%%%%%%%%%%%%%%%%%%%%%%%%%%%%%%%%%%%%%%%%%%%%%%%%%%%%%%%%%%%%%%
\section{Beyond Minimax Robust Detection}
\label{sec:beyond_minimax}
%%%%%%%%%%%%%%%%%%%%%%%%%%%%%%%%%%%%%%%%%%%%%%%%%%%%%%%%%%%%%%%%%%%%%%%%%%%%%%%%

The discussion in the previous sections showed that minimax robust detection has its distinct strengths and weaknesses. On the positive side, it is a well-motivated, principled approach that results in detectors with strict performance guarantees over large, non-parametric families of distributions. Moreover, a variety of tractable uncertainty models have been shown to exist, making minimax detectors flexible enough to guard against gross outliers, subtle model mismatches, and violations of common assumptions, such as symmetry, zero-mean, or Gaussianity. In addition, at least for the binary case, minimax detectors can often be implemented by simply clipping, compressing, or censoring the test statistic of the optimal non-robust test, making them a low-complexity, drop-in replacement. On the other hand, the design of minimax detectors can quickly become prohibitively complex when dealing with high-dimensional distributions, sequential data acquisition, or multiple hypotheses, all of which are commonly encountered in practice. Hence, it is safe to say that, for the foreseeable future, there will remain many applications for which minimax robust detectors simply do not exist---be it because they are not known, such as the minimax fixed-sample-size detector for multiple hypotheses, or because their design requires excessive computational resources, such as the minimax sequential detector. 

The fact that minimax detectors run into complexity limitations at some point, however, does not imply that no robust detectors exist or can be designed for such scenarios. Besides purely heuristic or highly specialized, application driven approaches, there are several avenues that can followed to extend the range of application of robust detectors. 

First, instead of looking for strictly minimax detectors, one can relax this requirement and look for \emph{asymptotically} minimax detectors, whose design is usually significantly simpler. Without going into details, the reason for this is that in the asymptotic regime, that is, for very large sample sizes, the error probabilities of any detector are determined by the KL divergences of the distributions; recall the discussion in Sec.~\ref{ssec:design_and_implementation}. As a consequence, the asymptotically least favorable distributions no longer have to minimize \emph{all} $f$-divergences (or $f$-dissimilarities in the multi-hypothesis case), but only a single one, namely the KL divergence (or a weighted sum of KL divergences). This makes finding least favorable distributions and in turn designing robust tests significantly simpler. Several examples of this approach, especially for sequential detectors, can be found in the literature \cite{Holm1975, Dragalin1988, Martin1992, Basu1998, Gao2018, Goldenshluger2015, Juditsky2015, Cao2017}. 

However, resorting to asymptotic measures of robustness is not without disadvantages. First, since asymptotically minimax detectors are still based on least favorable distributions, they inherit their ancestors' limitations when it comes to high dimensional data and non-standard uncertainty models. Second, as briefly discussed in the FAQ in Sec.~\ref{ssec:design_and_implementation}, there are conceptual problems with asymptotic minimax robustness. In the non-asymptotic sample size regime, there are no strict performance guarantees, while in the asymptotic regime the sample size is likely to be large enough to reduce the uncertainty to a point where robust methods stop being attractive.

An alternative to the asymptotic approach, which is of a similar spirit, is to extend the techniques that have been identified in minimax robust detection to more complex scenarios. For example, the idea of clipping, compressing or censoring the test statistic in order to make a detector robust with respect to outliers or model uncertainties often shows excellent results in practice, even if the techniques are no longer provably optimal. It is clear, however, that such detectors, although being well-motivated, are based on a speculative generalization. Nevertheless, as of today, transforming the test statistic of an existing detector is often the most viable option, given that it is easy to implement, theoretically backed and well-tried in practice. 

The idea of transforming a nominal test statistic is reminiscent of a popular class of robust \emph{estimators}, namely, Huber's M-estimators \cite{Huber1981, Zoubir2018, Zoubir2012}. The estimates obtained by M-estimators can be interpreted as maximizers of a transformed version of the log-likelihood function. Huber's loss function, for example, corresponds to the regular squared loss with a clipped derivative. More elaborate loss functions can be found in the literature. In general, there are clear guidelines what properties a loss function needs to admit such that the resulting estimator is robust. Moreover, a variety of measures exist to make the notion of robustness precise, such as the breakdown point, the influence function, and the sensitivity curve, to name just a few. This split between flexible yet theoretically-backed design guidelines on the one hand and independent yet rigorous and well-motivated robustness measures on the other hand has proved highly useful in robust estimation. 

In robust detection, the situation is very different. Although various design guidelines and robustness measures have been proposed in the literature, none of them---except for the minimax approach---is as well-studied, as useful in practice, and as universally accepted as their counterparts in robust estimation. In order to close this gap and to make progress towards a more unified theory of robust statistics, researchers in robust detection will need to find answers to questions such as: What is the detection equivalent of an influence function? Can ideas similar to those underlying, for example, the class of M-estimators be applied in detection as well? And how can complexity constraints be incorporated in the design process? Questions like these may sound obvious, but are hard to answer or even formulate. Often enough, although strong connections between estimation and detection clearly exist, the different nature of the two problems makes a straightforward transfer of concepts difficult; recall the example of censoring discussed in Sec.~\ref{ssec:design_and_implementation}. 

Based on the discussion in the previous chapters, we conjecture that statistical similarity measures might offer a possible path towards a more unified theory of robust detection and estimation. In particular, defining distances on the distribution space instead of the parameter space makes it possible to formulate detection and estimation problems within the same framework, with the KL divergence playing the role of a ``nominal'' distance, in analogy to the squared error loss in estimation. An M-procedure, for example, could then be defined by choosing different, robustness inducing distances. Results pointing in this direction have been published throughout the last decades, however, they remain somewhat isolated and have not been unified into a coherent, canonical framework \cite{Poor1980_Distance, Park2003, Toma2010, Toma2011, Unnikrishnan2011a}. An interesting connection between Bayesian and minimax inference based on Wasserstein distance uncertainty sets has recently been shown in \cite{ShafieezadehAbadeh2018} and \cite{Nguyen2019}. Some preliminary results on robust joint detection and estimation can be found in \cite{Reinhard2016, Reinhard2018}. 

In order to illustrate how similarity measures bridge different areas of statistics, consider the example of the KL divergence. In information theory, it provides a measure for the randomness or predictability of a random variable in relation to a reference random variable. Hence, it usually goes by the name \emph{relative entropy} in this context. In detection, it determines the asymptotic rate at which the error probabilities decrease when the sample size increases; compare \eqref{eq:err_exp_0} and \eqref{eq:err_exp_1}. In estimation, the KL divergence between two distributions of the same parametric family is locally approximated by the Fisher information, which provides a fundamental bound on the accuracy of estimators. Moreover, the KL divergence has been shown to be closely related to the mean square error when estimating a random variable in additive Gaussian noise.

The latter connection is worth exploring in some more detail. To this end, consider the additive Gaussian channel
\begin{equation*}
  Y_\gamma = \sqrt{\gamma}X + W,
\end{equation*}
where $\gamma > 0$ denotes the SNR and $X \sim P$, $W \sim \mathcal{N}(0, 1)$. The KL divergence can now be written as \cite{Guo2013}
\begin{equation}
  D_\text{KL}(P \Vert Q) = \frac{1}{2} \int_0^\infty \mse_Q(P, \gamma) - \mse_P(P,\gamma) \, \mathrm{d}\gamma,
  \label{eq:immse}
\end{equation}
where
\begin{equation}
  \mse_Q(P, \gamma) = \mathbb{E}_P\Bigl[ \left( X - E_Q\bigl[X \vert Y_\gamma \bigr] \right)^2 \Bigr].
\end{equation}
That is, the KL divergence can be obtained by accumulating the increase in the mean squared error (MSE) when using a mismatched MMSE estimator over all SNR values. The identity in \eqref{eq:immse} is reminiscent of the spectral representation of $f$-divergences in \eqref{eq:f-div_spectrum}. In fact, the latter has a similar interpretation: the weighted total variation distance in \eqref{eq:weighted_tv_distance} can be shown to correspond to the decrease in the Bayesian error of a detector when making a decision based on the posterior distributions instead of the prior distributions. Hence, in both detection and estimation there exist representations that express the KL divergence as an accumulated difference in performance of an optimal and a sub-optimal procedure. Finally, using the standard definition of entropy, $H(P) = -E_P[\log p(X)]$, and cross-entropy, $H(P, Q) = -E_Q[\log p(X)]$, one can arrive at the following triangle:
\begin{gather*}
  H(P) - H(P,Q) \\
  \mathbin{\rotatebox[origin=c]{45}{$=$}} \quad \mathbin{\rotatebox[origin=c]{-45}{$=$}} \\
  \frac{1}{2} \int_0^\infty \Delta \mse_Q(P,\gamma) \, \mathrm{d}\gamma \; = \; 
  \int_0^\infty D_\text{TV}(P \Vert \lambda^{-1} Q) \, \mathrm{d}\lambda
\end{gather*}
This triangle connects central measures in information theory (the Shannon entropy), estimation theory (the MMSE), and detection theory (the Bayes error) with the KL divergence being the central linchpin. Similar connections can be shown to exist around other divergence measures as well \cite{Guo2013, Sason2018}.

These and related findings suggest that convex similarity measures may provide a solid foundation for a unified theory of robustness in detection, estimation, and information theory, where ``least favorable'' seamlessly translates to ``most similar'' with respect to a certain divergence, the latter inherently defining a performance and robustness measure. 

In summary, although there has been steady progress in both areas, there are still lose ends in robust detection and estimation and future research is called for to identify ways of tying them together, both in theory and practice.

%%%%%%%%%%%%%%%%%%%%%%%%%%%%%%%%%%%%%%%%%%%%%%%%%%%%%%%%%%%%%%%%%%%%%%%%%%%%%%%%
\section{Conclusion}
\label{sec:conclusion}
%%%%%%%%%%%%%%%%%%%%%%%%%%%%%%%%%%%%%%%%%%%%%%%%%%%%%%%%%%%%%%%%%%%%%%%%%%%%%%%%

The paper has discussed classic results as well as recent advances in minimax robust detection. After having introduced the minimax principle, the two-hypothesis case has been studied in detail. Three criteria for the characterization of least favorbale distributions have been presented and discussed. In this context, the importance of $f$-divergences in robust detection has been highlighted and explained. Three types of uncertainty models have been presented in more detail, namely $\varepsilon$-contamination, probability density bands, and $f$-divergence balls, and it has been shown how their properties translate to clipping, censoring, or compressing the test statistic of the corresponding minimax detectors. The second part of the paper has been concerned with robust testing for multiple hypotheses (classification), starting with a discussion of why it poses a much more challenging problem than the binary case. Sequential detection has then been introduced as a technique that enables strictly minimax optimal tests in the multi-hypothesis case. Finally, the usefulness of robust detectors in practice has been demonstrated using the example of ground penetrating radar and an outlook on robust detection beyond the minimax principle has been given. In conclusion, robust tests for two hypotheses are well-researched, yet arguably underused in practice, while robust tests for multiple hypotheses are an active area of research with potential future applications in many areas of applied statistics.

%%%%%%%%%%%%%%%%%%%%%%%%%%%%%%%%%%%%%%%%%%%%%%%%%%%%%%%%%%%%%%%%%%%%%%%%%%%%%%%%
\section*{Acknowledgment}
%%%%%%%%%%%%%%%%%%%%%%%%%%%%%%%%%%%%%%%%%%%%%%%%%%%%%%%%%%%%%%%%%%%%%%%%%%%%%%%%

The authors would like to thank the Associate Editor and the anonymous reviewers for their constructive feedback which greatly helped to improve the quality of the paper.

%%%%%%%%%%%%%%%%%%%%%%%%%%%%%%%%%%%%%%%%%%%%%%%%%%%%%%%%%%%%%%%%%%%%%%%%%%%%%%%%
\bibliographystyle{IEEEtran}
\bibliography{bibliography}
%%%%%%%%%%%%%%%%%%%%%%%%%%%%%%%%%%%%%%%%%%%%%%%%%%%%%%%%%%%%%%%%%%%%%%%%%%%%%%%%

\begin{IEEEbiography}[{\includegraphics[width=1in, height=1.25in, clip, keepaspectratio]{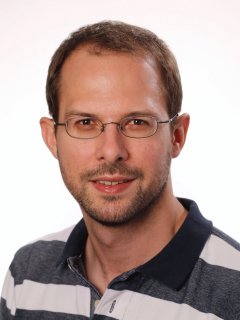}}]{Michael Fau{\ss}}
  Michael Fau{\ss} received the Dipl.-Ing.~degree from Technische Universit\"at M\"unchen in 2010 and the Dr.-Ing.~degree from Technische Universit\"at Darmstadt in 2016, both in electrical engineering. In November 2011, he joined the Signal Processing Group at Technische Universit\"at Darmstadt, and in 2017 he received the dissertation award of the German Information Technology Society for his PhD thesis on robust sequential detection. In September 2019 he joined Prof.~H.~Vincent Poor's group at Princeton University as a postdoc on a research grant by the German Research Foundation (DFG). His current research interests include statistical robustness, sequential detection and estimation, and the role of similarity measures in statistical inference.
\end{IEEEbiography}

\begin{IEEEbiography}[{\includegraphics[width=1in, height=1.25in, clip, keepaspectratio]{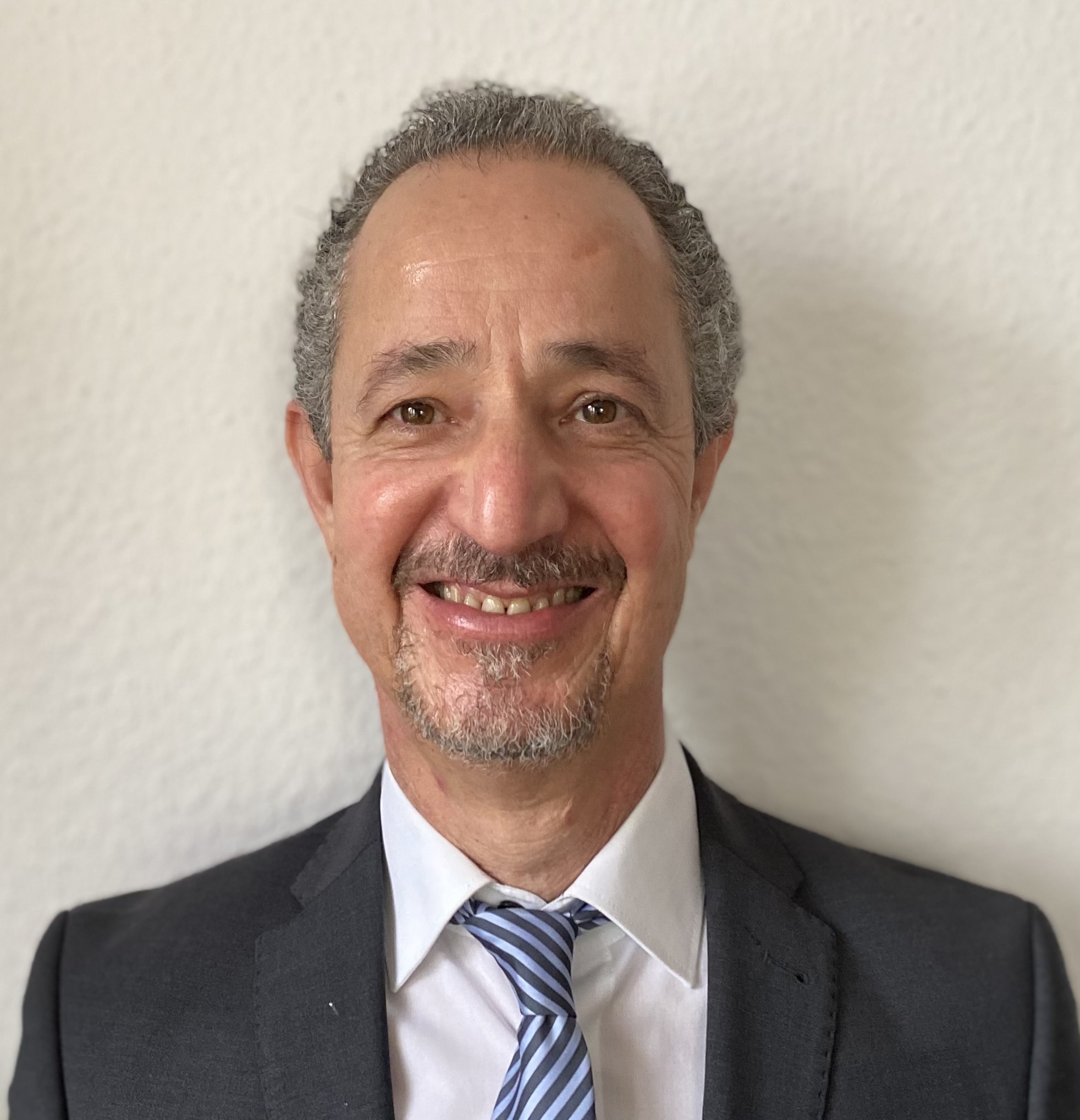}}]{Abdelhak M.~Zoubir}
  Abdelhak M.~Zoubir received the Dr.-Ing.~degree in Electrical Engineering from Ruhr-Universit\"at Bochum, Germany, in 1992 and has been Professor of Signal Processing and Head of the Signal Processing Group at Technische Universitä{\"a}t Darmstadt, Germany since 2003. His research interest lies in statistical methods for signal processing with emphasis on bootstrap techniques, robust detection and estimation and array processing applied to radar, sonar, telecommunications, automotive monitoring and safety, and biomedicine. He published over 400 journal and conference papers on the above areas. Dr.~Zoubir served as General Chair and Technical Chair of numerous international IEEE conferences and workshops, most notably as Technical Chair of ICASSP-14. He also served on publication boards of various journals, notably as Editor-in-Chief of the IEEE Signal Processing Magazine (2012-2014). Dr.~Zoubir was the Chair (2010-2011) of the IEEE Signal Processing Society (SPS) Technical Committee Signal Processing Theory and Methods (SPTM) and served on the Board of Governors of the IEEE SPS as a Member-at-Large (2015-2017). He was the president of the European Association of Signal Processing (EURASIP) from 2017 to 2018. He is a Fellow of the IEEE and an IEEE Distinguished Lecturer (Class 2010-2011). He received several best paper awards, and the 2018 IEEE Leo L.~Beranek Meritorious Service Award.
\end{IEEEbiography}

%\newpage

\begin{IEEEbiography}[{\includegraphics[width=1in, height=1.25in, clip, keepaspectratio]{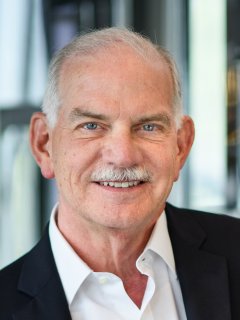}}]{H.~Vincent Poor}
  (S’72, M’77, SM’82, F’87) received the Ph.D.~degree in EECS from Princeton University in 1977. From 1977 until 1990, he was on the faculty of the University of Illinois at Urbana-Champaign. Since 1990 he has been on the faculty at Princeton, where he is the Michael Henry Strater University Professor of Electrical Engineering. From 2006 until 2016, he served as Dean of Princeton’s School of Engineering and Applied Science. He has also held visiting appointments at several other institutions, including most recently at Berkeley and Cambridge. His research interests are in the areas of information theory, machine learning and network science, and their applications in wireless networks, energy systems and related fields. Among his publications in these areas is the forthcoming book Machine Learning and Wireless Communications (Cambridge University Press, 2021).

  Dr.~Poor is a member of the National Academy of Engineering and the National Academy of Sciences, and is a foreign member of the Chinese Academy of Sciences, the Royal Society and other national and international academies. He received the Technical Achievement and Society Awards of the IEEE Signal Processing Society in 2007 and 2011, respectively. Recent recognition of his work includes the 2017 IEEE Alexander Graham Bell Medal, the 2019 ASEE Benjamin Garver Lamme Award, a D.Sc.~honoris causa from Syracuse University, awarded in 2017, and a D.Eng.~honoris causa from the University of Waterloo, awarded in 2019.
\end{IEEEbiography}

\end{document}